%% file: master.tex
\documentclass[manuscript,screen,nonacm]{acmart} 

\usepackage{mdframed}
\usepackage{layouts}
\usepackage{todonotes}
\usepackage{pdfpages}
\usepackage{soul}
\mdfsetup{skipabove=1em,skipbelow=1.5em}
\newmdenv[topline=false,rightline=false,bottomline=false,linewidth=4pt,linecolor=lightgray]{highlightbox}
\newmdenv[linewidth=0pt,linecolor=white,innertopmargin=9,innerbottommargin=9,backgroundcolor=lightgray]{searchbox}

\newcommand{\qref}[1]{\hyperref[tab: guiding questions]{#1}}

\setcopyright{none}
\copyrightyear{}
\acmYear{}
\acmDOI{}

\begin{document}

\title{EarXplore: An Open Research Database on Earable Interaction}

\author{Jonas Hummel}
\email{jonas.hummel@kit.edu}
\orcid{0009-0005-8563-6175}
\affiliation{%
  \institution{Karlsruhe Institute of Technology}
  \city{Karlsruhe}
  \country{Germany}
}

\author{Tobias R{\"o}ddiger}
\email{tobias.roeddiger@kit.edu}
\orcid{0000-0002-4718-9280}
\affiliation{%
  \institution{Karlsruhe Institute of Technology}
  \city{Karlsruhe}
  \country{Germany}
}

\author{Valeria Zitz}
\email{valeria.zitz@kit.edu}
\orcid{0009-0004-1158-861X}
\affiliation{%
  \institution{Karlsruhe Institute of Technology}
  \city{Karlsruhe}
  \country{Germany}
}

\author{Philipp Lepold}
\email{philipp.lepold@kit.edu}
\orcid{0009-0003-0391-588X}
\affiliation{%
  \institution{Karlsruhe Institute of Technology}
  \city{Karlsruhe}
  \country{Germany}
}

\author{Michael K{\"u}ttner}
\email{michael.kuettner@kit.edu}
\orcid{0009-0000-9021-0359}
\affiliation{%
  \institution{Karlsruhe Institute of Technology}
  \city{Karlsruhe}
  \country{Germany}
}

\author{Marius Prill}
\email{uwptv@student.kit.edu}
\orcid{0009-0007-3152-4383}
\affiliation{%
  \institution{Karlsruhe Institute of Technology}
  \city{Karlsruhe}
  \country{Germany}
}

\author{Christopher Clarke}
\email{cjc234@bath.ac.uk}
\orcid{0009-0004-1158-861X}
\affiliation{%
  \institution{University of Bath}
  \city{Bath}
  \country{United Kingdom}
}

\author{Hans Gellersen}
\email{h.gellersen@lancaster.ac.uk}
\orcid{0000-0003-2233-2121}
\affiliation{%
  \institution{Lancaster University}
  \city{Lancaster}
  \country{United Kingdom}}
   \affiliation{%
    \institution{Aarhus University}
   \city{Aarhus}
   \country{Denmark}
}

\author{Michael Beigl}
\email{michael.beigl@kit.edu}
\orcid{0000-0001-5009-2327}
\affiliation{%
  \institution{Karlsruhe Institute of Technology}
  \city{Karlsruhe}
  \country{Germany}
}

\renewcommand{\shortauthors}{Hummel et al.}

\begin{abstract}
    Interaction with sensor-augmented earphones, referred to as earables or hearables, represents a major area of earable research. Proximate to the head and reachable by hand, earables support diverse interactions and can detect multiple inputs simultaneously. Yet this diversity has fragmented research, complicating the tracking of developments. To address this, we introduce \textit{EarXplore}, a curated, interactive online database on earable interaction research. Designed through a question-centered approach that guided the development of 33 criteria applied to annotate 118 studies and the structure of the platform, \textit{EarXplore} comprises four integrated views: a Tabular View for structured exploration, a Graphical View for visual overviews, a Similarity View for conceptual links, and a Timeline View for scholarly trends. We demonstrate how the platform supports tailored exploration and filtering, and we leverage its capabilities to discuss gaps and opportunities. With community update mechanisms, \textit{EarXplore} evolves with the field, serving as a living resource to accelerate future research.

\end{abstract}

\ccsdesc[500]{General and reference~Surveys and overviews}
\ccsdesc[500]{Human-centered computing~Interaction devices}
\ccsdesc[500]{Information systems~Information retrieval systems}

\keywords{Earables, Hearables, Interaction, Interactive Online Database, Question-Centered Design Approach}
\begin{teaserfigure}
  \centering\includegraphics[width=1\textwidth]{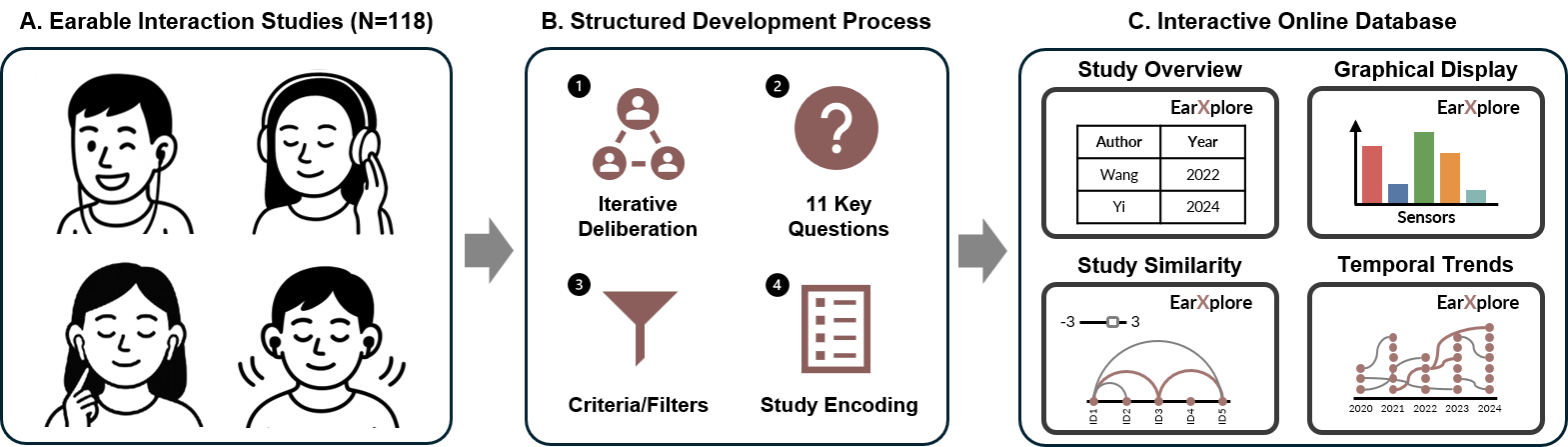}
  \caption{Overview of the \textit{EarXplore} development pipeline and platform. A. We identified 118 papers on earable interaction, covering a diverse range of input modalities such as head, hand, and facial gestures.; B. Iterative deliberation among the authors informed the derivation of key questions and corresponding criteria, which were then applied to systematically encode each study.; C. The resulting interactive platform, \textit{EarXplore}, supports exploration through four interconnected views (Tabular, Graphical, Similarity, Timeline), each offering a distinct perspective on the literature.}
  \Description{Overview of the \textit{EarXplore} development pipeline and platform. A. We identified 118 papers on earable interaction, covering a diverse range of input modalities such as head, hand, and facial gestures. The picture shows four examples of these gestures. The first is a man blinking while wearing earphones. The seocnd is a woman with headphones closing her eyes and touching one of the lids of the headphone. The third is a woman wearing earbuds with eyes closed and touching her cheek with a finger. The fourth is a man shaking his head with eyes closed and earphones on; B. Through a structured development approach, expert input informed the derivation of key questions and corresponding criteria, which were then applied to systematically encode each study. The picture shows the four key parts of this process (1st: Expert Input, 2nd: Key Questions, 3rd: Criteria/Filters, 4th: Study Encoding; C. The resulting interactive platform, \textit{EarXplore}, supports exploration through four interconnected views (Tabular, Graphical, Similarity, Timeline), each offering a distinct perspective on the literature. The views are presented schematically. The first one is the Study Overview, which is represented by a small table. The second is the Graphical Display, which is represented by a small bar chart showing the distribution of sensors schematically. The third one is the Study Similarity, which is represented by a graph showing the connections between the studies, some of which are highlighted. The fourth one is the Temporal Trends, showing a schematic timeline with nodes representing papers for each year. There are connections between the nodes of which some are highlighted.}
  \label{fig: teaser}
\end{teaserfigure}

\maketitle

\input{sections/1_Introduction}
\input{sections/2_Related_Work}
\input{sections/3_Methodology}
\input{sections/4_EarXplore}

\input{sections/5_Insights}

\input{sections/6_Discussion}

\input{sections/7_Conclusion}

\begin{acks}

Funded by the German Research Foundation (Deutsche Forschungsgemeinschaft -- DFG) -- GRK2739/1 -- Project Nr. 447089431 -- Research Training Group: KD$^2$School -- Designing Adaptive Systems for Economic Decisions.

\end{acks}

\bibliographystyle{ACM-Reference-Format}
\bibliography{sample-base}

\includepdf[pages=-]{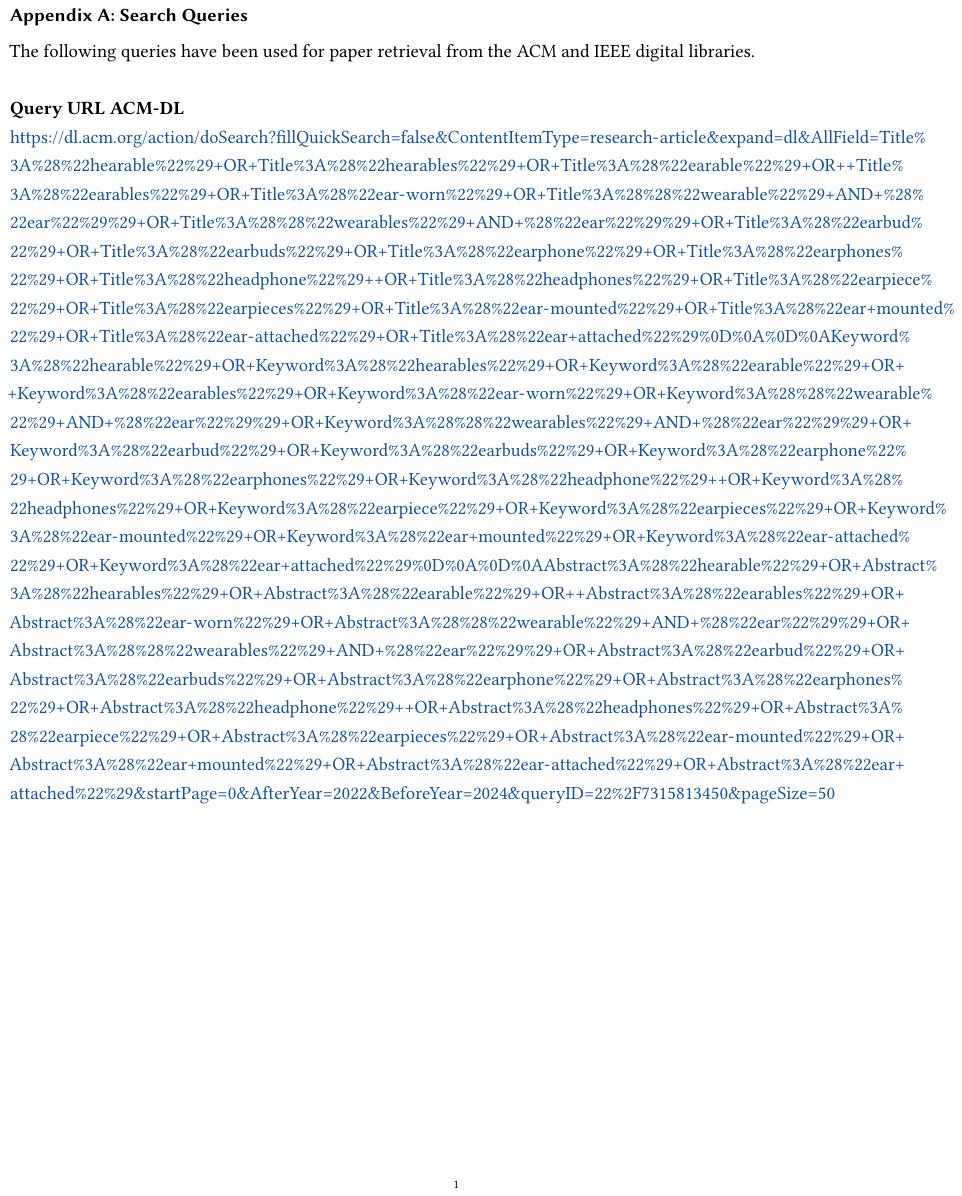}




\end{document}

%% file: sections/1_Introduction.tex
\section{Introduction}\label{sec: Intro}

The human ear has emerged as a compelling site for sensing and interaction. Not only is this reflected in everyday behaviors such as the omnipresent double tap on earphones but also in a growing body of research exploring interaction with earables -- ear-worn devices equipped with additional sensors \cite{roddiger_sensing_2022, hu_survey_2025}. Compared to other wearables, earables offer a unique combination of advantages: they are compact, socially acceptable, in close proximity to vital physiological signals (e.g., carotid pulse, jaw movement, brain activity), and typically tethered to smartphones, providing additional computing power. These characteristics have opened up a broad and growing space of novel interaction modalities. Prior work has demonstrated a variety of inputs, including but not limited to hand gestures \cite{ID22_kikuchi_eartouch_2017, ID639_suzuki_earhover_2024, ID608_alkiek_eargest_2022}, facial muscle activity \cite{ID231_verma_expressear_2021, ID8_amesaka_facial_2019, ID278_li_eario_2022}, tongue movements \cite{ID64_maag_barton_2017, ID65_nguyen_tyth-typing_2018, ID60_taniguchi_earable_2018}, and even mental gestures \cite{ID69_merrill_classifying_2016}. Remarkably, earables have been shown to be able to reliably recognize multiple interactive inputs at once \cite{ID449_ronco_tinyssimoradar_2024, ID354_yang_maf_2024, ID17_ando_canalsense_2017}. These attributes open up application opportunities in contexts where individuals must operate or coordinate multiple devices, such as in industrial settings, medical applications, or fieldwork. Earables could function as a “third hand,” or even a “fifth limb,” particularly when both hands and feet are already engaged. Research has already demonstrated applications of earable interaction in diverse domains, including augmented reality for museum exhibitions \cite{ID611_yang_ear-ar_2020}, controlling robotic systems \cite{ID78_odoemelem_using_2020}, and sign language recognition and translation into spoken language \cite{ID226_jin_sonicasl_2021,  ID342_jin_smartasl_2023}. With the recent emergence of commercially available devices featuring multimodal sensing capabilities \cite{roddiger_openearable_2025, kawsar_earables_2018, montanari_omnibuds_2024}, interaction with earables is moving beyond the research lab toward real-world applications. As the field grows rapidly in scale and diversity, there is a need for a platform that organizes, connects, and makes accessible the wealth of research on earable interaction, helping researchers to identify patterns and discover gaps, thus accelerating innovation in the field. 

To this end, we introduce \textit{EarXplore}: a curated, structured, and interactive online database consolidating published work on interaction with earables. Beyond merely listing studies, \textit{EarXplore} supports in-depth exploration through four dynamic and integrated views that enable filtering, comparison, and visualization. The \textit{Tabular View} presents a structured overview of all included studies, allowing users to filter and query based on multiple categories and criteria. The \textit{Graphical View} provides a visual summary of key study characteristics, supporting quick comparisons and high-level insights. The \textit{Similarity View} highlights connections between studies that share similar attributes, helping users discover related work. Finally, the \textit{Timeline View} visualizes the temporal evolution of the field, revealing trends, citation links, and author networks. These views were shaped through a question-centered design approach that identified eleven key research questions guiding the structure and functionality of \textit{EarXplore}. This way, the platform aims to enhance the accessibility, consistency, and comprehensiveness of literature research in the rapidly evolving earable interaction domain. Moreover, through mechanisms incorporated directly in the online platform itself as well as the corresponding open source GitHub repository, researchers can actively contribute their works to the platform, thus keeping it a living community resource that stays up to date, even long after publication.

Prior work identified interaction as one of the four major areas of earable usage in a comprehensive survey on earable sensing \cite{roddiger_sensing_2022}. To extend this line of work and ensure continuity, we build directly on its theoretical and methodological foundations by adopting its literature extraction pipeline and adapt its taxonomy of eight interaction locations (e.g., \textit{Ear and Earable}, \textit{Mouth}). This ensures our study remains aligned with an established line of research and comparable to earlier findings. Within this shared framework, we position our work in three key ways. First, as a follow-up, our study extends the work of the earlier survey, which has grown increasingly outdated in light of the rapid pace of developments in earable interaction research. Second, as an in-depth examination, we focus specifically on the subfield of interaction, offering a more detailed, fine-grained view on individual studies. Third, we address the limitations of static surveys: while valuable, they are inherently fixed in scope and presentation, making it difficult for researchers to explore relationships and generate tailored insights. As a result, readers are largely restricted to the connections pointed out by the original authors and the conclusions explicitly drawn by them. Furthermore, even comprehensive surveys are already becoming outdated by the time they are published and lack mechanisms for continuous updates or community involvement.

All in all, our key contributions are: (i) a comprehensive corpus of 118 studies on interaction with earables, categorized across 33 structured criteria; (ii) the development of \textit{EarXplore}, an interactive online database that provides four distinct yet integrated views (\textit{Tabular}, \textit{Graphical}, \textit{Similarity}, and \textit{Timeline}) on earable interaction research, designed through a question-centered approach; (iii) a discussion of research gaps and future opportunities in earable interaction, grounded in the analysis enabled by \textit{EarXplore}, and offering insights into the field’s current state and open challenges.



%% file: sections/2_Related_Work.tex
\section{Related Work}\label{sec: Related Work}

In this section, we review existing surveys on earables (\autoref{sec: surveys on earables}), with particular emphasis on the insights on earable interaction presented in \citet{roddiger_sensing_2022} (\autoref{sec: roeddiger_interaction}) before summarizing the works on already established interactive online databases (\autoref{sec: Interactive Online Survey Databases}).

\subsection{Surveys on Earables}\label{sec: surveys on earables}

Earables have emerged as a rapidly growing research area, with most publications appearing within the last decade \cite{roddiger_sensing_2022}. To date, six major reviews have been published on the topic, reflecting the increasing scholarly interest.

Three of these surveys addressed the general state of earable literature. The earliest one by \citet{plazak_survey_2018} focused on the conceptual affordances of earables as wearable computing interfaces, outlining early possibilities and future directions. The most comprehensive review on the field was published by \citet{roddiger_sensing_2022}, who conducted a systematic literature review of 271 publications and synthesized the findings into a detailed taxonomy of 47 sensing phenomena. These are grouped into four major umbrella categories, including interaction with earables. Most recently, \citet{hu_survey_2025} released an update which reviews over 100 studies published since 2022 along the framework established by \citet{roddiger_sensing_2022}, analyzing recent developments, emerging applications, and open challenges.

The other three reviews focus on specific application areas. \citet{mase_hearables_2020} examined earables in the context of physiological monitoring, particularly temperature, heart rate, and oxygen saturation in sport, medicine, and extreme environments. \citet{choi_health-related_2022} reviewed the use of earables for health-related monitoring, describing the devices with respect to health outcomes, biomarkers, sensors, and their utility for disease prevention. Closely related is the narrative review by \citet{ne_hearables_2021}, which analyzed biosignal acquisition with earables, covering both methodological characteristics (e.g., study setting) and technical limitations (e.g., battery life).

As this work builds on the foundations established by \citet{roddiger_sensing_2022}, we summarize its insights on earable interaction research below.


\subsection{Interaction with Earables}\label{sec: roeddiger_interaction}
\citet{roddiger_sensing_2022} present a comprehensive high-level overview of the state of the art on sensing with earables, with earable interaction comprising a major section in the survey. The section is organized around a taxonomy of eight interaction locations, such as \textit{Ear and Earable}, \textit{Head Gestures and Pointing}, and \textit{Mouth}, which we also adopt to maintain conceptual continuity. These locations form the backbone of their analysis, and for each, the authors provide a concise narrative summary of the existing research on interaction with earables. Within each location, the authors further delineate relevant subdomains, organizing diverse approaches into coherent thematic clusters. For instance, the survey includes focused summaries on silent speech input, teeth gestures, and different actuation mechanisms, each reviewed and discussed while being contextualized within the broader research landscape. This structure enables researchers to quickly grasp the composition of the field, navigate its constituent subdomains, and identify recurring methodological patterns, sensing strategies, and open research challenges.

While the work of \citet{roddiger_sensing_2022} provides a strong foundation, it necessarily remains broad in scope and does not offer in-depth analysis of individual studies or their specific characteristics. Additionally, as a static, text-based review, it offers limited support for tailored exploration: readers are constrained to the structure and interpretations provided by the original authors, with no mechanism to dynamically filter, compare, or update the underlying corpus. This contrasts with more recent efforts in human-computer interaction \cite{seifi_haptipedia_2019, di-luca_locomotive_2021, bhatia_text_2025}, which have introduced interactive platforms to support flexible, multi-perspective engagement with complex research spaces. 



\subsection{Interactive Online Survey Databases}\label{sec: Interactive Online Survey Databases}

Commercial product catalogs have long demonstrated interactive exploration capabilities through gallery and tabular, filter-based views of inventory (e.g., \cite{mcmasterCarr, wurthRedexpert, digikey}). In recent years, this paradigm has begun to influence academic research, where making survey results available as interactive online databases has emerged as a complementary approach to traditional written reviews, enabling more dynamic, user-driven exploration of complex research landscapes. One of the earliest examples is \textit{VibViz} \cite{seifi_vibviz_2015}, which introduced an interactive platform for browsing a library of vibrotactile stimuli along five different taxonomies. Another pioneering project is \textit{TimeViz Browser} \cite{tominski_timeviz_2023}, an online platform that allows users to view different visualization techniques in a gallery panel and interactively access related information. 

This concept was significantly extended by \textit{Haptipedia} \cite{seifi_haptipedia_2019}, which combines a taxonomy-driven approach with a rich, publicly accessible database of over 100 grounded force-feedback devices. Designed for both interaction and device designers, the platform allows exploration through multiple views, each representing a different angle on the data. Conceptually aligned, \textit{LocomotionVault} \cite{di-luca_locomotive_2021} organizes over 100 VR locomotion techniques along multiple dimensions such as hardware, simulation sickness, and accessibility in a filterable gallery view, while also introducing a similarity metric to aid navigation. A related initiative is \textit{LoCoMoTe} \cite{croucher_locomote_2024}, a dashboard built around a structured taxonomy for categorizing natural walking experiments in VR. Relying on the same filterable gallery as LocomotionVault, \textit{TEXT} \cite{bhatia_text_2025}, a database of 176 text entry techniques, allows researchers to analyze trends and performance metrics and to examine how design attributes affect user experience.

Additional efforts have demonstrated the utility of interactive online supplements to scientific reviews and databases. \citet{seitz_impact_2024} accompanied their literature review on video meeting systems with an interactive morphological box, allowing readers to explore how different features in video meetings impact psychological user states. Similarly, \citet{aaltonen_datastudiesbibliographyorg_2025} augmented their online database on data visualizations with several interactive figures.

Prior interactive survey databases demonstrate the value of visual, filterable access to research corpora, although they often focus on devices or techniques rather than empirical studies. Building on these prior efforts, \textit{EarXplore} brings the concept of interactive survey databases to the domain of earable interaction. By supporting multifaceted exploration at the level of individual studies through a coordinated multi-view interface, it serves as the first interactive online platform for synthesizing technical, methodological, and conceptual advances in earable research.




%% file: sections/3_Methodology.tex
\section{Methodology}\label{sec: Methodology}

The development of \textit{EarXplore} involved two primary steps: a systematic literature search and the iterative design of the interactive online database. The literature search was based on the approach described by \citet{roddiger_sensing_2022}, with adaptations to suit the scope of interaction with earables detailed in \autoref{sec: Selection Criteria, Filtering, and Paper Retrieval}. The design and implementation of the database followed a structured yet iterative process, informed by collaborative deliberation among the authors and grounded in a set of key questions guiding the research. This process is further described in ~\autoref{sec: Development of EarXplore}.



\subsection{Study Identification and Paper Retrieval}\label{sec: Selection Criteria, Filtering, and Paper Retrieval}

To extract the relevant papers for the literature corpus of \textit{EarXplore}, it was first necessary to delineate the area of research. As \citet{roddiger_sensing_2022}, we define:

\begin{highlightbox}
    \textbf{Earables} are devices that attach in, on, or in the immediate vicinity of the ear to offer functionalities beyond basic audio in- and output.
\end{highlightbox}

The present study focuses specifically on interaction with earables. However, no universally accepted definition of interaction in HCI exists \cite{hornbaek_what_2017}. Therefore, building on \citet{roddiger_sensing_2022}, we define:

\begin{highlightbox}
    \textbf{Interaction with Earables} is the engagement between a user and devices that attach in, on, or in the immediate vicinity of the ear through intentional control inputs and/or non-auditory feedback modalities.
\end{highlightbox}

The objective of the platform is to incorporate all papers within the scope of this definition. To cover the full breadth of the field, we also include elicitation studies that may not include any sensing (e.g., simulated input to a conceptual ~device).

\begin{figure*}[t]
    \centering
    \includegraphics[width=1\linewidth]{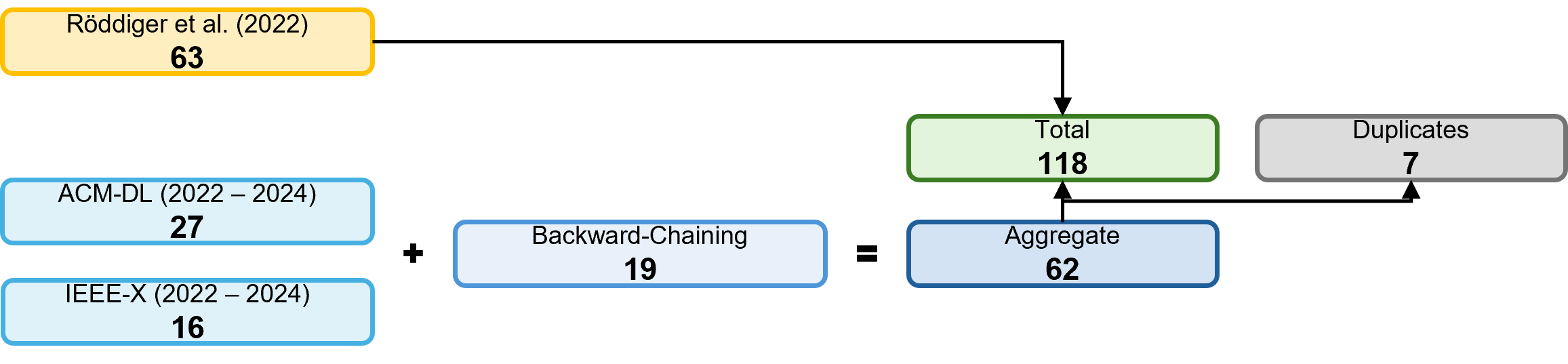} 
    \caption{Literature Retrieval Process. 63 papers were drawn from the literature corpus of \citet{roddiger_sensing_2022}. To update the corpus, we conducted a follow-up search (2022–2024) in the ACM-DL and IEEE-X using the same criteria, and applied backward-chaining afterwards. This yielded 62 additional papers. After removing duplicates, the final corpus comprises 118 peer-reviewed studies on earable interaction.}
    \Description{Literature was extracted from two sources. First, the literature corpus of the  \citet{roddiger_sensing_2022} paper led to a selection of 63 papers. Second, as they only included papers until January 21st, 2022, the same queries were used to search the ACM-DL and IEEE-X libraries for 2022-20224, leading to another 43 papers. Of these, 27 were coming from the ACM Digital Library and 16 from the IEEE Xplore. Backsearch yielded another 19 papers, which aggregated to 62 papers. Seven of these papers were duplicate versions of other papers (e.g., excerpts), which we excluded for further analysis, leading to a corpus of 118 papers. The whole process is described in more detail in \autoref{sec: Selection Criteria, Filtering, and Paper Retrieval}.}
    \label{fig: Literature Sources Flow Chart}
\end{figure*}

To further refine the scope of \textit{EarXplore}, we exclude papers that

\begin{enumerate}
    \item are not peer-reviewed, e.g., workshop proposals, theses, patents, and technical reports.
    \item are larger head-worn or off-body systems (e.g., VR headsets).
    \item are technical designs of audio earphones.
    \item are not written in English.
\end{enumerate}

Our literature search focused on the ACM Digital Library (ACM-DL) and IEEE Xplore (IEEE-X) libraries, as these two databases are widely regarded as the most relevant and comprehensive sources of literature related to wearable and HCI publications. We leveraged the existing interaction corpus by \citet{roddiger_sensing_2022}, who had already undertaken a systematic literature search in these two databases. The first and third author re-examined their selection, identifying a set of 71 studies and verifying all but eight of the studies to be within the scope of our criteria, resulting in a final selection of 63 studies. These studies were disregarded for having no clear interaction paradigm \cite{ID57_salzar_improving_2008, ID249_pfreundtner_wearable_2021, 2014_looney_ear-eeg_2014, ID70_bleichner_concealed_2017, ID86_bedri_detecting_2015, ID9_pham_wake_2020}, not employing an earable \cite{ID223_cao_earphonetrack_2020}, or for being purely conceptional work \cite{ID21_choi_toning_2020}.

As this corpus only includes studies published until January 21st, 2022, we updated the search using the same keywords (listed below). These keywords were then matched against the titles, abstracts, and author keywords of all studies published in the two libraries from January 21st, 2022 to December 31st, 2024. This search yielded 148 papers in the ACM-DL and 275 studies in IEEE-X. For reference, the original queries are provided in Appendix A.

\begin{figure}[!h]
    \vspace{-0.6cm}
    \centering
    \Description{query target: Title, Keywords, Abstract (ACM-DL) / Document Title, Index Terms, Abstract (IEEE-X) \\
                    keywords: earable(s), hearable(s), ear-worn, ear AND wearable(s), earbud(s), earphone(s), \\ \textcolor{lightgray}{.}\hspace{13mm} headphone(s), earpiece(s),  ear(-)mounted, ear(-)attached, ear(-)based\\
                    filter: Research Article OR Short Paper (ACM-DL) / Conferences, Journals (IEEE-X)}
    \begin{searchbox}
        \small{
            \texttt{query target: Title, Keywords, Abstract (ACM-DL) / Document Title, Index Terms, Abstract (IEEE-X) \\
                    keywords: earable(s), hearable(s), ear-worn, ear AND wearable(s), earbud(s), earphone(s), \\ \textcolor{lightgray}{.}\hspace{13mm} headphone(s), earpiece(s),  ear(-)mounted, ear(-)attached, ear(-)based\\
                    filter: Research Article OR Short Paper (ACM-DL) / Conferences, Journals (IEEE-X)
            }
        }
    \end{searchbox}
    \vspace{-0.6cm}

\end{figure}

The combined 423 results were subsequently reviewed by the first two authors who independently read the titles and abstracts before screening the papers checking if they were within the scope of our criteria. This yielded an initial set of 43 papers (27 ACM-DL, 16 IEEE-X). Subsequently, backward-chaining was applied to the selected papers to account for publications that are not available in the ACM-DL and IEEE-X library or were missed by the keywords, resulting in the inclusion of papers from other publishers including Springer, MDPI, and Sage. The first author scanned the references of the papers published from 2022 to 2024 (2,296 including duplicates), leading to the identification of additional papers, which were then subjected to backward-chaining themselves. The second author subsequently confirmed the resulting selection of 19 papers. No additional papers that the authors were aware of had to be added, yielding a total of 62 papers. Upon further review of the first two authors, seven papers were excluded because they were identified to be alternative versions of papers that were already included (e.g., excerpts). All in all, this led to the final corpus of 118 papers, listed in Appendix B. The retrieval process is also depicted in \autoref{fig: Literature Sources Flow Chart}.

To ensure a comprehensive and inclusive representation of the field, we deliberately included all relevant studies on earable interaction, regardless of citation count, publication venue, or study size. The resulting set of studies formed the literature corpus of \textit{EarXplore}, and hence the empirical basis for its design process.

\begin{figure*}[t]
    \centering
    \includegraphics[width=1\linewidth]{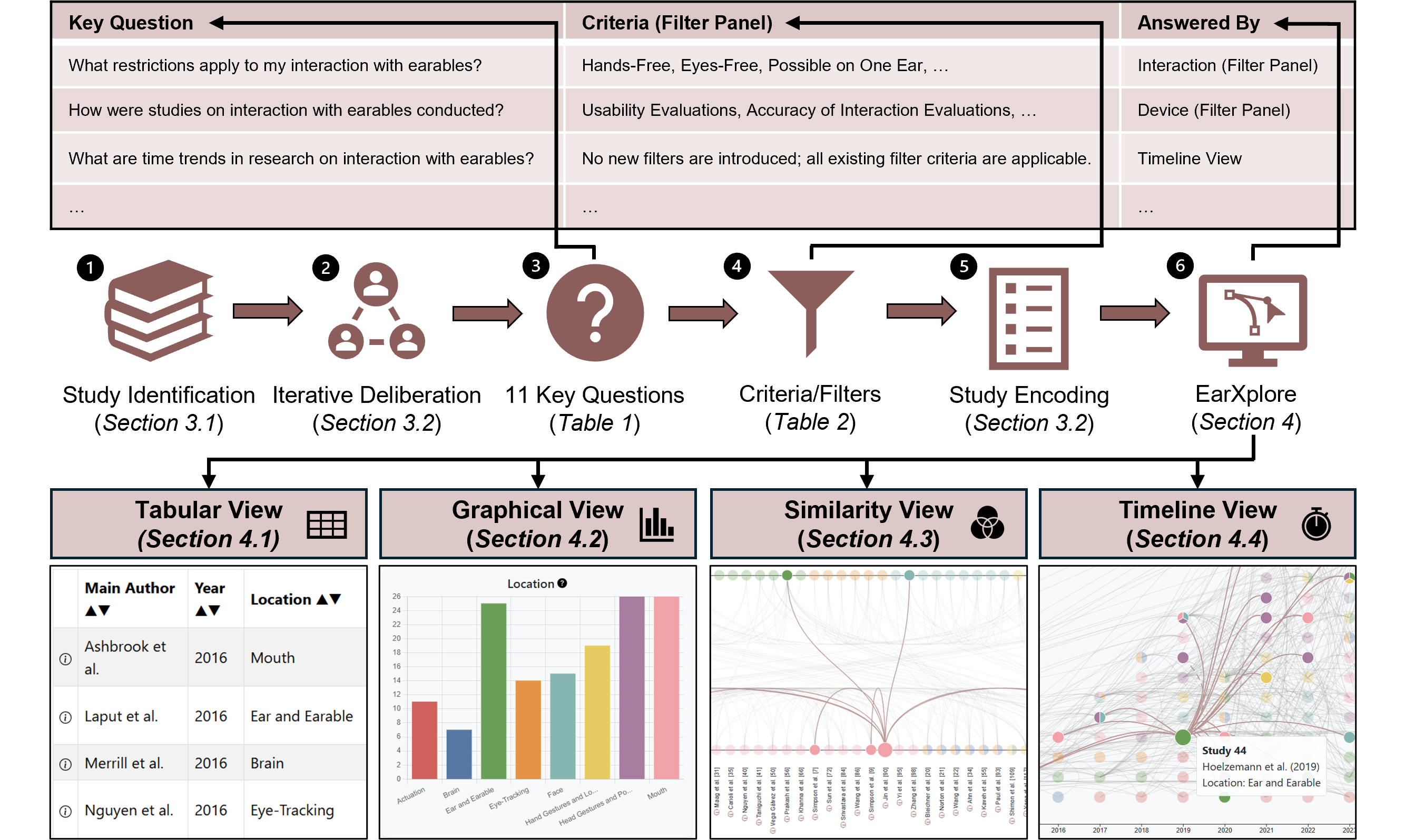} 
    \caption{Design and development process of \textit{EarXplore}. It is illustrated how the question-centered design approach informed the structure and implementation of the platform’s four integrated views. The images below each view are excerpts that illustrate key functionality and visual characteristics of the respective interface components.}
    \Description{Design and development process of \textit{EarXplore}. It is illustrated how the question-centered design approach informed the structure and implementation of the platform’s four integrated views. Its six steps were (1) Study Identification  (see \autoref{sec: Selection Criteria, Filtering, and Paper Retrieval}), (2) Iterative Deliberation (see \autoref{sec: Development of EarXplore}), (3) 11 Key Questions (see \autoref{tab: guiding questions}), (4) Criteria/Filters (see \autoref{tab: criteria_descriptions}), (5) Study Encoding (see \autoref{sec: Development of EarXplore}), and (6) EarXplore (see \autoref{sec: EarXplore}). The structure of the platform is built around steps (3), (4), and (6) in particular. The box above the development process shows the different Key Question - Criteria (Filter Panel) - Answered By triplets, that map to step (3), (4), and (6) of the process. These triplets are:

    Key Question: What restrictions apply to my interaction with earables? - Criteria (Filter Panel): Hands-Free, Eyes-Free, Possible on One Ear, ... - Answered By: Interaction (Filter Panel)
    Key Question: How were studies on interaction with earables conducted? - Criteria (Filter Panel): Usability Evaluations, Accuracy of Interaction Evaluations, ... - Answered By: Device (Filter Panel)
    Key Question: What are time trends in research on interaction with earables? - Criteria (Filter Panel): Now new filters are introduced; all existing filter criteria are applicable. - Answered By: Timeline View
    
    The images below each view (Tabular, Graphical, Similarity, Timeline) are excerpts that illustrate key functionality and visual characteristics of the respective interface components. The first shows an excerpt of the table of the Tabular View. The second shows the bar chart for Location as an example. The third shows an excerpt of the similarity graph of the Similarity View that is colored and that has highlighted the connections of one paper. The fourth shows and excerpt of the timeline of the Timeline View with the connections of one paper highlighted.}
    \label{fig: EarXplore Scheme}
\end{figure*}


\subsection{Design and Development Process}\label{sec: Development of EarXplore}

The design and development of \textit{EarXplore} were guided by an in-depth exploration of the key questions that researchers typically ask when engaging with the literature on interaction with earables. These questions shaped both the underlying structure and the interface of the platform. While development followed an iterative and exploratory approach, we describe its core stages systematically below. An overview of the entire development process is depicted in \autoref{fig: EarXplore Scheme}.

\begin{table*}[t]
    \centering
    \footnotesize
    \caption{Overview of the eleven key questions that guided the design of \textit{EarXplore}. Each question is mapped to the corresponding interface view or filter panel, along with the specific criteria used as filters to answer it. Question \textit{Q1-8} are addressed through filter panels, while additional views are introduced for \textit{Q9} and \textit{Q10}. In contrast to the other ten questions, \textit{Q11} requires consideration of the full database functionalities. Detailed descriptions of each criterion are provided in \autoref{tab: criteria_descriptions} and Appendix C.}
    \begin{tabular}{p{5cm}p{2cm}p{7.4cm}}
    \toprule
    \textbf{Key Question} & \textbf{Answered By} & \textbf{Criteria (Filter Panels)} \\
    \midrule
    \textbf{Q1} What interactions with earables have been performed? & General Information (Filter Panel) & Main Author, Year, Location, Input Body Part, Gesture \\
        \addlinespace[5pt] 
    \textbf{Q2} What restrictions apply to my interaction with earables? & Interaction (Filter Panel) & Number of Selected Gestures, Resolution, Hands-Free, Eyes-Free, Possible on One Ear, Adaptation of the Interaction Detection Algorithm to User, Discreetness of Interaction Techniques, Accuracy of Interaction Techniques, Robustness of Interaction Detection \\
        \addlinespace[5pt] 
    \textbf{Q3} What interactions can I perform with my particular earable, given its sensors? & Sensing (Filter Panel) & Sensors, No Additional Sensing \\
        \addlinespace[5pt] 
    \textbf{Q4} How were studies on interaction with earables conducted? & Study (Filter Panel) & Elicitation Study, Usability Evaluations, Cognitive Ease Evaluations, Discreetness of Interactions Evaluations, Social Acceptability of Interactions Evaluations, Accuracy of Interaction Evaluations, Alternative Interaction Validity Evaluations, Evaluation of Different Settings, Evaluation of Different Conditions (User-Related), Evaluation of Different Conditions (Environment-Related) \\
        \addlinespace[5pt] 
    \textbf{Q5} What devices and processing architectures did researchers use for interaction with earables? & Device (Filter Panel) & Earphone Type, Development Stage, Real-Time Processing, On-Device Processing \\
        \addlinespace[5pt] 
    \textbf{Q6} What motivations were behind the study on interaction with earables? & Motivations (Filter Panel) & Motivations \\
        \addlinespace[5pt] 
    \textbf{Q7} What are the intended applications for interactions with earables according to the researchers? & Applications (Filter Panel) & Intended Applications \\
        \addlinespace[5pt] 
    \textbf{Q8} What keywords can researchers use when searching for interaction with earables? & Keywords (Filter Panel) & Keywords \\
    \midrule
    \textbf{Q9} How can I find similar work on a specific earable interaction paper? & Similarity View & No new filters are introduced; all existing filter criteria are applicable. \\
        \addlinespace[5pt] 
    \textbf{Q10} What are time trends in research on interaction with earables? & Timeline View & No new filters are introduced; all existing filter criteria are applicable. \\
    \midrule
    \textbf{Q11} Where are the gaps in the literature on interaction with earables? & Whole Database & No new filters are introduced; all existing filter criteria are applicable. \\
    \bottomrule
    \end{tabular}
    \label{tab: guiding questions}
\end{table*}

\begin{table*}[p]
    \centering
    \footnotesize
    \caption{Descriptions of \textit{EarXplore} criteria including how they were encoded (\textit{B} = binary, \textit{O} = ordinal, or \textit{C} = categorical, \textit{N} = numeric). More detailed descriptions of the criteria together with their answer options are listed in Appendix C.}
    \begin{tabular}{p{3cm}p{11.4cm}}
    \toprule
    \textbf{Criterion} & \textbf{Description} \\
    \midrule
    Main Author & The main author of the paper, written as in an in-text citation. \\
    \addlinespace[2pt] 
    Year & Publication year of the study.  \\
    \addlinespace[2pt] 
    Location & The specific location(s) on the body where the interactions with the earable device occur. As in \cite{roddiger_sensing_2022}. (\textit{C}) \\
    \addlinespace[2pt] 
    Input Body Part & The part(s) of the body used to elicit the interactions with the earable device. (\textit{C}) \\
    \addlinespace[2pt] 
    Gesture & The specific gesture(s), movement(s), or action(s) the user must perform to initiate interactions with the earable device. Gestures only present in elicitation studies are not considered. (\textit{C}) \\
    \addlinespace[2pt] 
    Number of Selected Gestures & The number of gestures selected for interaction. Gestures only present in elicitation studies are not counted. If the same gesture can be performed on different ears, it is only counted once. None-type-gestures are not considered. (\textit{N})\\
    \addlinespace[2pt] 
    Resolution & Whether interactions involve a single distinct input (Semantic) or are continuous, with distinct steps (Coarse) or subtle, unnoticeable steps (Fine). (\textit{C})\\
    \addlinespace[2pt] 
    Hands-Free & Whether the hands play an active role in the interactions with the earable. (\textit{C})\\
    \addlinespace[2pt] 
    Eyes-Free & Whether the eyes play an active role in the interactions with the earable (e.g., gaze direction). Visual Attention may be necessary without an active interaction role of the eyes. (\textit{C})\\
    \addlinespace[2pt] 
    Possible on One Ear & Whether the interaction detection presented in the study is (also) possible using an earable that covers only one ear. ~(\textit{C})\\
    \addlinespace[2pt] 
    Adaptation of the Interaction Detection Algorithm to User & Whether it has been demonstrated that the interaction detection algorithm can be adapted or fine-tuned to an individual user. (\textit{B}) \\
    \addlinespace[2pt] 
    Discreetness of Interactions & Extent to which interaction techniques remain unseen, unheard, or undetectable in public settings. Excluding device design. For multiple techniques, the rating reflects the predominant level. Rated by two authors ($\kappa_w = .72$). (\textit{O}) \\
    \addlinespace[2pt] 
    Accuracy of Interaction Detection & The system’s ability to accurately detect and interpret interactions, considering only the most basic reported condition and setting (e.g., sitting in a lab) for consistency. Only applies to studies reporting accuracies. (\textit{O}) \\
    \addlinespace[2pt] 
    Robustness of Interaction Detection & Demonstrated robustness of interaction detection performance across additional settings (e.g., a café, outdoors) or conditions (e.g., walking, listening to music) that may cause interference, defined as an accuracy drop of less than 10\%. Only applies to studies reporting accuracies. (\textit{O}) \\
    \addlinespace[2pt] 
    Sensors & The specific sensor(s) the earable needs to enable (e.g., speaker for ultrasound) and/or recognize (e.g., a microphone) the interactions. Only sensors that are part of the earable. (\textit{C})\\
    \addlinespace[2pt] 
    No Additional Sensing & No additional devices like smartphones, rings etc. are needed for sensing the interaction presented in the study. (\textit{B}) \\
    \addlinespace[2pt] 
    Elicitation Study & An elicitation study on interaction with earables has been performed. (\textit{B}) \\
    \addlinespace[2pt] 
    Usability Evaluations & Whether the usability of the interactions was assessed. (\textit{B})\\
    \addlinespace[2pt] 
    Cognitive Ease Evaluations & Whether the cognitive ease of the interactions was assessed. (\textit{B})\\
    \addlinespace[2pt] 
    Discreetness of Interactions Evaluations & Whether the discreetness of the interactions was assessed. (\textit{B})\\
    \addlinespace[2pt] 
    Social Acceptability of Interactions Evaluations & Whether the social acceptability of the interactions was assessed. (\textit{B})\\
    \addlinespace[2pt] 
    Accuracy of Interactions Evaluations & Whether the accuracy of the system’s interactions detection was assessed. (\textit{B})\\
    \addlinespace[2pt] 
    Alternative Interaction Validity Evaluations & Whether a metric other than accuracy of interaction detection was used to assess the validity of the system’s interaction detection (e.g., angular error). (\textit{B})\\
    \addlinespace[2pt] 
    Evaluation of Different Settings & The different settings the interactions were evaluated in (e.g., Lab, Cafe). (\textit{C})\\
    \addlinespace[2pt] 
    Evaluation of Different Conditions (User-Related) & The different user-related conditions the interactions were evaluated in (e.g., Standing, Walking). (\textit{C})\\
    \addlinespace[2pt] 
    Evaluation of Different Conditions (Environment-Related) & The different environment-related conditions the interactions were evaluated in (e.g., Music, Noise). Only deliberately induced environmental conditions not inherently part of the setting are listed, to avoid redundancy. (\textit{C})\\
    \addlinespace[2pt] 
    Earphone Type & The specific type of earphone used in the study. (\textit{C})\\
    \addlinespace[2pt] 
    Development Stage & Whether a commercial earable was used or a research prototype. (\textit{C})\\
    \addlinespace[2pt] 
    Real-Time Processing & The system is demonstrated to be able to immediately detect and respond to a user’s performed interaction. (\textit{B})\\
    \addlinespace[2pt] 
    On-Device Processing & Excluding initial training, interaction detection occurs solely on the earable itself, without external device support. Only processing directly on the device attached in, on, or the immediate vicinity of the ear qualifies, excluding smartphones or connected microcomputers. (\textit{B})\\
    \addlinespace[2pt] 
    Motivations & The motivations the authors explicitly outline for their study on interaction with earables. Only six core motivations recurring across studies were extracted, to keep the coding scheme concise and consistent. (\textit{C})\\
    \addlinespace[2pt] 
    Intended Applications & The applications the authors explicitly suggest for their proposed interaction with earables. (\textit{C})\\
    \addlinespace[2pt] 
    Keywords & The keywords selected by the authors. (\textit{C})\\
    \bottomrule
    \end{tabular}
    \label{tab: criteria_descriptions}
\end{table*}

\paragraph{Identifying Key Questions}

To inform the design of both the database and user interface, we first identified the key questions that \textit{EarXplore} should help researchers answer. The process was grounded in in-depth discussions among all but one of the eight contributing authors, all of whom are active in earable research. The discussions focused on the types of inquiries peers in the field might pose, ranging from questions of technical feasibility to those concerning study design and application domains. These discussions led to a preliminary list of candidate questions, which the first author synthesized into eleven guiding questions. These questions formed the conceptual backbone of \textit{EarXplore}, informing both the system’s content and structure. An overview is presented in \autoref{tab: guiding questions}.

\paragraph{Defining Criteria and Filters}

Building on the guiding questions, we next determined how each could be operationalized within the platform. Eight of the eleven questions were best addressed through structured criteria that could later serve as interactive filters in the interface. The remaining three required advanced visualizations, discussed further in \textit{Structuring the Platform}. In line with prior work \cite{bhatia_text_2025, seifi_haptipedia_2019}, the criteria were derived through iterative discussions among the authors and grounded in terminology commonly used across the literature corpus represented in the database. For each criterion, appropriate answer types (binary, ordinal, categorical, numeric) were determined to support meaningful and precise filtering. For example, to answer \textit{Which sensors are present in a given earable device?}, we extracted the list of sensing modalities reported in the reviewed studies. This systematic mapping of questions to criteria ensured that user queries could be efficiently translated into structured searches. The final set of criteria and their corresponding descriptions are presented in \autoref{tab: criteria_descriptions}.

\paragraph{Study Encoding and Data Normalization}
With the criteria established, all studies in the database were systematically annotated according to these dimensions. This involved detailed review and coding of each study’s relevant attributes. To improve filter applicability and reduce redundancy, the first author consolidated similar categorical terms under unified labels. This normalization process was essential to maintain consistency and facilitate accurate filtering. After normalization, the annotations were verified to ensure consistency with the contents of the respective studies.

\paragraph{Structuring the Platform}

To ensure the interface design aligned with user needs, we analyzed how each of the guiding questions could be most effectively supported through specific interface components. This step was critical for aligning the system’s interaction design with the nature of user inquiries, ranging from precise data retrieval to broader exploratory analysis. As outlined earlier, eight of the questions (e.g., \textit{What restrictions apply to my interaction with earables?}) could be effectively addressed through structured filtering mechanisms. Two questions (\textit{How can I find similar work on a specific earable interaction paper?} and \textit{What are time trends in research on interaction with earables?}) were not answerable merely by filtering the database and thus required the introduction of additional tailored views. Finally, one question aiming for the identification of research gaps necessitated analytical synthesis across the whole capabilities of the database. These mappings informed the design of four interconnected views, each tailored to support different types of inquiries. We detail these views in the following section.

%% file: sections/4_EarXplore.tex
\section{EarXplore}\label{sec: EarXplore}

\begin{figure*}[t]
    \centering
    \includegraphics[width=\linewidth]{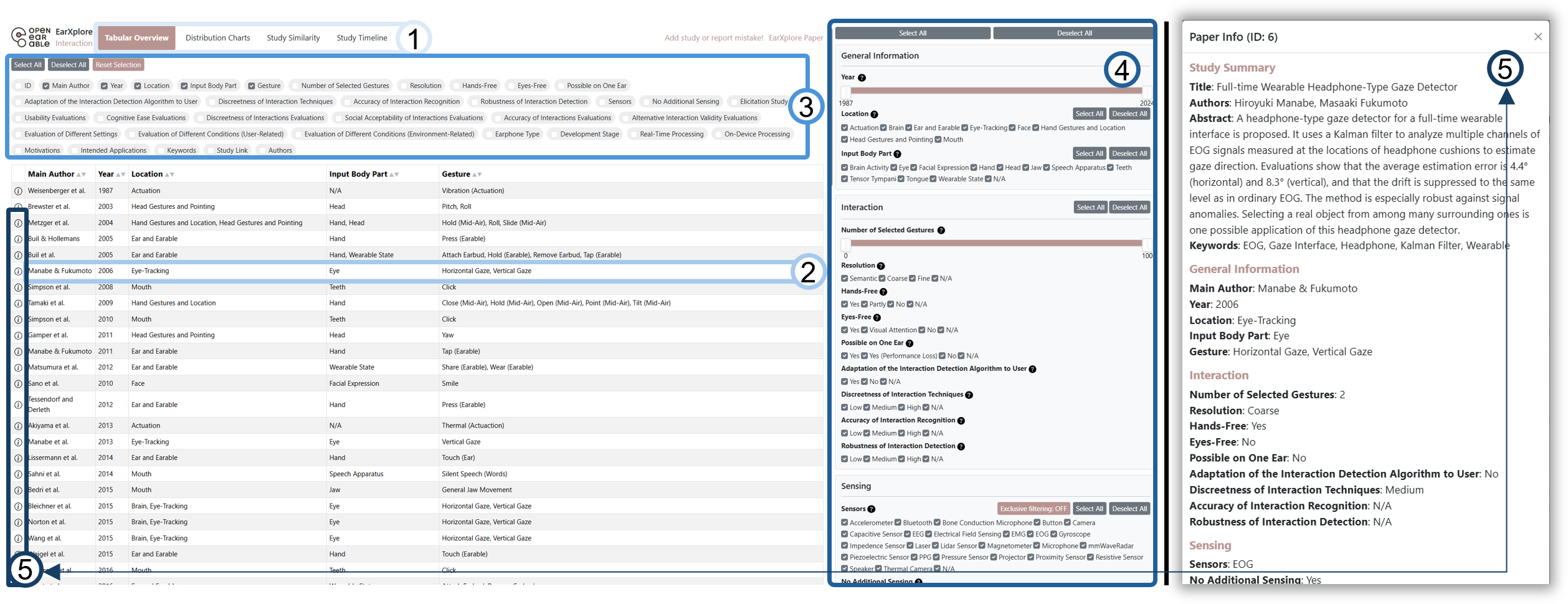} 
    \caption{Tabular View -- (1) The Tabular View serves as landing page and can also be selected via the view selection menu. (2) By default, key information on each study (\textit{Main Author}, \textit{Year}, \textit{Location}, \textit{Input Body Part}, and \textit{Gesture}) is displayed. (3) The top toggle menu allows users to show or hide columns with additional information. (4) Filters in the sidebar enable users to refine the database by including or excluding specific attribute values. (5) Clicking the info icons at the beginning of each row opens a modal overlay that displays all available information for the selected study. (6 - not visible) The entire dataset or a selected subset can be downloaded as a CSV file.}
    \Description{Tabular View -- The figure shows a screenshot of the Tabular View. Its components, functionalities, and interaction potentials are further explained in the following. (1) The Tabular View serves as landing page and can also be selected via the view selection menu at the top of the page. (2) By default, key information on each study (\textit{Main Author}, \textit{Year}, \textit{Location}, \textit{Input Body Part}, and \textit{Gesture}) is displayed in a table structure that fills most of the space of the page. (3) The top toggle menu allows users to show or hide columns with additional information. All of the filter criteria as well as some additional information (e.g., Study Link) are listed here. (4) Filters in the sidebar enable users to refine the database by including or excluding specific attribute values. (5) Clicking the info icons at the beginning of each row opens a modal overlay that displays all available information for the selected study. This modal overlay is visualized on the right. It features all the information on each study and begins with a Study Summary (Title, Authors, Abstract, Keywords) and then lists all the different criteria together with the answer that applies to the paper (6 - not visible) The entire dataset or a selected subset can be downloaded as a CSV file.}
    \label{fig: Tabular View}
\end{figure*}

We introduce \textit{EarXplore}\footnote{
see here: \href{https://earxplore.teco.edu/}{https://earxplore.teco.edu/}
}, an online interactive database for exploring research on earable interaction. The platform allows users to browse, filter, and compare earable interaction studies through four complementary views: a \textit{Tabular View}, a \textit{Graphical View}, a \textit{Similarity View}, and a \textit{Timeline View}. These views are interconnected, enabling users to explore data from multiple perspectives while maintaining context through features such as persistent filtering across all views.

The structure of \textit{EarXplore} is organized around eleven key questions that reflect core concerns in the field, ranging from technical affordances and study design to broader research trends. \autoref{tab: guiding questions} presents all eleven questions, indicating how each is supported by specific interface components. For those eight questions addressed through filtering options, it also shows how these are operationalized through corresponding database criteria, which are listed and described in \autoref{tab: criteria_descriptions}.
In the following, we introduce the four main views featured in \textit{EarXplore} and describe how they map to the key questions defined above.

\subsection{Tabular View}\label{sec: Tabular View}


The \textit{Tabular View} (\autoref{fig: Tabular View}) serves as the entry point for engaging with the database in a structured, detail-oriented manner. It addresses researchers who seek precise, record-level access to study information. In particular, it supports researchers addressing the eight guiding questions best addressed through filters (\qref{Q1–Q8}). A tabular format was chosen for its familiarity, readability, and compatibility with filter-based exploration.

Each study is presented as a row in a sortable table, with a default display of key information (\textit{Main Author}, \textit{Year}, \textit{Location}, \textit{Input Body Part}, \textit{Gesture}), chosen to surface the main characteristics at a glance. Through the toggle menu, users can reveal additional columns based on their specific information needs, such as \textit{Earphone Type} or \textit{Intended Applications}. This customization supports both high-level comparison and detailed inspection. The sidebar filters allow users to narrow down the selection of studies by multiple criteria simultaneously, supporting exploratory synthesis and targeted search. The info icon at the start of each row reveals a modal overlay with all available information for that study, helping users quickly assess relevance without leaving the interface. In addition, download functionality enables both full and filtered datasets to be exported as CSV files, facilitating integration into external analyzes.

\begin{figure*}[t]
    \centering
    \includegraphics[width=\linewidth]{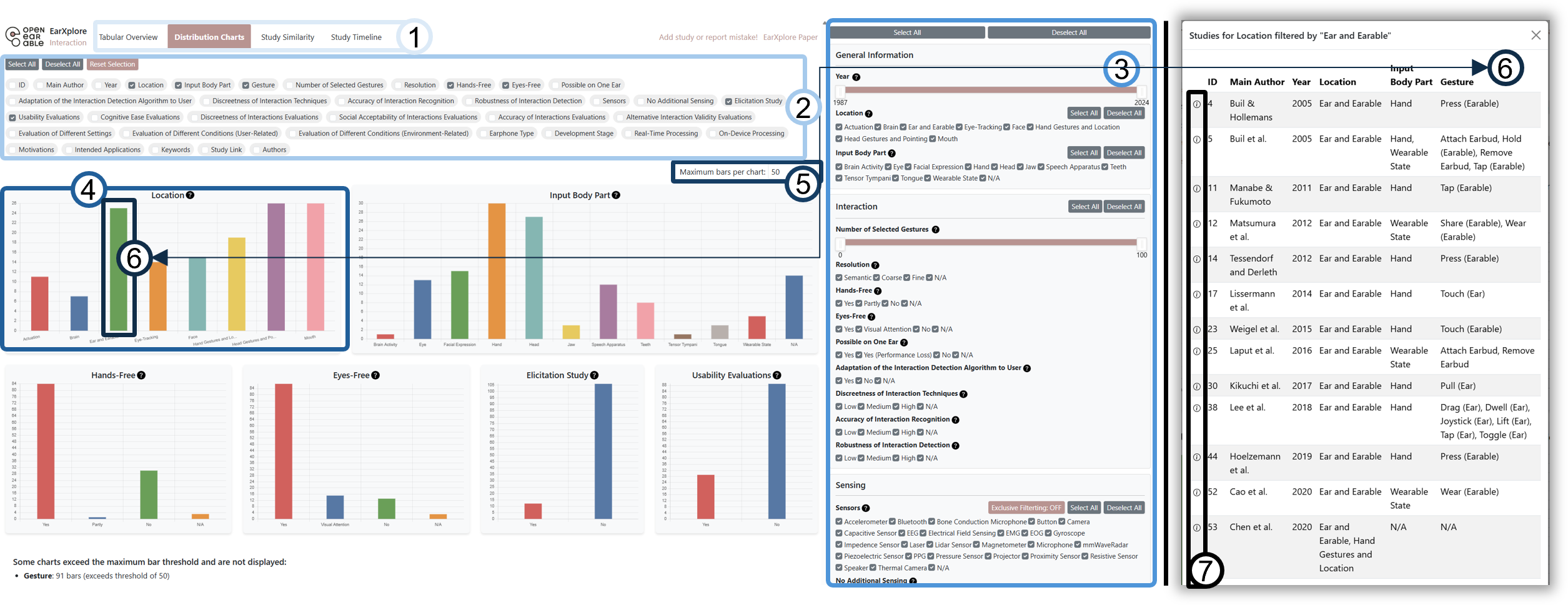} 
    \caption{Graphical View -- (1) The Graphical View can be selected via the view selection menu. (2) The top toggle menu allows users to show or hide bar charts with additional information. (3) Filters in the sidebar enable users to refine the database by including or excluding specific attribute values. (4) For each selected criterion, a bar chart displays the distribution of answer options. Chart size automatically adapts to the number of bars. (5) Users can adjust the threshold for the maximum number of bars shown per chart. (6) Clicking on a bar opens a modal overlay showing key information on all studies represented by that bar. (7) Clicking on the info icons at the beginning of each row within the modal overlay reveals the full information modal overlay for the respective study.}
    \Description{
    Graphical View -- The figure shows a screenshot of the Graphical View. Its components, functionalities, and interaction potentials are further explained in the following. (1) The Graphical View can be selected via the view selection menu at the top of the page. (2) The top toggle menu allows users to show or hide columns with additional information. All of the filter criteria as well as some additional information (e.g., Study Link) are listed here. (3) Filters in the sidebar enable users to refine the database by including or excluding specific attribute values. (4) For each selected criterion, a bar chart displays the distribution of answer options. Chart size automatically adapts to the number of bars. These bar charts fill most of the page and have the criteria name as a heading and colored bars representing the answer options, each with its corresponding label. In the picture, the bar charts for Location, Hands-Free, Eyes-Free, Possible on One Ear, Elicitation Study, and Usuability Evaluations are selected. (5) Users can adjust the threshold for the maximum number of bars shown per chart using a dropdown menu. (6) Clicking on a bar opens a modal overlay showing key information on all studies represented by that bar. This modal overlay is organized around the key information on each study, thereby showing the info icon, ID, Main Author, Year, Location, Input Body Part, and Gesture for each of the studies in that bar. (7) Clicking on the info icons at the beginning of each row within the modal overlay reveals the full information modal overlay for the respective study.}
    \label{fig: Graphical View}
\end{figure*}

\subsection{Graphical View}\label{sec: Graphical View}


The \textit{Graphical View} (\autoref{fig: Graphical View}) complements the Tabular View by offering an aggregated visual overview of the dataset, presenting an alternative perspective on the first eight guiding questions (\qref{Q1–Q8}). It addresses the need for high-level insights, particularly valuable during early-stage exploration and when identifying dominant characteristics or gaps in the literature. By visualizing the distribution of values within each criterion, it enables researchers to interpret patterns across the full or filtered dataset at a glance.

Each selected criterion is represented as a bar chart that shows the frequency of its corresponding answer options. This format was chosen for its accessibility, interpretability, and compatibility with multi-attribute filtering. A dynamic layout ensures that chart size adapts to the number of bars, preserving readability. By default, key information (\textit{Year}, \textit{Location}, \textit{Input Body Part}, and \textit{Gesture Type}) is visualized, with additional bar charts available via a toggle menu. Sidebar filters allow users to refine their selection interactively, showing how attribute distributions shift based on specific selections. To support a smooth transition from high-level insights to detailed study information, clicking on any bar opens a modal overlay showing key information for all studies it represents. Full study records can be accessed via info icons within the overlay. This interaction design enables fluid movement between aggregate analysis and in-depth inspection, supporting both broad and focused research workflows.

\subsection{Similarity View}\label{sec: Similarity View}



\begin{figure*}[t]
    \centering
    \includegraphics[width=\linewidth]{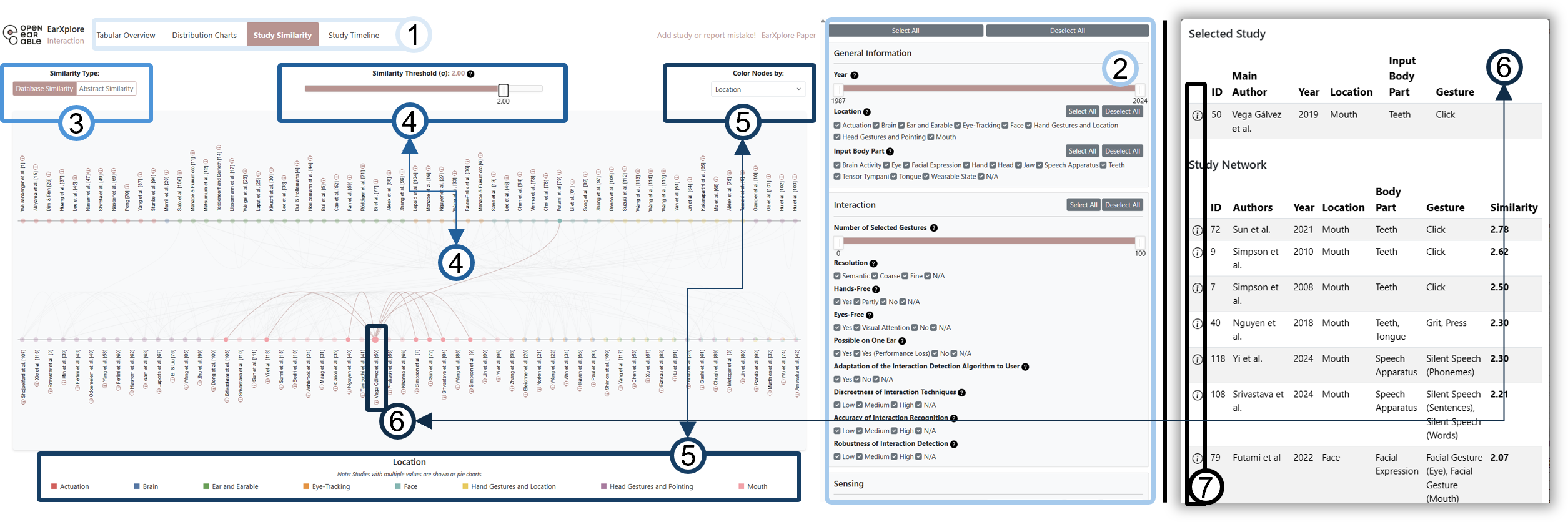} 
    \caption{Similarity View -- (1) The Similarity View can be selected via the view selection menu. (2) Filters allow the user to refine the database along several criteria. (3) The user can choose between \textit{Database Similarity} and \textit{Abstract Similarity}. (4) A threshold slider controls which similarity connections are displayed. (5) The nodes representing the studies can be colored and sorted along several criteria. (6) Clicking on a node opens a modal overlay showing key information on all studies that meet the similarity threshold with the selected study. (7) Clicking the info icons at the beginning of each row within the modal overlay reveals the full information modal overlay for the respective study. The full information view can also be displayed via the info icons attached to each study node.}
    \Description{Similarity View -- The figure shows a screenshot of the Similarity View. Its components, functionalities, and interaction potentials are further explained in the following. (1) The Similarity View can be selected via the view selection menu at the top of the page. (2) Filters at the sidebar allow the user to refine the database along several criteria. (3) The user can choose between \textit{Database Similarity} and \textit{Abstract Similarity}. (4) A threshold slider controls which similarity connections are displayed. This threshold slider goes from -3 to 3 (representing standard deviation units in a Gaussian distribution). (5) The nodes representing the studies can be colored and sorted along several criteria. These can be selected via a dropdown menu. (6) Clicking on a node opens a modal overlay showing key information on all studies that meet the similarity threshold with the selected study ordered by their similarity and the selected study itself. These key information are ID, Main Author, Year, Location, Input Body Part, Gesture, and Similarity (only for the similar studies, not the study itself). (7) Clicking the info icons at the beginning of each row within the modal overlay reveals the full information modal overlay for the respective study. The full information view can also be displayed via the info icons attached to each study node.}
    \label{fig: Similarity View}
\end{figure*}

The \textit{Similarity View} (\autoref{fig: Similarity View}) was developed to support researchers in identifying thematic and methodological relationships across the literature, especially when connections between studies are not discoverable through shared keywords or simple filtering. It addresses \qref{Q9} by visualizing pairwise similarity among studies, enabling users to explore clusters of conceptually or structurally related work. This view is particularly valuable for uncovering overlooked connections, surfacing adjacent research areas, or identifying prior work that may inform new directions.

To support this goal, the interface offers two similarity modes: \textit{\hyperref[par: database_similarity]{Database Similarity}}, computed from the data on the 33 criteria, and \textit{\hyperref[par: abstract_similarity]{Abstract Similarity}}, derived from natural language processing of the study abstracts (both described in detail below). This dual approach enables both structured and semantic comparison, reflecting different dimensions of relatedness. A similarity threshold slider lets users focus on stronger relationships by limiting the view to connections above a chosen threshold, reducing visual clutter and enabling targeted exploration. Each study is represented as a node labeled with its study ID and the main author's name. Users can refine their selection using sidebar filters and customize the visual layout by coloring or sorting nodes based on categorical attributes (e.g., \textit{Input Body Part} or \textit{Gesture}). Clicking on a node opens a modal overlay listing all studies connected above the similarity threshold, along with key information. Full information on each study can be accessed via info icons within the modal or directly on the label next to the nodes, allowing seamless transition from network-level insight to individual study inspection.

In the following, we describe the underlying computation of the two similarity metrics in more detail.

\begin{figure*}[t]
    \centering
    \includegraphics[width=\linewidth]{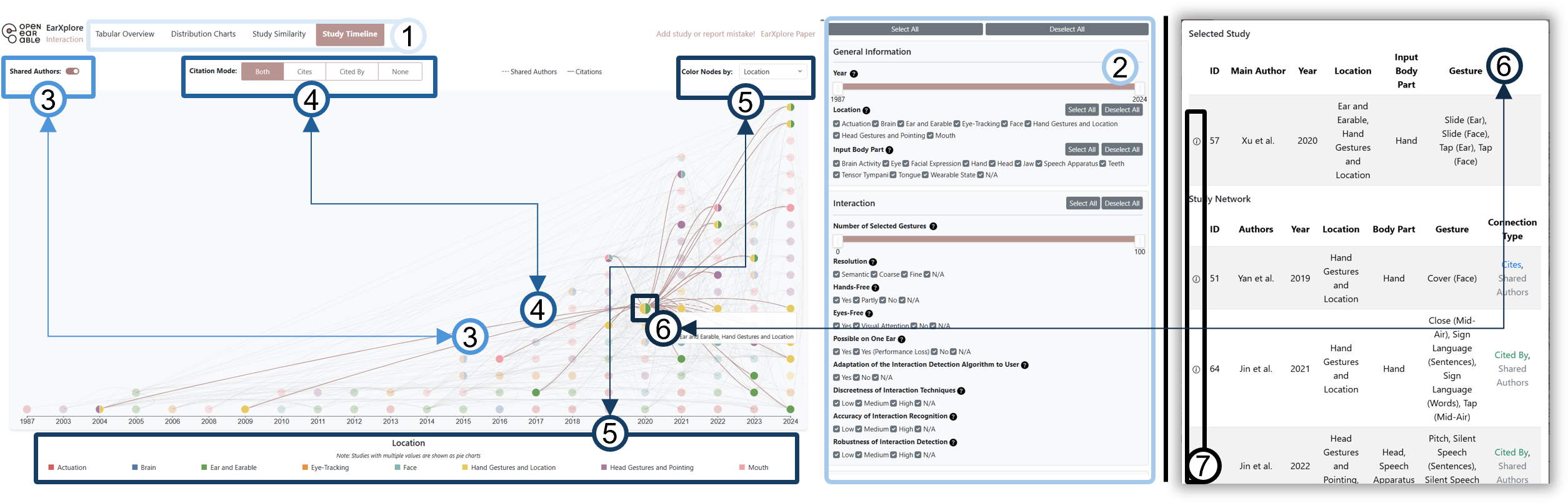} 
    \caption{Timeline View -- (1) The Timeline View can be selected via the view selection menu. (2) Filters allow the user to refine the database along several criteria. (3) The user can display shared author connections as dashed lines. (4) Citation connections, including their direction, can be shown as solid lines. (5) The nodes representing the studies can be colored and sorted along several criteria. (6) Clicking on a node opens a modal overlay displaying key information on all studies connected to the selected study through shared authorship or citations based on the user's selection. (7) Clicking the info icons at the beginning of each row within the modal overlay reveals the full information modal overlay for the respective study.}
    \Description{Timeline View -- The figure shows a screenshot of the Timeline View. Its components, functionalities, and interaction potentials are further explained in the following. (1) The Timeline View can be selected via the view selection menu at the top of the page. (2) Filters at the sidebar allow the user to refine the database along several criteria. (3) The user can display shared author connections as dashed lines. (4) Citation connections, including their direction, can be shown as solid lines. (5) The nodes representing the studies can be colored and sorted along several criteria using a dropdown menu. (6) Clicking on a node opens a modal overlay displaying key information on all studies connected to the selected study through shared authorship or citations based on the user's selection as well as the selected study itself. These key information are ID, Main Author, Year, Location, Input Body Part, Gesture, and Connection Type (only for the connected studies, not the study itself). (7) Clicking the info icons at the beginning of each row within the modal overlay reveals the full information modal overlay for the respective study.}
    \label{fig: Timeline View}
\end{figure*}

\paragraph{Database Similarity}\label{par: database_similarity}

Analogous to \citet{di-luca_locomotive_2021} and \citet{seifi_haptipedia_2019}, we computed pairwise similarities between studies based on the criteria defined in our database. To enable consistent comparison across heterogeneous data types, we first normalized all ordinal single-value columns by mapping their values to the [0, 1] range (e.g., Low = 0, Medium = 0.5, High = 1). The only numerical columns \textit{Number of Selected Gestures} was log-transformed prior to normalization to account for exponential differences in gesture counts and then min-max normalized. Using these normalized values, we calculated how similar two studies were by first taking the absolute difference between them for each criterion, subtracting that result from 1 to get a similarity score per criterion, and then adding up those scores across all applicable criteria. For categorical fields (with multiple values), we used an adjusted Jaccard index, defined as the size of the intersection divided by the geometric mean of the cardinalities of both sets. This adjustment mitigates bias towards larger sets. Our approach ensures that all criteria have the same weight in the overall measure, as it maps their values to the [0, 1] range. Note that comparisons involving missing or non-applicable (N/A) values were treated as maximally dissimilar (i.e., similarity = 0). To ensure interpretability, we standardized the resulting similarity scores by converting them to \textit{z}-scores relative to the distribution of all pairwise similarities in the corpus.

\paragraph{Abstract Similarity}\label{par: abstract_similarity}

To enable study comparison independent of the criteria defined in the database, we introduce the concept of abstract similarity. Prior work has used textual similarity to compare research content (e.g., via document embeddings \cite{ostendorff_specialized_2022, liu_measuring_2017}). To the best of our knowledge, however, using text embeddings of study abstracts to compute similarity has not been previously applied, particularly within a structured database context. We extracted the abstracts from all studies in our database and retrieved their text embedding representation using the \texttt{gemini-embedding-exp-03-07} model \cite{kilpatrick_state-2025}. This model was selected based on its top-ranking performance in the Massive Text Embedding Benchmark (MTEB) leaderboard \cite{enevoldsen_mmteb_2025} as of March 26, 2025. We configured the model for the \texttt{clustering} task, which aligns closely with our goal of measuring study similarity\footnote{see here: \href{https://ai.google.dev/gemini-api/docs/embeddings}{https://ai.google.dev/gemini-api/docs/embeddings}}.
Using these embeddings, we computed the pairwise cosine similarity between all abstracts in the database. To ensure interpretability, we standardized the resulting similarity scores by converting them to \textit{z}-scores relative to the distribution of all pairwise similarities in the corpus.

\subsection{Timeline View}\label{sec: Timeline View}


The \textit{Timeline View} (\autoref{fig: Timeline View}) was developed to provide a temporal and relational perspective on the body of literature. Individual studies are situated within broader historical, collaborative, and citation-driven trends, thus directly addressing \qref{Q10}. It supports longitudinal exploration of how research on interaction with earables has evolved over time, how scholarly communities form and shift, and how ideas propagate through citation networks.

To support this, each study is visualized as a node placed along a horizontal timeline based on its publication year. Researchers can apply filters to narrow their selection and dynamically observe how trends unfold over time within a selected subset. Two types of relationships can be overlaid: shared authorship (dashed lines) and citation links (solid lines), with the option to visualize citation directionality. These features enable the identification of research clusters, influential publications, or isolated works. Nodes can also be colored or sorted by categorical attributes (e.g., \textit{Input Body Part}, \textit{Gesture Type}), allowing researchers to discover temporal or thematic patterns at a glance. Clicking on a node reveals a modal with key information about the study and its direct connections, with full information accessible via info icons. This allows users to fluidly shift between high-level trends and study-specific detail.

In the following, we describe the methods used to extract shared author and citation data.

\paragraph{Shared Authors and Cross-Citation Extraction}\label{sec: Shared Author and Cross-Citation Extraction}

To identify shared authorship across studies, we manually extracted author lists and determined which studies included overlapping authors using scripts. For citations, we followed an approach similar to \citet{seifi_haptipedia_2019}, leveraging GROBID \cite{grobid_2025}, a machine learning library for parsing bibliographic metadata from PDFs. We first compiled BibTeXs for all studies in the corpus and then parsed each study’s reference list, generating BibTeXs for each referenced work. Using fuzzy matching techniques, we compared these entries against our corpus to identify intra-corpus citations. Citations with low matching confidence were flagged and manually reviewed to ensure accuracy.

%% file: sections/5_Insights.tex
\section{Discovering Research through EarXplore}\label{sec: Insights}

Having introduced \textit{EarXplore}, we turn to providing an overview of its data. We begin with a snapshot of the platform’s current content, organized along its structured set of criteria (\autoref{sec: snap_shot}). Next, we present a scenario-based walkthrough to illustrate potential use cases and research workflows (\autoref{sec: earxplore scenario}). Finally, we outline how the platform is designed to remain relevant and sustainable as a community-driven resource (\autoref{sec: earxplore update}).

\subsection{Overview of Earable Interaction}\label{sec: snap_shot}

\begin{figure*}[t]
    \centering
    \includegraphics[width=\linewidth]{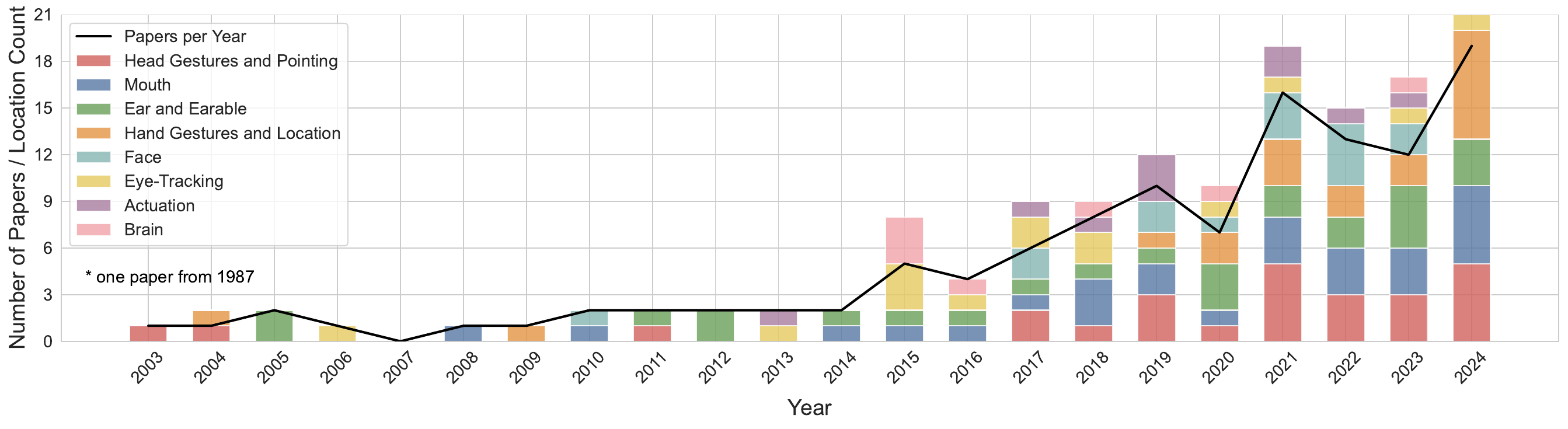} 
    \caption{Temporal trends in earable interaction research. The total publications per year are indicated by the black line and the distribution of interaction \textit{Locations} \cite{roddiger_sensing_2022} within those years by the colored bars. Individual studies may contribute to multiple location categories.}
    \Description{Temporal trends in earable interaction research. The black line indicates the total number of studies published per year. Colored bars represent the frequency of locations present in those studies, with one study potentially contributing to multiple location categories. One can see that there is a steady upward trend in publications starting around 2015-2017. Here is the data the figure contains as listing:
    
    # Location Timeline Listings

    Timeline Locations (Totals 2003–2024): "Head Gestures and Pointing" = 26, "Mouth" = 26, "Ear and Earable" = 25, "Hand Gestures and Location" = 19, "Face" = 15, "Eye-Tracking" = 14, "Actuation" = 10, "Brain" = 7
    
    # Per-Year Breakdown
    
    2003: "Head Gestures and Pointing" = 1
    2004: "Head Gestures and Pointing" = 1, "Hand Gestures and Location" = 1
    2005: "Ear and Earable" = 2
    2006: "Eye-Tracking" = 1
    2008: "Mouth" = 1
    2009: "Hand Gestures and Location" = 1
    2010: "Mouth" = 1, "Face" = 1
    2011: "Head Gestures and Pointing" = 1, "Ear and Earable" = 1
    2012: "Ear and Earable" = 2
    2013: "Eye-Tracking" = 1, "Actuation" = 1
    2014: "Mouth" = 1, "Ear and Earable" = 1
    2015: "Mouth" = 1, "Ear and Earable" = 1, "Eye-Tracking" = 3, "Brain" = 3
    2016: "Mouth" = 1, "Ear and Earable" = 1, "Eye-Tracking" = 1, "Brain" = 1
    2017: "Head Gestures and Pointing" = 2, "Mouth" = 1, "Ear and Earable" = 1, "Face" = 2, "Eye-Tracking" = 2, "Actuation" = 1
    2018: "Head Gestures and Pointing" = 1, "Mouth" = 3, "Ear and Earable" = 1, "Eye-Tracking" = 2, "Actuation" = 1, "Brain" = 1
    2019: "Head Gestures and Pointing" = 3, "Mouth" = 2, "Ear and Earable" = 1, "Hand Gestures and Location" = 1, "Face" = 2, "Actuation" = 3
    2020: "Head Gestures and Pointing" = 1, "Mouth" = 1, "Ear and Earable" = 3, "Hand Gestures and Location" = 2, "Face" = 1, "Eye-Tracking" = 1, "Brain" = 1
    2021: "Head Gestures and Pointing" = 5, "Mouth" = 3, "Ear and Earable" = 2, "Hand Gestures and Location" = 3, "Face" = 3, "Eye-Tracking" = 1, "Actuation" = 2
    2022: "Head Gestures and Pointing" = 3, "Mouth" = 3, "Ear and Earable" = 2, "Hand Gestures and Location" = 2, "Face" = 4, "Actuation" = 1
    2023: "Head Gestures and Pointing" = 3, "Mouth" = 3, "Ear and Earable" = 4, "Hand Gestures and Location" = 2, "Face" = 2, "Eye-Tracking" = 1, "Actuation" = 1, "Brain" = 1
    2024: "Head Gestures and Pointing" = 5, "Mouth" = 5, "Ear and Earable" = 3, "Hand Gestures and Location" = 7, "Eye-Tracking" = 1}
    \label{fig: Location Timeline}
\end{figure*}

To demonstrate the analytical potential of \textit{EarXplore}'s curated content, we examine it along the set of guiding questions answerable by filters (\qref{Q1}-\qref{Q8}). This analysis reveals both the current state of earable interaction research and demonstrates how the platform can surface field-wide insights that might otherwise remain scattered across individual studies. To establish temporal context for our analysis, we begin by presenting the field's publication trajectory in \autoref{fig: Location Timeline}. While the earliest study in our database dates to 1987, earable interaction research has gained substantial momentum only recently. 85\% of studies were published since 2015 and approximately half of the entire corpus appeared within the last five years alone. Within this rapidly expanding body of work, our analysis reveals several key patterns in interaction approaches and methodologies.

\begin{figure*}[t]
    \centering
    \includegraphics[width=\linewidth]{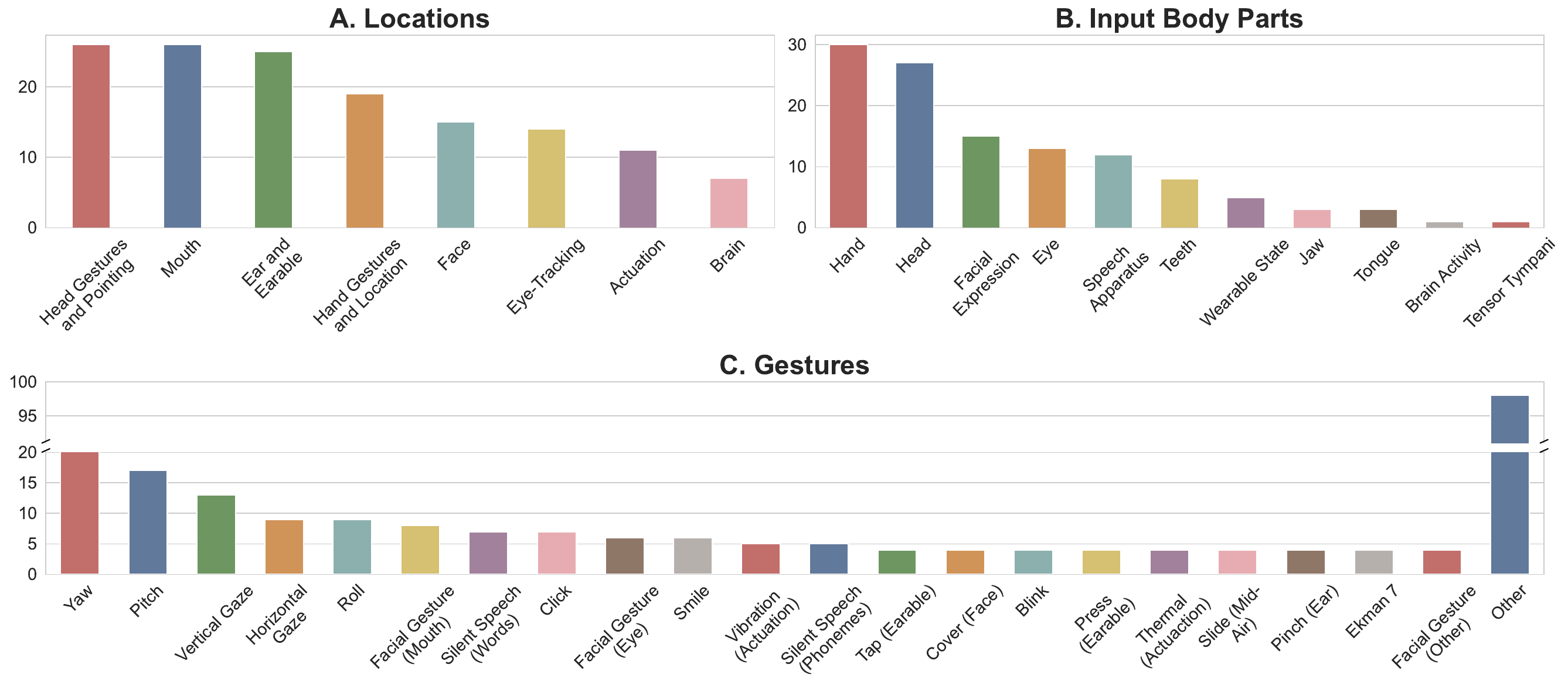} 
    \caption{Interaction types and input modalities (\qref{Q1}). A. The specific locations \cite{roddiger_sensing_2022} on the body where the interactions with the earable device occur.; B. The parts of the body used to elicit the interactions with the earable device.; C. The specific actions the user must perform to initiate interactions with the earable device. Gestures only present in elicitation studies are not considered.}
    \Description{Interaction types and input modalities (\qref{Q1}). Notable observations and additional information are listed in \autoref{sec: snap_shot}. A. The specific locations \cite{roddiger_sensing_2022} on the body where the interactions with the earable device occur.; B. The parts of the body used to elicit the interactions with the earable device.; C. The specific actions the user must perform to initiate interactions with the earable device. Gestures only present in elicitation studies are not considered. Here is the data the figure contains as a listing:
    
    A. Locations: "Head Gestures and Pointing" = 26, "Mouth" = 26, "Ear and Earable" = 25, "Hand Gestures and Location" = 19, "Face" = 15, "Eye-Tracking" = 14, "Actuation" = 11, "Brain" = 7
    B. Input Body Parts: "Hand" = 30, "Head" = 27, "Facial Expression" = 15, "Eye" = 13, "Speech Apparatus" = 12, "Teeth" = 8, "Wearable State" = 5, "Jaw" = 3, "Tongue" = 3, "Brain Activity" = 1, "Tensor Tympani" = 1
    C. Gestures: "Yaw" = 21, "Pitch" = 17, "Vertical Gaze" = 13, "Horizontal Gaze" = 9, "Roll" = 9, "Facial Gesture (Mouth)" = 8, "Silent Speech (Words)" = 7, "Click" = 7, "Facial Gesture (Eye)" = 6, "Smile" = 6, "Vibration (Actuation)" = 5, "Silent Speech (Phonemes)" = 5, "Tap (Earable)" = 4, "Cover (Face)" = 4, "Blink" = 4, "Press (Earable)" = 4, "Thermal (Actuaction)" = 4, "Slide (Mid-Air)" = 4, "Pinch (Ear)" = 4, "Ekman 7" = 4, "Facial Gesture (Other)" = 4, "Other" = 98}
    \label{fig: Summary Q1}
\end{figure*}

Investigating the types of interactions performed (\qref{Q1}, \autoref{fig: Summary Q1}), the most prominent \textit{Locations} where interaction occurs, categorized by \citet{roddiger_sensing_2022}, are centered around the head, mouth, and the ear or earable itself. Hand gestures around the ear, on the other hand, have gained recent momentum, with more than a third of studies in 2024 publishing within that area. Consequently, the most frequently used \textit{Input Body Parts} were the \textit{Head} and \textit{Hand}. Notable unique input modalities include a study directly utilizing \textit{Brain Activity} \cite{ID69_merrill_classifying_2016} and another employing the \textit{Tensor Tympani}, a small muscle in the ear \cite{ID227_roddiger_earrumble_2021}. In total, 67 distinct gesture types were identified, underscoring the diversity of the field, with gestures comprising for instance or \textit{Twisting}, \textit{Gripping}, \textit{Covering the Face}, or a \textit{Calling Gesture}. The most common gestures were \textit{Roll}, \textit{Pitch}, and \textit{Yaw}, indicating both their widespread use and the relatively narrow variation within head-related gestures.

\begin{figure*}[t]
    \centering
    \includegraphics[width=\linewidth]{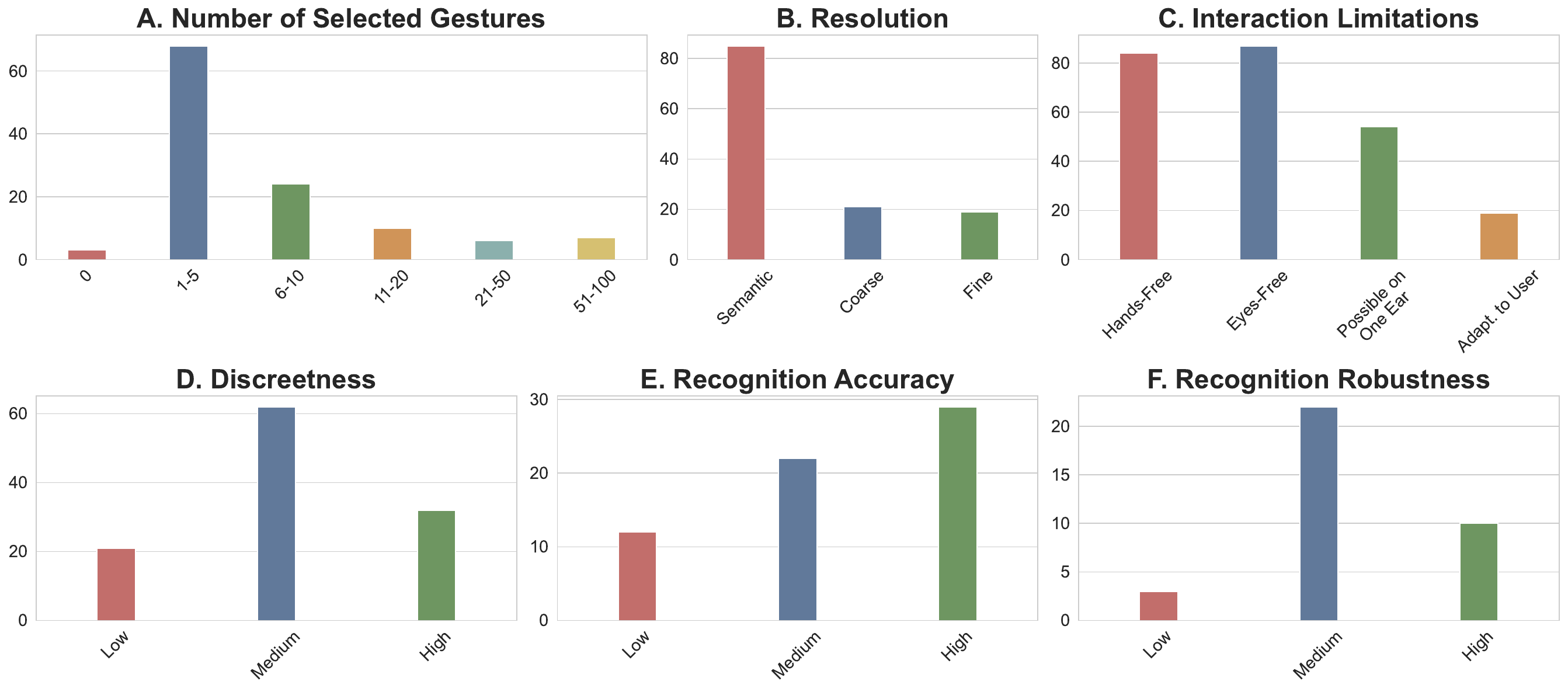} 
    \caption{Interaction characteristics and restrictions (\qref{Q2}). A. Number of gestures selected for interaction. Gestures only present in elicitation studies are not counted.; B. Whether interactions involve a single distinct input (Semantic) or are continuous (distinct steps = Coarse, unnoticeable steps = Fine).; C. Whether the (1) hands or (2) eyes play an active role in the interactions with the earable. (3) Whether the interaction detection is (also) possible using an earable covering only one ear. (4) Whether it has been demonstrated that the interaction detection algorithm can be adapted to a user.; D. Extent to which interaction techniques remain undetectable in public settings.; E. The system’s ability to accurately detect interactions. Accuracy-reporting studies only.; F. Demonstrated robustness of interaction detection performance (accuracy drop $\leq 10\%$) across additional settings (e.g., a café) or conditions (e.g., walking) that may cause interference. Accuracy-reporting studies only.}
    \Description{Interaction characteristics and restrictions (\qref{Q2}). Notable observations and additional information are listed in \autoref{sec: snap_shot}. A. Number of gestures selected for interaction. Gestures only present in elicitation studies are not counted.; B. Whether interactions involve a single distinct input (Semantic) or are continuous (distinct steps = Coarse, unnoticeable steps = Fine).; C. Whether the (1) hands or (2) eyes play an active role in the interactions with the earable. (3) Whether the interaction detection is (also) possible using an earable covering only one ear. (4) Whether it has been demonstrated that the interaction detection algorithm can be adapted to a user.; D. Extent to which interaction techniques remain undetectable in public settings.; E. The system’s ability to accurately detect interactions. Accuracy-reporting studies only.; F. Demonstrated robustness of interaction detection performance across additional settings (e.g., a café) or conditions (e.g., walking) that may cause interference ($\leq 10\%$ accuracy drop). Accuracy-reporting studies only. Here is the data the figure contains as a listing:
    
    A. Number of Selected Gestures: "0" = 3, "1-5" = 68, "6-10" = 24, "11-20" = 10, "21-50" = 6, "51-100" = 7
    B. Resolution: "Semantic" = 85, "Coarse" = 21, "Fine" = 19
    C. Interaction Limitations: "Hands-Free" = 84, "Eyes-Free" = 87, "Possible on One Ear" = 54, "Adapt. to User" = 19
    D. Discreetness: "Low" = 21, "Medium" = 62, "High" = 32
    F. Recognition Accuracy: "Low" = 12, "Medium" = 22, "High" = 29
    G. Recognition Robustness: "Low" = 3, "Medium" = 22, "High" = 10}
    \label{fig: Summary Q2}
\end{figure*}

Data on interaction restrictions (\qref{Q2}, \autoref{fig: Summary Q2}) show that most studies employ a relatively small \textit{Number of Selected Gestures}. The three studies reporting zero selected gestures \cite{ID296_rateau_leveraging_2022, ID234_chen_exploring_2020, ID275_shimon_exploring_2024} are pure elicitation studies. In contrast, those featuring more than fifty gestures are predominantly focused on silent speech \cite{ID226_jin_sonicasl_2021, ID315_jin_earcommand_2022, ID307_srivastava_muteit_2022, ID346_dong_rehearsse_2024, ID640_srivastava_unvoiced_2024, ID506_sun_earssr_2024}, alongside one actuation study employing vibrotactile feedback \cite{ID388_yang_customized_2022}. Most studies favor single, distinct inputs (\textit{Semantic}) over continuous ones. Furthermore, approximately two-thirds support \textit{Hands-Free} or \textit{Eyes-Free} interaction, with half of studies supporting both. Addressed in slightly less than half of studies are interactions that are \textit{Possible on One Ear}, and fewer than 20\% of studies incorporate \textit{Adaptation of the Interaction Detection Algorithm to the User} -- though this aspect has shown an upward trend in recent years. Approximately a quarter of studies scores \textit{High} on \textit{Discreetness}. Only half of the studies were assigned a value for \textit{Accuracy of Interaction Recognition} and \textit{Robustness of Interaction Detection} as the other half does not report accuracy values. While recognition accuracy is often \textit{High}, most studies do not successfully demonstrate the robustness of their interactions in both a different setting and context.

\begin{figure*}[t]
    \centering
    \includegraphics[width=\linewidth]{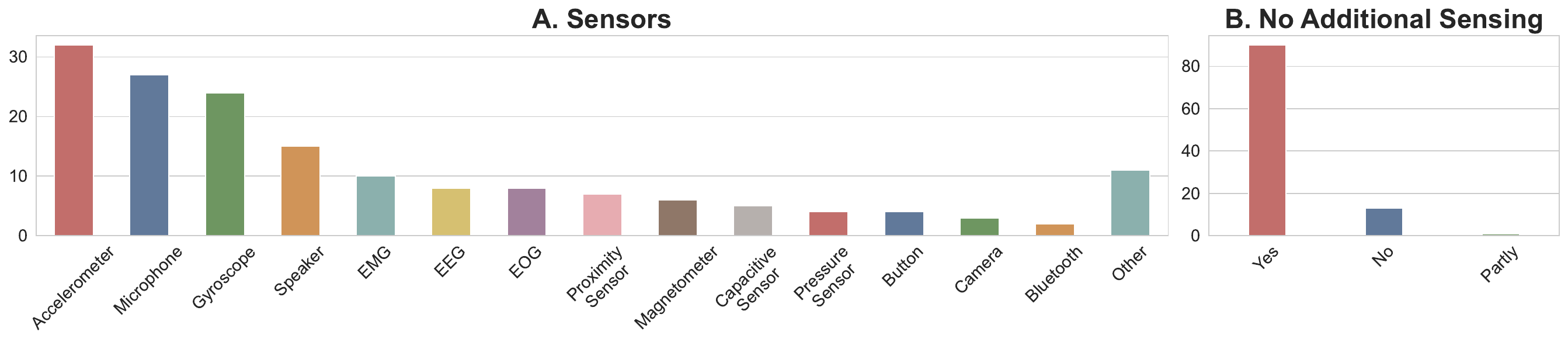} 
    \caption{Sensing modalities (\qref{Q3}). A. The specific sensors the earable needs to enable (e.g., speaker for ultrasound) and/or recognize (e.g., a microphone) the interactions.; B. No additional devices are needed for sensing the interaction presented in the study.}
    \Description{Sensing modalities (\qref{Q3}). Notable observations and additional information are listed in \autoref{sec: snap_shot}. A. The specific sensors the earable needs to enable (e.g., speaker for ultrasound) and/or recognize (e.g., a microphone) the interactions.; B. No additional devices are needed for sensing the interaction presented in the study. Here is the data the figure contains as a listing: 
    
    A. Sensors: "Accelerometer" = 32, "Microphone" = 27, "Gyroscope" = 24, "Speaker" = 15, "EMG" = 10, "EEG" = 8, "EOG" = 8, "Proximity Sensor" = 7, "Magnetometer" = 6, "Capacitive Sensor" = 5, "Pressure Sensor" = 4, "Button" = 4, "Camera" = 3, "Bluetooth" = 2, "Other" = 11
    B. No Additional Sensing: "Yes" = 90, "No" = 13, "Partly" = 1}
    \label{fig: Summary Q3}
\end{figure*}

\begin{figure*}[h]
    \centering
    \includegraphics[width=\linewidth]{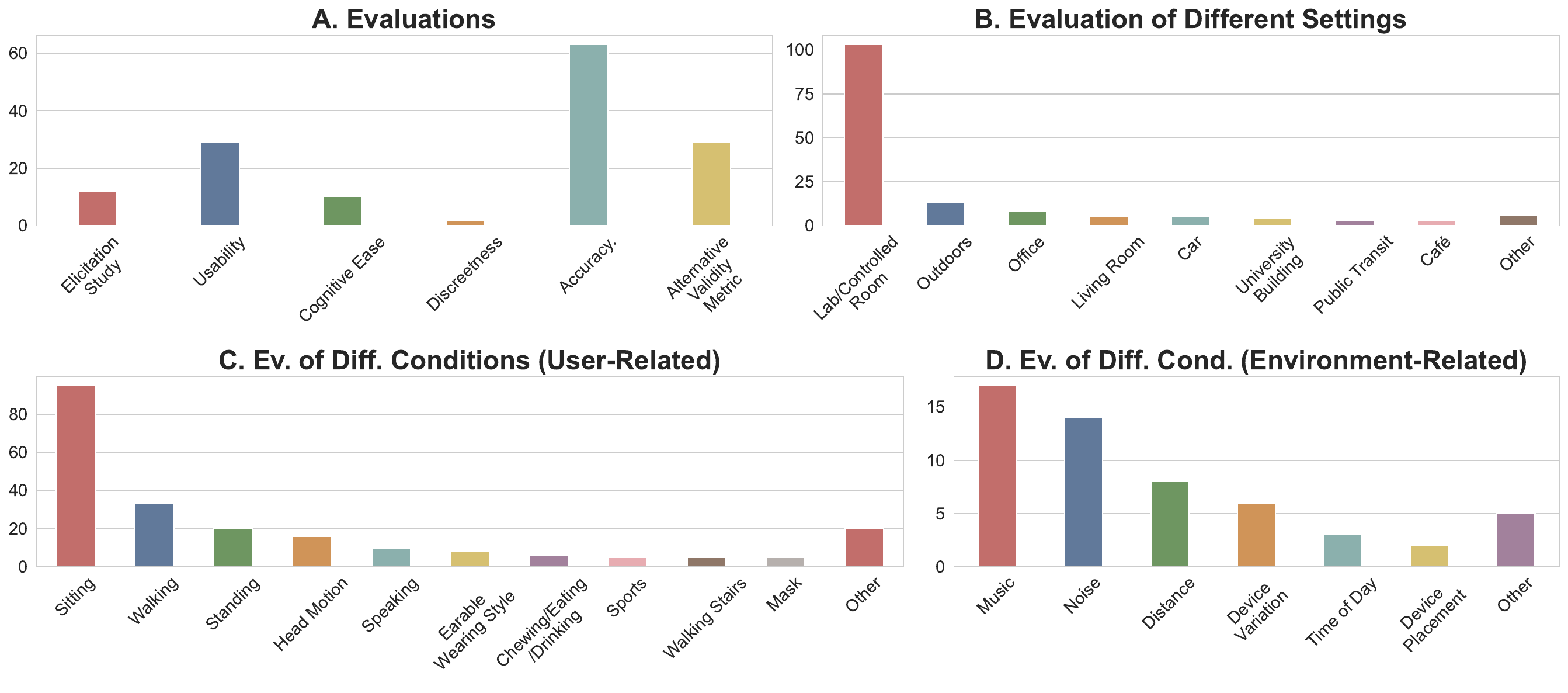} 
    \caption{Study characteristics and evaluations (\qref{Q4}). A. (1) An elicitation study on interaction with earables has been performed. Whether the (2) usability, (3) cognitive ease, (4) discreetness, or (5) social acceptability of the interactions was assessed. Whether the (6) accuracy or (7) an alternative validity metric of the system’s interactions detection was assessed.; B. The different settings evaluated.; C. The different user-related conditions evaluated.; D. The different environment-related conditions evaluated. Only deliberately induced environmental conditions not inherently part of the setting are listed, to avoid redundancy.}
    \Description{Study characteristics and evaluations (\qref{Q4}). Notable observations and additional information are listed in \autoref{sec: snap_shot}. A. (1) An elicitation study on interaction with earables has been performed. Whether the (2) usability, (3) cognitive ease, (4) discreetness, or (5) social acceptability of the interactions was assessed. Whether the (6) accuracy or (7) an alternative validity metric of the system’s interactions detection was assessed.; B. The different settings evaluated.; C. The different user-related conditions evaluated.; D. The different environment-related conditions evaluated. Only deliberately induced environmental conditions not inherently part of the setting are listed, to avoid redundancy. Here is the data the figure contains as a listing:
    
    A. Evaluations: "Elicitation Study" = 12, "Usability Evaluations" = 29, "Cognitive Ease Evaluations" = 10, "Discreetness of Interactions Evaluations" = 2, "Social Acceptability of Interactions Evaluations" = 11, "Accuracy of Interactions Evaluations" = 63, "Alternative Interaction Validity Evaluations" = 29
    B. Evaluation of Different Settings: "Lab/Controlled Room" = 103, "Outdoors" = 13, "Office" = 8, "Living Room" = 5, "Car" = 5, "University Building" = 4, "Public Transit" = 3, "Café" = 3, "Other" = 6
    C. Ev. of Diff. Conditions (User-Related): "Sitting" = 95, "Walking" = 33, "Standing" = 20, "Head Motion" = 16, "Speaking" = 10, "Earable Wearing Style" = 8, "Chewing/Eating/Drinking" = 6, "Sports" = 5, "Walking Stairs" = 5, "Mask" = 5, "Other" = 20
    D. Ev. of Diff. Cond. (Environment-Related): "Music" = 17, "Noise" = 14, "Distance" = 8, "Device Variation" = 6, "Time of Day" = 3, "Device Placement" = 2, "Other" = 5}
    \label{fig: Summary Q4}
\end{figure*}

\begin{figure*}[h]
    \centering
    \includegraphics[width=\linewidth]{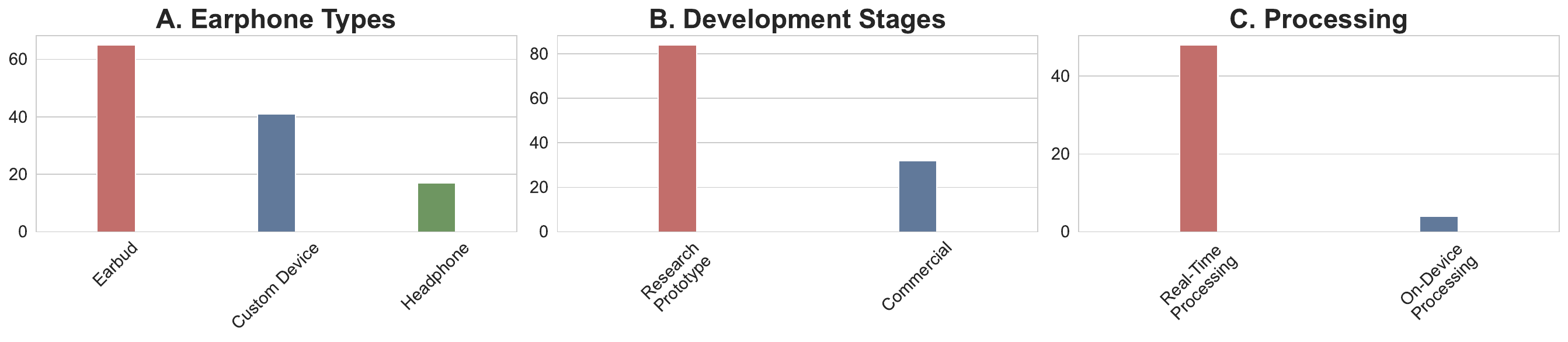} 
    \caption{Devices and processing architectures (\qref{Q5}). A. Type of earphone used.; B. Commercial device or research prototype employed.; C. The system is demonstrated to (1) immediately respond to a user’s interaction and (2) interaction detection occurs on the earable.}
    \Description{Devices and processing architectures (\qref{Q5}). Notable observations and additional information are listed in \autoref{sec: snap_shot}. A. Type of earphone used.; B. Commercial device or research prototype employed.; C. The system is demonstrated to (1) immediately respond to a user’s interaction and (2) interaction detection occurs on the earable. Here is the data the figure contains as a listing: 
    
    A. Earphone Types: "Earbud" = 65, "Custom Device" = 41, "Headphone" = 17
    B. Development Stages: "Research Prototype" = 84, "Commercial" = 32
    C. Processing: "Real-Time Processing" = 48, "On-Device Processing" = 4}
    \label{fig: Summary Q5}
\end{figure*}

\begin{figure*}[t]
    \centering
    \includegraphics[width=\linewidth]{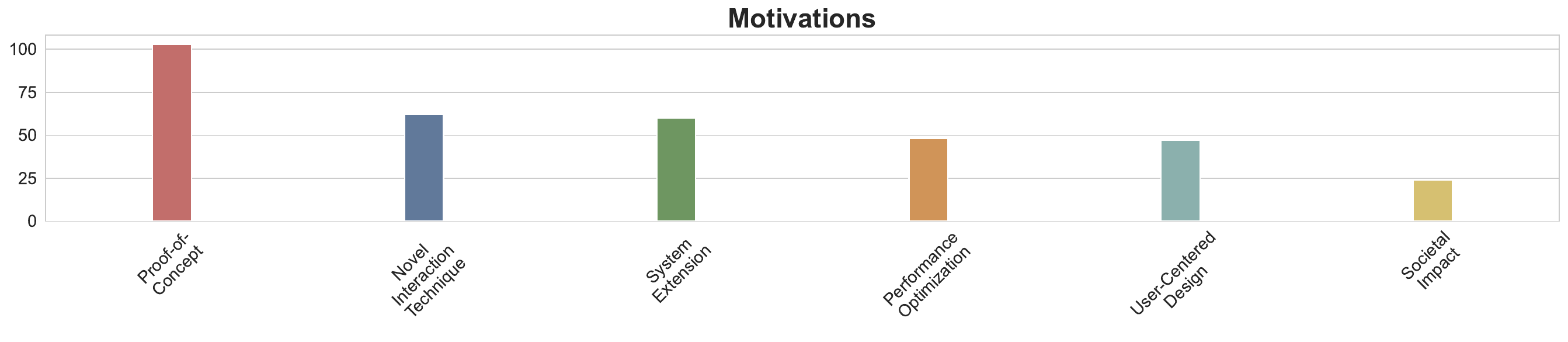} 
    \caption{Study motivations (\qref{Q6}). The motivations the authors explicitly outline for their study on interaction with earables. Only six core motivations recurring across studies were extracted, to keep the coding scheme concise and consistent.}
    \Description{Study motivations (\qref{Q6}). Notable observations and additional information are listed in \autoref{sec: snap_shot}. The motivations the authors explicitly outline for their study on interaction with earables. Only six core motivations recurring across studies were extracted, to keep the coding scheme concise and consistent. Here is the data the figure contains as a listing: 
    
    Motivations: "Proof-of-Concept" = 103, "Novel Interaction Technique" = 62, "System Extension" = 60, "Performance Optimization" = 48, "User-Centered Design" = 47, "Societal Impact" = 24}
    \label{fig: Summary Q6}
\end{figure*}

\begin{figure*}[t]
    \centering
    \includegraphics[width=\linewidth]{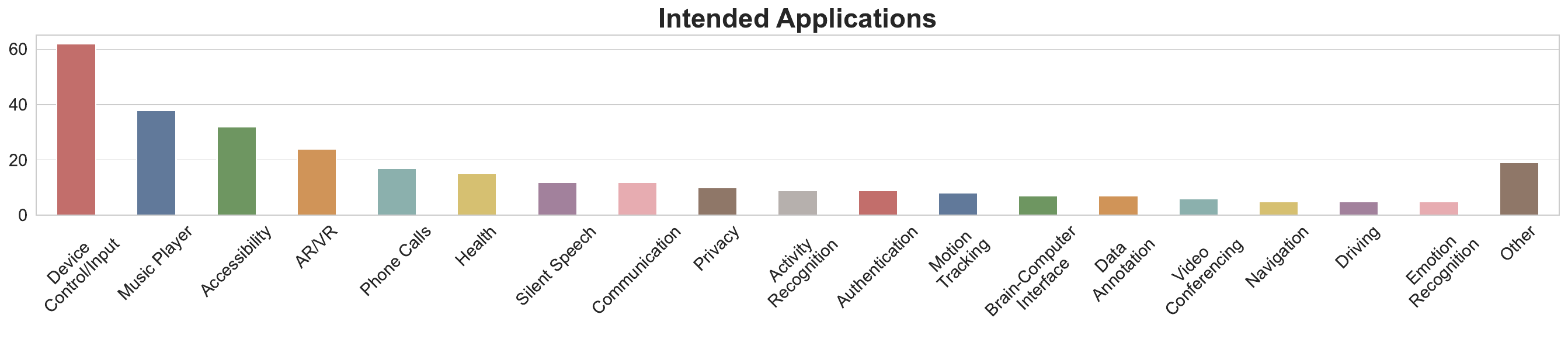} 
    \caption{Intended Applications (\qref{Q7}). The applications the authors explicitly suggest for their proposed interaction with earables.}
    \Description{Intended Applications (\qref{Q7}). Notable observations and additional information are listed in \autoref{sec: snap_shot}. The applications the authors explicitly suggest for their proposed interaction with earables. Here is the data the figure contains as a listing: 
    
    Intended Applications: "Device Control/Input" = 62, "Music Player" = 38, "Accessibility" = 32, "AR/VR" = 24, "Phone Calls" = 17, "Health" = 15, "Silent Speech" = 12, "Communication" = 12, "Privacy" = 10, "Activity Recognition" = 9, "Authentication" = 9, "Motion Tracking" = 8, "Brain-Computer Interface" = 7, "Data Annotation" = 7, "Video Conferencing" = 6, "Navigation" = 5, "Driving" = 5, "Emotion Recognition" = 5, "Other" = 19}
    \label{fig: Summary Q7}
\end{figure*}

An analysis of the \textit{Sensors} used (\qref{Q3}, \autoref{fig: Summary Q3}) reveals that most studies rely on either \textit{IMUs} or \textit{Microphones}, with use of the latter steadily increasing over time, often in combination with \textit{Speakers}. In total, 25 different sensing modalities have been employed, reflecting the diversity of the field. Additionally, some studies incorporate supplementary devices such as smartphones \cite{ID509_hu_headtrack_2024, ID482_hu_combining_2024}, speakers \cite{ID498_ge_ehtrack_2024}, or additional wearables \cite{ID376_li_enabling_2023} to enable interaction.

Most studies on earable interaction were conducted (\qref{Q4}, \autoref{fig: Summary Q4}) with limited evaluation of their proposed systems. Specifically, only 28\% of studies conducted at least two types of evaluations, and just 12\% reported three or more. This aligns with the observation that only 28\% of studies evaluated interactions outside of a \textit{Lab} or \textit{Controlled Room} setting. A similar proportion examined the effects of varying \textit{Environment-Related Conditions}, while only half of studies conducted any \textit{Evaluation of User-Related Conditions} beyond basic \textit{Sitting}. However, time trends show increased attention toward evaluation in additional settings and conditions. In 2024, over one-third of studies reported testing in multiple real-world settings. A similar pattern is seen for \textit{Evaluations in Different Conditions}, with more than two-thirds of studies published in 2024 evaluating their systems under at least one additional condition beyond the standard.

In examining the devices and processing techniques employed (\qref{Q5}, \autoref{fig: Summary Q5}), the \textit{Earbud} emerges as the favorite device. While most studies rely on custom-built research prototypes, several use commercial solutions, with the eSense platform \cite{kawsar_earables_2018} being the most frequently adopted \cite{ID626_ma_oesense_2021, ID608_alkiek_eargest_2022, ID284_alkiek_earbender_2023, ID80_ferlini_head_2019, ID217_hashem_leveraging_2021, ID617_hossain_human_2019, ID622_islam_exploring_2021, ID78_odoemelem_using_2020}. \textit{Real-Time Processing} is present in over a third of studies; however, only a few have demonstrated \textit{On-Device Processing} \cite{ID40_buil_acceptable_2005, ID85_matsumura_universal_2012, ID94_tessendorf_ear-worn_2012, ID449_ronco_tinyssimoradar_2024}.

The motivations behind the studies (\qref{Q6}, \autoref{fig: Summary Q6}) are predominantly centered on showing novelty. Most are driven by some form of \textit{Proof-of-Concept}, with roughly half aiming to introduce \textit{Novel Interaction Techniques}. These are followed by motivations focused on optimizing existing systems. A smaller portion of studies is guided by specific user needs -- such as hands-free interaction -- or by broader goals related to \textit{Societal Impact}, particularly in areas like accessibility (e.g., \cite{ID65_nguyen_tyth-typing_2018, ID624_simpson_evaluation_2010}) or health (e.g., \cite{ID618_kakaraparthi_facesense_2021, ID443_paul_versatile_2023}).

The range of \textit{Intended Applications} identified in the dataset (\qref{Q7}, \autoref{fig: Summary Q7}) reflects the growing versatility of earable technologies, spanning conventional use cases such as \textit{Music Players} and \textit{Phone Calls} to innovative applications in contexts around \textit{Driving} and \textit{Gaming}. A significant portion of studies focuses on \textit{Accessibility}, itself a diverse category that includes, for example, sign language recognition \cite{ID226_jin_sonicasl_2021} and ear-based Braille reading \cite{ID388_yang_customized_2022}.

\begin{figure*}[t]
    \centering
    \includegraphics[width=\linewidth]{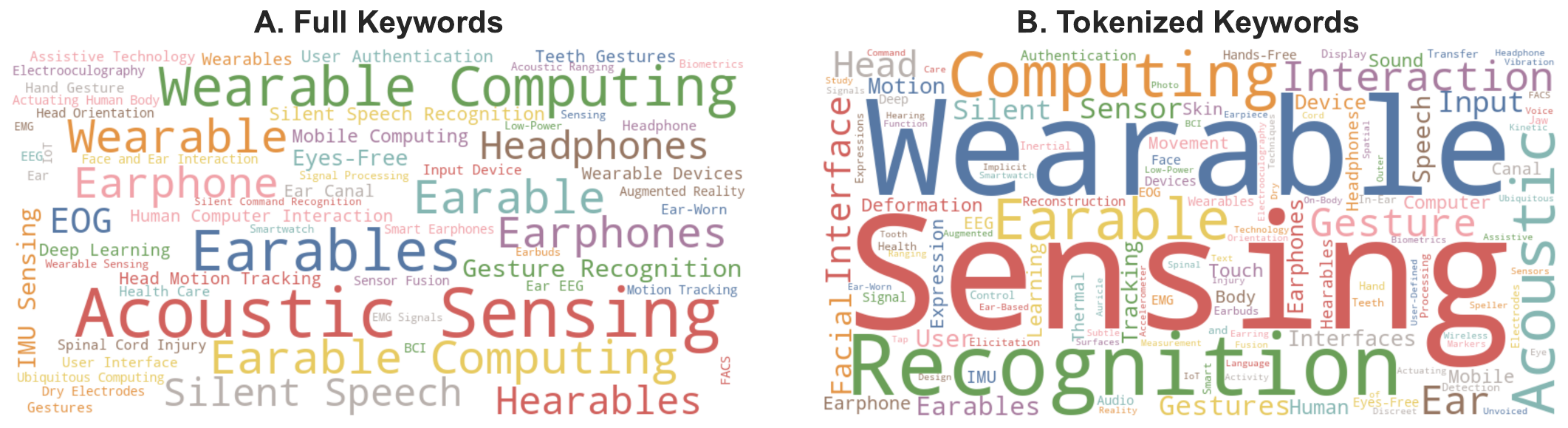} 
    \caption{Author-provided keywords (\qref{Q8}). A. Complete keywords as provided by the authors. B. Tokenized keywords, split at the word level to reveal underlying term frequency.}
    \Description{Keywords used in studies on interaction with earables. A. Complete keywords as provided by the authors. The keywords used most often are Acousting Sensing, Earables, and Wearable Computing. B. Tokenized keywords, split at the word level to reveal underlying term frequency. The keywords used most often are Sensing, Wearables, and Recognition. Here is the data the figure contains as a listing:
    
    ## Full Keywords (≥2)
    "Acoustic Sensing" = 16, "Earables" = 11, "Wearable Computing" = 10, "Wearable" = 7, "Earable" = 7, "Earable Computing" = 7, "Earphones" = 6, "Earphone" = 6, "Headphones" = 5, "Silent Speech" = 5, "Hearables" = 5, "EOG" = 4, "Gesture Recognition" = 4, "IMU Sensing" = 4, "Eyes-Free" = 3, "Silent Speech Recognition" = 3, "Mobile Computing" = 3, "Human Computer Interaction" = 3, "Wearable Devices" = 3, "Ear Canal" = 3, "Head Motion Tracking" = 3, "Teeth Gestures" = 3, "Deep Learning" = 3, "Wearables" = 3, "User Authentication" = 3, "User Interface" = 2, "Headphone" = 2, "Assistive Technology" = 2, "Spinal Cord Injury" = 2, "Hand Gesture" = 2, "Input Device" = 2, "Ear-Worn" = 2, "Ear EEG" = 2, "Gestures" = 2, "Health Care" = 2, "Ubiquitous Computing" = 2, "Electrooculography" = 2, "Actuating Human Body" = 2, "Augmented Reality" = 2, "Ear" = 2, "FACS" = 2, "Motion Tracking" = 2, "BCI" = 2, "Dry Electrodes" = 2, "EEG" = 2, "Face and Ear Interaction" = 2, "Sensor Fusion" = 2, "Smart Earphones" = 2, "Head Orientation" = 2, "IoT" = 2, "Wearable Sensing" = 2, "Sensing" = 2, "Low-Power" = 2, "Earbuds" = 2, "Smartwatch" = 2, "Signal Processing" = 2, "Acoustic Ranging" = 2, "Biometrics" = 2, "EMG" = 2, "EMG Signals" = 2, "Silent Command Recognition" = 2
    
    ## Tokenized Words (≥2)
    "Sensing" = 39, "Wearable" = 27, "Recognition" = 24, "Computing" = 22, "Acoustic" = 21, "Earable" = 20, "Interaction" = 19, "Gesture" = 17, "Ear" = 17, "Head" = 14, "Interface" = 13, "Input" = 13, "Sensor" = 12, "Facial" = 12, "Silent" = 11, "Gestures" = 11, "Earables" = 11, "User" = 10, "Speech" = 10, "Interfaces" = 8, "Earphones" = 8, "Motion" = 8, "Tracking" = 8, "Device" = 7, "Human" = 7, "Learning" = 7, "Touch" = 6, "Computer" = 6, "Earphone" = 6, "Mobile" = 6, "Deformation" = 6, "Expression" = 6, "Sound" = 6, "Headphones" = 5, "Thermal" = 5, "EEG" = 5, "Skin" = 5, "Movement" = 5, "Body" = 5, "Canal" = 5, "Hearables" = 5, "IMU" = 5, "Authentication" = 5, "EOG" = 4, "Audio" = 4, "Eyes-Free" = 4, "Devices" = 4, "Wearables" = 4, "Hands-Free" = 4, "Deep" = 4, "Reconstruction" = 4, "Face" = 4, "Signal" = 4, "EMG" = 4, "Control" = 3, "Hand" = 3, "Display" = 3, "Jaw" = 3, "Health" = 3, "In-Ear" = 3, "Elicitation" = 3, "Transfer" = 3, "Smart" = 3, "Teeth" = 3, "and" = 3, "Electrodes" = 3, "Earbuds" = 3, "Inertial" = 3, "Expressions" = 3, "Detection" = 3, "Processing" = 3, "Headphone" = 2, "Assistive" = 2, "Technology" = 2, "Spinal" = 2, "Cord" = 2, "Injury" = 2, "Tap" = 2, "Implicit" = 2, "Eye" = 2, "Ear-Based" = 2, "Ear-Worn" = 2, "Outer" = 2, "Speller" = 2, "Auricle" = 2, "Text" = 2, "On-Body" = 2, "Subtle" = 2, "Tooth" = 2, "Techniques" = 2, "Care" = 2, "Vibration" = 2, "Photo" = 2, "Sensors" = 2, "Ubiquitous" = 2, "of" = 2, "Surfaces" = 2, "Electrooculography" = 2, "Hearing" = 2, "Actuating" = 2, "Language" = 2, "Earpiece" = 2, "Augmented" = 2, "Reality" = 2, "Kinetic" = 2, "Measurement" = 2, "Function" = 2, "Earring" = 2, "FACS" = 2, "Discreet" = 2, "Voice" = 2, "User-Defined" = 2, "BCI" = 2, "Dry" = 2, "Wireless" = 2, "Fusion" = 2, "Spatial" = 2, "Orientation" = 2, "IoT" = 2, "Activity" = 2, "Accelerometer" = 2, "Unvoiced" = 2, "Low-Power" = 2, "Study" = 2, "Smartwatch" = 2, "Ranging" = 2, "Biometrics" = 2, "Markers" = 2, "Design" = 2, "Signals" = 2, "Command" = 2}
    \label{fig: Keyword_Wordcloud}
\end{figure*}

An analysis of author-provided \textit{Keywords} (\qref{Q8}, \autoref{fig: Keyword_Wordcloud}) shows that 351 distinct terms have been used. Interestingly, only 53\% of the studies include a keyword that contains the term \textit{Ear}, and only 41\% employ a keyword referencing some kind of earable device (e.g., \textit{Earbud}, \textit{Headphones}). This illustrates the challenges researchers face when searching for related work as the field encompasses a wide range of terminologies.

\subsection{Scenario-Based Exploration}\label{sec: earxplore scenario}

To showcase how \textit{EarXplore} can support researchers in navigating and building upon prior work, we demonstrate how the platform can be used from the perspective of a potential user.

A researcher interested in biopotential sensing seeks new directions for advancing earable platforms. Recalling the OpenEarable ExG \cite{ID625_lepold_openearable_2024} -- a device for measuring EEG, EOG, and EMG signals at the ear -- discovered during a recent workshop, they turn to \textit{EarXplore} for further exploration.

Upon entering the platform, the researcher navigates to the \textit{Similarity View}, using the OpenEarable ExG study entry as a starting point. To narrow the scope, they apply filters, selecting only studies that incorporate the available biopotential sensors (\textit{EEG}, \textit{EOG}, \textit{EMG}). Within the similarity visualization, clicking the \textit{node} corresponding to the OpenEarable ExG paper opens a modal overlay with key details on the study and the most similar ones, ranked by their \textit{Database Similarity}. They note the three most similar studies \cite{ID20_manabe_conductive_2013, ID91_favre-felix_improving_2018, ID61_manabe_full-time_2006} and switch to \textit{Abstract Similarity}, highlighting additional related studies and again noting the three most similar ones \cite{ID20_manabe_conductive_2013, ID12_matthies_earfieldsensing_2017, ID43_nguyen_lightweight_2016}.

Observing that most of these studies are relatively dated, the researcher shifts to the \textit{Timeline View} to investigate temporal trends. This reveals that this specific subdomain has seen limited recent activity. To investigate current developments, the researcher checks separately the three studies published in 2024. To support that, they use the \textit{Color Nodes By} option with \textit{Input Body Part}, revealing that the two studies \cite{ID527_yi_mordo2_2024, ID654_srivastava_whispering_2024} besides the OpenEarable ExG paper employed the speech apparatus as input modality. Clicking these studies' \textit{nodes} opens a modal overlay, revealing that both focused on silent speech -- a topic beyond the researcher's current scope.

To broaden the search, the researcher moves to the \textit{Graphical View} and \textit{Selects All} available bar chart visualizations. While scanning these figures, a study employing direct brain activity as an input modality \cite{ID69_merrill_classifying_2016} catches their attention. Intrigued, they click the corresponding \textit{bar} revealing key details on the study in a modal overlay. They then further click on the \textit{info icon} next to the entry, opening a modal with full study details. Finding the abstract particularly compelling, they use the \textit{Open Paper} feature to access the complete publication.

Following a review of the paper, the researcher returns to \textit{EarXplore}, navigating to the \textit{Tabular View} to check whether other studies also employ brain activity as input modality. By \textit{deselecting all} previous filters and applying a filter for the input modality \textit{Brain Activity}, they confirm that this is the only study currently employing this approach. Considering this, the researcher contemplates potential integration with existing platforms, either leveraging current hardware or through future enhancements.

This scenario demonstrates how \textit{EarXplore} supports open-ended inquiry, highlights underexplored research areas, and encourages innovation for researchers working with earable technologies. While this walkthrough reflects one potential use case, the platform supports a broad range of exploratory workflows. For instance, a researcher could filter for studies compatible with the sensor configuration of eSense \cite{kawsar_earables_2018}, a commercially available device frequently used in earable interaction \cite{ID626_ma_oesense_2021, ID608_alkiek_eargest_2022, ID284_alkiek_earbender_2023, ID80_ferlini_head_2019, ID217_hashem_leveraging_2021, ID617_hossain_human_2019, ID622_islam_exploring_2021, ID78_odoemelem_using_2020}. Alternatively, they might trace back scholarly lineages of these works through shared authorship or citation patterns to identify their blind spots, thus discovering opportunities to extend and optimize them. This personalized, multi-faceted mode of inquiry is the core motivation behind \textit{EarXplore} and the defining feature that differentiates it from conventional review tools, positioning it as a distinct contribution within the earable research landscape. In the following, we outline how we ensure its future relevance.

\subsection{EarXplore as a Sustainable Community-Driven Resource}\label{sec: earxplore update}

\textit{EarXplore} is designed as a dynamic, extensible resource that evolves alongside ongoing developments in earable interaction research. Its architecture includes updating mechanisms that support flexible data integration. Both the structure of the website and its individual views enable seamless incorporation of new content. In addition to updates made by the maintainers, \textit{EarXplore} supports two mechanisms for community contributions. First, researchers can suggest the incorporation of additional studies via the contact form embedded in the platform. Second, they can fork the public GitHub repository\footnote{
\href{https://github.com/OpenEarable/earXplore}{https://github.com/OpenEarable/earXplore}
} and propose additions directly. In both cases, submissions undergo review by the maintainers before being automatically integrated into all views of the database via backend scripts. These mechanisms position \textit{EarXplore} as a sustainable, community-driven resource that supports the ongoing curation, comparison, and exploration of earable interaction research.



%% file: sections/6_Discussion.tex
\section{Discussion}\label{sec: Discussion}

We present \textit{EarXplore}, a comprehensive, structured database comprising 118 studies on interaction with earables, categorized across 33 criteria. We detailed the development process, introduced its four distinct yet integrated views, and demonstrated how its resources can inform research in the field. In the following, we identify key research gaps and opportunities in \autoref{sec: research gaps} before discussing this work's limitations and directions for future research in \autoref{sec: limitations_future work}.

\subsection{Research Gaps and Opportunities}\label{sec: research gaps}

The earable landscape is undergoing rapid development, with emerging platforms now offering a wide array of integrated sensors \cite{roddiger_openearable_2025, montanari_omnibuds_2024}. This progress signals a shift from advanced earables being purely research prototypes to becoming viable tools for everyday use. Nearly two decades after the introduction of the first iPhone \cite{apple_iphone_2007}, an event that fundamentally reshaped how we think about mobile phones, earphones may be poised for a similar transformation. With growing hardware maturity and increasingly sophisticated interaction techniques, the field has enormous potential to expand into broader real-world applications. This raises the question: How might earable interaction evolve in everyday settings? And what gaps remain in the literature that must be addressed to enable this transition? We identify several promising directions in which research can advance earable technologies, bridge existing gaps, and support their broader adoption in everyday contexts.

\paragraph{Going beyond Proof-of-Concept}

Most studies featured in the database introduce some form of new system. While earlier works usually stress the novelty of their technique, more recent ones rather add new features or drive up performance of already existing solutions. That said, whereas most studies present functional research prototypes and report their initial performance, few are followed by subsequent publications detailing refined or extensively developed versions of the same systems. With some notable exceptions (e.g., \cite{ID527_yi_mordo2_2024, ID624_simpson_evaluation_2010}), such continued development remains largely undocumented in the literature. The earable community should express greater appreciation for the iterative development and real-world validation of these systems. Maturing prototypes into deployable technologies will be key to transitioning from laboratory novelty to practical application. 

\paragraph{User-Centered Evaluations}

To seamlessly integrate earable interaction into everyday life, strong technical performance alone is not enough. Interaction techniques must also feel natural, comfortable, and intuitive to users -- otherwise, real-world adoption is unlikely. Yet in current research, these dimensions are rarely addressed together. Only about a quarter of studies in our corpus include usability evaluations, and even fewer assess factors such as cognitive load or social acceptability (see \autoref{sec: snap_shot}). This highlights a critical shortfall, particularly given that these figures only reflect the presence of evaluations, not their depth or quality. Addressing these challenges requires that the research community places greater value on evaluation methodologies and not only on the novelty of the device or interaction technique. While the field is diverse, adhering to accepted evaluation standards is important to enable meaningful comparison across studies. Platforms like \textit{EarXplore}, which provide structured and filterable access to prior work, offer a promising step toward this goal. 

\paragraph{Ecological Validity}

One of the most compelling arguments for earables is that they build on a platform that is already routinely used in everyday life. However, real-world usage typically occurs in dynamic environments. It is therefore a significant shortcoming that interactions are rarely evaluated outside of controlled laboratory settings. Although there is growing awareness of this gap in the literature (see \autoref{sec: snap_shot}), many studies still fail to incorporate naturalistic settings into their evaluations. There are however positive examples, including deployments in public transport \cite{ID231_verma_expressear_2021, ID452_bi_smartear_2022, ID307_srivastava_muteit_2022}, a supermarket \cite{ID306_wang_toothsonic_2022}, or daily use over a two-week period \cite{ID628_peng_wigglears_2021}. We therefore recommend that future research defines plausible use-case scenarios and systematically evaluates systems within these real-world contexts.

\paragraph{Robust Input Methods}

For many professional applications, a critical prerequisite is the development of robust and highly reliable input mechanisms for earable interaction. At present, innovation is often demonstrated through systems capable of recognizing a wide range of gestures (e.g., \cite{ID231_verma_expressear_2021, ID12_matthies_earfieldsensing_2017, ID221_sun_teethtap_2021}). While such breadth advances interaction design, it may compromise reliability in high-stakes environments. In domains such as healthcare or industrial control, even minor recognition errors can carry serious consequences. In these cases, a smaller set of highly dependable input methods may be preferable. As earables gain traction in professional settings, we anticipate a shift in research priorities toward minimizing error rates and optimizing interaction robustness under real-world conditions.

\paragraph{Multimodal Sensing}

One promising path toward greater robustness in earable interaction is the integration of multiple sensing modalities, such as combining a microphone with an IMU, into the interaction detection pipeline. This allows the system to compensate for noise or failure in one sensor through complementary input from another. Most studies however rely on a single sensor or a homogeneous sensor unit (e.g., a 6-axis IMU). Despite the literature showcasing a wide variety of sensing opportunities, multiple modalities are often only applied to separate phases of interaction \cite{ID315_jin_earcommand_2022} or to detect distinct gesture types \cite{ID308_panda_beyond_2023}. Genuine multimodal sensing is rare, with only six such studies in our corpus \cite{ID93_sahni_tongue_2014, ID65_nguyen_tyth-typing_2018, ID618_kakaraparthi_facesense_2021, ID221_sun_teethtap_2021, ID345_choi_ppgface_2022, ID654_srivastava_whispering_2024}. This scarcity can be attributed in part to the lack of devices tailored for earable interaction, with researchers required to develop their own hardware or rely on limited commercial devices -- an issue closely tied to the next opportunity we identify.

\paragraph{Advances in Hardware}

For years, eSense \cite{kawsar_earables_2018}, equipped with a 6-axis IMU and a microphone, stood out as the only research platform used across studies on earable interaction \cite{ID626_ma_oesense_2021, ID608_alkiek_eargest_2022, ID284_alkiek_earbender_2023, ID80_ferlini_head_2019, ID217_hashem_leveraging_2021, ID617_hossain_human_2019, ID622_islam_exploring_2021, ID78_odoemelem_using_2020}. More recently, however, the field has seen the emergence of further advanced and openly accessible earable platforms. Omnibuds \cite{ montanari_omnibuds_2024} integrates a 9-axis IMU, two microphones, a PPG and a temperature sensor. OpenEarable 2.0 \cite{roddiger_openearable_2025} features all of these capabilities and, in addition, a bone-conduction microphone and a pressure sensor. The community should embrace these tools to explore richer multimodal sensing approaches better suited to complex, real-world scenarios. Converging to shared platforms would also improve the comparability of results, foster reuse, and accelerate progress by building on each other's work. Moreover, as these platforms integrate onboard processing capabilities, researchers are encouraged to design self-contained systems that manage both sensing and computation locally, an approach particularly relevant for ensuring data privacy.

\paragraph{Multi-Device Sensing}

Apart from leveraging multiple sensing options within one device, it could also be beneficial to explore how different wearable devices might be combined to make interaction detection more reliable and to uncover entirely new forms of interaction. For instance, even seemingly simple questions, such as determining which hand was used to perform an interaction, can be difficult or impossible to answer using a standalone earable. While research has begun to demonstrate the promise of multi-device integration recently \cite{ID654_srivastava_whispering_2024, ID342_jin_smartasl_2023, ID376_li_enabling_2023}, this area remains underexplored and holds substantial potential for future investigation.

\subsection{Limitations and Future Work}\label{sec: limitations_future work}

As outlined, designing an interactive online platform offers several advantages over a traditional survey paper. However, it also introduces certain limitations worth acknowledging. 
First, although our criteria-based approach enhances comparability across studies, it inevitably introduces a degree of simplification. Two studies marked with the same criteria may still differ in detail, execution, or methodological rigor. Additionally, the criteria reflect the presence but not the quality or depth of elements such as evaluation procedures. To support consistent filtering, certain abstractions were necessary -- for example, grouping slightly different \textit{Intended Applications} under broader categories such as \textit{Device Control and Input}, which encompass a range of submodalities.
Second, while our goal was to comprehensively capture the field of earable interaction, our scope did not extend to all research that might be applicable to interaction. For example, work on HAR or facial action coding systems (FACS) could, in principle, be adapted for use with earable interaction. However, our focus was on studies that explicitly address interaction through earables, rather than those that may only be repurposed for such use.
Third, our review is limited to peer-reviewed research and does not include commercially available earables. Many commercial devices are now equipped with additional sensors such as IMUs, thus featuring basic interaction methods like taps or slides. These capabilities are only included in \textit{EarXplore} if they have been explicitly studied and evaluated in peer-reviewed publications, leaving most commercially driven capabilities outside of scope.

Despite these limitations, we believe that open, accessible, and interactive online databases offer a modern and impactful way of disseminating research. To support similar initiatives, we release the source code for \textit{EarXplore}\footnote{see here: \href{https://github.com/OpenEarable/earXplore}{https://github.com/OpenEarable/earXplore}
}. Since the website is dynamically generated, the \textit{Tabular} and \textit{Graphical Views} can be created from a structured CSV dataset with minimal setup. The \textit{Similarity} and \textit{Timeline Views} require minor additional configuration, for which we provide scripts and documentation. By sharing \textit{EarXplore} this way, we hope to establish a hub for research on earable interaction and to encourage broader adoption of open, exploratory dissemination practices across other domains of earable research and beyond.



%% file: sections/7_Conclusion.tex
\section{Conclusion}\label{sec: Conclusion}

In this work, we introduced \textit{EarXplore}, an interactive online database dedicated to interaction with earables featuring four integrated views. The platform enables targeted, fine-grained research queries that are not possible to conduct through traditional literature reviews -- but can be answered in seconds via combined filtering of metadata fields. Similarly, the platform enables novel forms of synthesis: by systematically comparing sensing locations, interaction modalities, and device types, users can spot trends and gaps in the field that may not be apparent from reading individual papers. The platform’s interactive visualizations also facilitate new discovery pathways. Instead of searching by keywords, users can explore semantically related work based on database or abstract similarity. Instead of tracing back citation chains of a single paper, they can uncover the whole evolution of the field at once and while displaying the studies' characteristics. This opens opportunities for sophisticated discovery and can aid in identifying unexpected connections across subfields. We therefore hope that \textit{EarXplore} serves as both a tool for advancing earable interaction research and as a blueprint for creating interactive, accessible platforms in other domains.

%% file: sample-base.bib
@String{Computing = "Computing" }

@String{Computer = "{IEEE} Computer" }

@String{Academic = "Academic Press" }

@String{Springer = "Springer-Verlag" }

@inproceedings{ID22_kikuchi_eartouch_2017,
	address = {New York, NY, USA},
	series = {{MobileHCI} '17},
	title = {{EarTouch}: turning the ear into an input surface},
	isbn = {978-1-4503-5075-4},
	shorttitle = {{EarTouch}},
	url = {https://dl.acm.org/doi/10.1145/3098279.3098538},
	doi = {10.1145/3098279.3098538},
	urldate = {2024-09-30},
	booktitle = {Proceedings of the 19th {International} {Conference} on {Human}-{Computer} {Interaction} with {Mobile} {Devices} and {Services}},
	publisher = {Association for Computing Machinery},
	author = {Kikuchi, Takashi and Sugiura, Yuta and Masai, Katsutoshi and Sugimoto, Maki and Thomas, Bruce H.},
	month = sep,
	year = {2017},
	pages = {1--6},
}

@inproceedings{ID227_roddiger_earrumble_2021,
	address = {New York, NY, USA},
	series = {{CHI} '21},
	title = {{EarRumble}: {Discreet} {Hands}- and {Eyes}-{Free} {Input} by {Voluntary} {Tensor} {Tympani} {Muscle} {Contraction}},
	isbn = {978-1-4503-8096-6},
	shorttitle = {{EarRumble}},
	url = {https://dl.acm.org/doi/10.1145/3411764.3445205},
	doi = {10.1145/3411764.3445205},
	urldate = {2024-09-30},
	booktitle = {Proceedings of the 2021 {CHI} {Conference} on {Human} {Factors} in {Computing} {Systems}},
	publisher = {Association for Computing Machinery},
	author = {Röddiger, Tobias and Clarke, Christopher and Wolffram, Daniel and Budde, Matthias and Beigl, Michael},
	month = may,
	year = {2021},
	pages = {1--14},
}

@inproceedings{ID40_buil_acceptable_2005,
	title = {Acceptable operating force for buttons on in-ear type headphones},
	url = {https://ieeexplore.ieee.org/document/1550805},
	doi = {10.1109/ISWC.2005.7},
	urldate = {2024-09-30},
	booktitle = {Ninth {IEEE} {International} {Symposium} on {Wearable} {Computers} ({ISWC}'05)},
	author = {Buil, V. and Hollemans, G.},
	month = oct,
	year = {2005},
	pages = {186--189},
        publisher = {IEEE}, 
        address = {Piscataway, NJ, USA}
}

@inproceedings{ID85_matsumura_universal_2012,
	address = {New York, NY, USA},
	series = {{IUI} '12},
	title = {Universal earphones: earphones with automatic side and shared use detection},
	isbn = {978-1-4503-1048-2},
	shorttitle = {Universal earphones},
	url = {https://dl.acm.org/doi/10.1145/2166966.2167025},
	doi = {10.1145/2166966.2167025},
	urldate = {2024-09-30},
	booktitle = {Proceedings of the 2012 {ACM} international conference on {Intelligent} {User} {Interfaces}},
	publisher = {Association for Computing Machinery},
	author = {Matsumura, Kohei and Sakamoto, Daisuke and Inami, Masahiko and Igarashi, Takeo},
	month = feb,
	year = {2012},
	pages = {305--306},
}

@inproceedings{ID223_cao_earphonetrack_2020,
	address = {New York, NY, USA},
	series = {{SenSys} '20},
	title = {{EarphoneTrack}: involving earphones into the ecosystem of acoustic motion tracking},
	isbn = {978-1-4503-7590-0},
	shorttitle = {{EarphoneTrack}},
	url = {https://dl.acm.org/doi/10.1145/3384419.3430730},
	doi = {10.1145/3384419.3430730},
	urldate = {2024-09-30},
	booktitle = {Proceedings of the 18th {Conference} on {Embedded} {Networked} {Sensor} {Systems}},
	publisher = {Association for Computing Machinery},
	author = {Cao, Gaoshuai and Yuan, Kuang and Xiong, Jie and Yang, Panlong and Yan, Yubo and Zhou, Hao and Li, Xiang-Yang},
	month = nov,
	year = {2020},
	pages = {95--108},
}

@article{ID234_chen_exploring_2020,
	title = {Exploring {User} {Defined} {Gestures} for {Ear}-{Based} {Interactions}},
	volume = {4},
	url = {https://dl.acm.org/doi/10.1145/3427314},
	doi = {10.1145/3427314},
	number = {ISS},
	urldate = {2024-09-30},
	journal = {Proc. ACM Hum.-Comput. Interact.},
	author = {Chen, Yu-Chun and Liao, Chia-Ying and Hsu, Shuo-wen and Huang, Da-Yuan and Chen, Bing-Yu},
	month = nov,
	year = {2020},
	pages = {186:1--186:20},
}

@article{ID226_jin_sonicasl_2021,
	title = {{SonicASL}: {An} {Acoustic}-based {Sign} {Language} {Gesture} {Recognizer} {Using} {Earphones}},
	volume = {5},
	shorttitle = {{SonicASL}},
	url = {https://dl.acm.org/doi/10.1145/3463519},
	doi = {10.1145/3463519},
	number = {2},
	urldate = {2024-09-30},
	journal = {Proc. ACM Interact. Mob. Wearable Ubiquitous Technol.},
	author = {Jin, Yincheng and Gao, Yang and Zhu, Yanjun and Wang, Wei and Li, Jiyang and Choi, Seokmin and Li, Zhangyu and Chauhan, Jagmohan and Dey, Anind K. and Jin, Zhanpeng},
	month = jun,
	year = {2021},
	pages = {67:1--67:30},
}

@inproceedings{ID78_odoemelem_using_2020,
	address = {New York, NY, USA},
	series = {{EarComp}'19},
	title = {Using the {eSense} {Wearable} {Earbud} as a {Light}-{Weight} {Robot} {Arm} {Controller}},
	isbn = {978-1-4503-6902-2},
	url = {https://dl.acm.org/doi/10.1145/3345615.3361138},
	doi = {10.1145/3345615.3361138},
	urldate = {2024-09-30},
	booktitle = {Proceedings of the 1st {International} {Workshop} on {Earable} {Computing}},
	publisher = {Association for Computing Machinery},
	author = {Odoemelem, Henry and Hölzemann, Alexander and Van Laerhoven, Kristof},
	month = feb,
	year = {2020},
	pages = {26--29},
}

@inproceedings{ID80_ferlini_head_2019,
	address = {New York, NY, USA},
	series = {{EarComp}'19},
	title = {Head {Motion} {Tracking} {Through} in-{Ear} {Wearables}},
	isbn = {978-1-4503-6902-2},
	url = {https://dl.acm.org/doi/10.1145/3345615.3361131},
	doi = {10.1145/3345615.3361131},
	urldate = {2024-09-30},
	booktitle = {Proceedings of the 1st {International} {Workshop} on {Earable} {Computing}},
	publisher = {Association for Computing Machinery},
	author = {Ferlini, Andrea and Montanari, Alessandro and Mascolo, Cecilia and Harle, Robert},
	month = feb,
	year = {2019},
	pages = {8--13},
}

@inproceedings{ID217_hashem_leveraging_2021,
	address = {New York, NY, USA},
	series = {{HotMobile} '21},
	title = {Leveraging {Earables} for {Natural} {Calibration}-{Free} {Multi}-{Device} {Identification} in {Smart} {Environments}},
	isbn = {978-1-4503-8323-3},
	url = {https://dl.acm.org/doi/10.1145/3446382.3448653},
	doi = {10.1145/3446382.3448653},
	urldate = {2024-09-30},
	booktitle = {Proceedings of the 22nd {International} {Workshop} on {Mobile} {Computing} {Systems} and {Applications}},
	publisher = {Association for Computing Machinery},
	author = {Hashem, Omar and Alkiek, Khaled and Youssef, Moustafa and Harras, Khaled A.},
	month = feb,
	year = {2021},
	pages = {92--98},
}

@inproceedings{ID249_pfreundtner_wearable_2021,
	title = {({W}){Earable} {Microphone} {Array} and {Ultrasonic} {Echo} {Localization} for {Coarse} {Indoor} {Environment} {Mapping}},
	url = {https://ieeexplore.ieee.org/document/9414356/?arnumber=9414356},
	doi = {10.1109/ICASSP39728.2021.9414356},
	urldate = {2024-09-30},
	booktitle = {{ICASSP} 2021 - 2021 {IEEE} {International} {Conference} on {Acoustics}, {Speech} and {Signal} {Processing} ({ICASSP})},
	author = {Pfreundtner, Felix and Yang, Jing and Sörös, Gábor},
	month = jun,
	year = {2021},
	pages = {4475--4479},
        publisher = {IEEE}, 
        address = {Piscataway, NJ, USA}
}

@inproceedings{ID94_tessendorf_ear-worn_2012,
	title = {Ear-worn reference data collection and annotation for multimodal context-aware hearing instruments},
	url = {https://ieeexplore.ieee.org/document/6346464},
	doi = {10.1109/EMBC.2012.6346464},
	urldate = {2024-09-30},
	booktitle = {2012 {Annual} {International} {Conference} of the {IEEE} {Engineering} in {Medicine} and {Biology} {Society}},
	author = {Tessendorf, Bernd and Derleth, Peter and Feilner, Manuela and Gravenhorst, Franz and Kettner, Andreas and Roggen, Daniel and Stiefmeier, Thomas and Tröster, Gerhard},
	month = aug,
	year = {2012},
	pages = {2468--2471},
        publisher = {IEEE}, 
        address = {Piscataway, NJ, USA}
}

@article{ID57_salzar_improving_2008,
	title = {Improving {Earpiece} {Accelerometer} {Coupling} to the {Head}},
	volume = {1},
	url = {https://saemobilus.sae.org/articles/improving-earpiece-accelerometer-coupling-head-2008-01-2978},
	doi = {10.4271/2008-01-2978},
	language = {English},
	number = {1},
	urldate = {2024-09-30},
	journal = {SAE International Journal of Passenger Cars - Mechanical Systems},
	author = {Salzar, Robert S. and Bass, Cameron R. ‘Dale’ and Pellettiere, Joseph A. and Salzar, Robert S. and Bass, Cameron R. ‘Dale’ and Pellettiere, Joseph A.},
	month = dec,
	year = {2008},
	pages = {1367--1381},
}

@inproceedings{ID17_ando_canalsense_2017,
	address = {New York, NY, USA},
	series = {{UIST} '17},
	title = {{CanalSense}: {Face}-{Related} {Movement} {Recognition} {System} based on {Sensing} {Air} {Pressure} in {Ear} {Canals}},
	isbn = {978-1-4503-4981-9},
	shorttitle = {{CanalSense}},
	url = {https://dl.acm.org/doi/10.1145/3126594.3126649},
	doi = {10.1145/3126594.3126649},
	urldate = {2024-09-30},
	booktitle = {Proceedings of the 30th {Annual} {ACM} {Symposium} on {User} {Interface} {Software} and {Technology}},
	publisher = {Association for Computing Machinery},
	author = {Ando, Toshiyuki and Kubo, Yuki and Shizuki, Buntarou and Takahashi, Shin},
	month = oct,
	year = {2017},
	pages = {679--689},
}

@article{ID60_taniguchi_earable_2018,
	title = {Earable {TEMPO}: {A} {Novel}, {Hands}-{Free} {Input} {Device} that {Uses} the {Movement} of the {Tongue} {Measured} with a {Wearable} {Ear} {Sensor}},
	volume = {18},
	copyright = {http://creativecommons.org/licenses/by/3.0/},
	issn = {1424-8220},
	shorttitle = {Earable {TEMPO}},
	url = {https://www.mdpi.com/1424-8220/18/3/733},
	doi = {10.3390/s18030733},
	language = {en},
	number = {3},
	urldate = {2024-09-30},
	journal = {Sensors},
	author = {Taniguchi, Kazuhiro and Kondo, Hisashi and Kurosawa, Mami and Nishikawa, Atsushi},
	month = mar,
	year = {2018},
	pages = {733},
}

@inproceedings{ID64_maag_barton_2017,
	title = {{BARTON}: {Low} {Power} {Tongue} {Movement} {Sensing} with {In}-{Ear} {Barometers}},
	shorttitle = {{BARTON}},
	url = {https://ieeexplore.ieee.org/document/8368342/?arnumber=8368342},
	doi = {10.1109/ICPADS.2017.00013},
	urldate = {2024-09-30},
	booktitle = {2017 {IEEE} 23rd {International} {Conference} on {Parallel} and {Distributed} {Systems} ({ICPADS})},
	author = {Maag, Balz and Zhou, Zimu and Saukh, Olga and Thiele, Lothar},
	month = dec,
	year = {2017},
	pages = {9--16},
        publisher = {IEEE}, 
        address = {Piscataway, NJ, USA}
}

@inproceedings{ID65_nguyen_tyth-typing_2018,
	address = {New York, NY, USA},
	series = {{MobiSys} '18},
	title = {{TYTH}-{Typing} {On} {Your} {Teeth}: {Tongue}-{Teeth} {Localization} for {Human}-{Computer} {Interface}},
	isbn = {978-1-4503-5720-3},
	shorttitle = {{TYTH}-{Typing} {On} {Your} {Teeth}},
	url = {https://dl.acm.org/doi/10.1145/3210240.3210322},
	doi = {10.1145/3210240.3210322},
	urldate = {2024-09-30},
	booktitle = {Proceedings of the 16th {Annual} {International} {Conference} on {Mobile} {Systems}, {Applications}, and {Services}},
	publisher = {Association for Computing Machinery},
	author = {Nguyen, Phuc and Bui, Nam and Nguyen, Anh and Truong, Hoang and Suresh, Abhijit and Whitlock, Matt and Pham, Duy and Dinh, Thang and Vu, Tam},
	month = jun,
	year = {2018},
	pages = {269--282},
}

@inproceedings{ID93_sahni_tongue_2014,
	address = {New York, NY, USA},
	series = {{ISWC} '14},
	title = {The tongue and ear interface: a wearable system for silent speech recognition},
	isbn = {978-1-4503-2969-9},
	shorttitle = {The tongue and ear interface},
	url = {https://dl.acm.org/doi/10.1145/2634317.2634322},
	doi = {10.1145/2634317.2634322},
	urldate = {2024-09-30},
	booktitle = {Proceedings of the 2014 {ACM} {International} {Symposium} on {Wearable} {Computers}},
	publisher = {Association for Computing Machinery},
	author = {Sahni, Himanshu and Bedri, Abdelkareem and Reyes, Gabriel and Thukral, Pavleen and Guo, Zehua and Starner, Thad and Ghovanloo, Maysam},
	month = sep,
	year = {2014},
	pages = {47--54},
}

@inproceedings{ID86_bedri_detecting_2015,
	address = {New York, NY, USA},
	series = {{ICMI} '15},
	title = {Detecting {Mastication}: {A} {Wearable} {Approach}},
	isbn = {978-1-4503-3912-4},
	shorttitle = {Detecting {Mastication}},
	url = {https://dl.acm.org/doi/10.1145/2818346.2820767},
	doi = {10.1145/2818346.2820767},
	urldate = {2024-09-30},
	booktitle = {Proceedings of the 2015 {ACM} on {International} {Conference} on {Multimodal} {Interaction}},
	publisher = {Association for Computing Machinery},
	author = {Bedri, Abdelkareem and Verlekar, Apoorva and Thomaz, Edison and Avva, Valerie and Starner, Thad},
	month = nov,
	year = {2015},
	pages = {247--250},
}

@inproceedings{ID221_sun_teethtap_2021,
	address = {New York, NY, USA},
	series = {{IUI} '21},
	title = {{TeethTap}: {Recognizing} {Discrete} {Teeth} {Gestures} {Using} {Motion} and {Acoustic} {Sensing} on an {Earpiece}},
	isbn = {978-1-4503-8017-1},
	shorttitle = {{TeethTap}},
	url = {https://dl.acm.org/doi/10.1145/3397481.3450645},
	doi = {10.1145/3397481.3450645},
	urldate = {2024-09-30},
	booktitle = {Proceedings of the 26th {International} {Conference} on {Intelligent} {User} {Interfaces}},
	publisher = {Association for Computing Machinery},
	author = {Sun, Wei and Li, Franklin Mingzhe and Steeper, Benjamin and Xu, Songlin and Tian, Feng and Zhang, Cheng},
	month = apr,
	year = {2021},
	pages = {161--169},
}

@inproceedings{ID8_amesaka_facial_2019,
	address = {New York, NY, USA},
	series = {{ISWC} '19},
	title = {Facial expression recognition using ear canal transfer function},
	isbn = {978-1-4503-6870-4},
	url = {https://dl.acm.org/doi/10.1145/3341163.3347747},
	doi = {10.1145/3341163.3347747},
	urldate = {2024-09-30},
	booktitle = {Proceedings of the 2019 {ACM} {International} {Symposium} on {Wearable} {Computers}},
	publisher = {Association for Computing Machinery},
	author = {Amesaka, Takashi and Watanabe, Hiroki and Sugimoto, Masanori},
	month = sep,
	year = {2019},
	pages = {1--9},
}

@article{ID231_verma_expressear_2021,
	title = {{ExpressEar}: {Sensing} {Fine}-{Grained} {Facial} {Expressions} with {Earables}},
	volume = {5},
	shorttitle = {{ExpressEar}},
	url = {https://dl.acm.org/doi/10.1145/3478085},
	doi = {10.1145/3478085},
	number = {3},
	urldate = {2024-09-30},
	journal = {Proc. ACM Interact. Mob. Wearable Ubiquitous Technol.},
	author = {Verma, Dhruv and Bhalla, Sejal and Sahnan, Dhruv and Shukla, Jainendra and Parnami, Aman},
	month = sep,
	year = {2021},
	pages = {129:1--129:28},
}

@inproceedings{ID12_matthies_earfieldsensing_2017,
	address = {New York, NY, USA},
	series = {{CHI} '17},
	title = {{EarFieldSensing}: {A} {Novel} {In}-{Ear} {Electric} {Field} {Sensing} to {Enrich} {Wearable} {Gesture} {Input} through {Facial} {Expressions}},
	isbn = {978-1-4503-4655-9},
	shorttitle = {{EarFieldSensing}},
	url = {https://dl.acm.org/doi/10.1145/3025453.3025692},
	doi = {10.1145/3025453.3025692},
	urldate = {2024-09-30},
	booktitle = {Proceedings of the 2017 {CHI} {Conference} on {Human} {Factors} in {Computing} {Systems}},
	publisher = {Association for Computing Machinery},
	author = {Matthies, Denys J. C. and Strecker, Bernhard A. and Urban, Bodo},
	month = may,
	year = {2017},
	pages = {1911--1922},
}

@inproceedings{ID9_pham_wake_2020,
	address = {New York, NY, USA},
	series = {{MobiSys} '20},
	title = {{WAKE}: a behind-the-ear wearable system for microsleep detection},
	isbn = {978-1-4503-7954-0},
	shorttitle = {{WAKE}},
	url = {https://dl.acm.org/doi/10.1145/3386901.3389032},
	doi = {10.1145/3386901.3389032},
	urldate = {2024-09-30},
	booktitle = {Proceedings of the 18th {International} {Conference} on {Mobile} {Systems}, {Applications}, and {Services}},
	publisher = {Association for Computing Machinery},
	author = {Pham, Nhat and Dinh, Tuan and Raghebi, Zohreh and Kim, Taeho and Bui, Nam and Nguyen, Phuc and Truong, Hoang and Banaei-Kashani, Farnoush and Halbower, Ann and Dinh, Thang and Vu, Tam},
	month = jun,
	year = {2020},
	pages = {404--418},
}

@inproceedings{ID20_manabe_conductive_2013,
	address = {New York, NY, USA},
	series = {{ISWC} '13},
	title = {Conductive rubber electrodes for earphone-based eye gesture input interface},
	isbn = {978-1-4503-2127-3},
	url = {https://dl.acm.org/doi/10.1145/2493988.2494329},
	doi = {10.1145/2493988.2494329},
	urldate = {2024-09-30},
	booktitle = {Proceedings of the 2013 {International} {Symposium} on {Wearable} {Computers}},
	publisher = {Association for Computing Machinery},
	author = {Manabe, Hiroyuki and Fukumoto, Masaaki and Yagi, Tohru},
	month = sep,
	year = {2013},
	pages = {33--40},
}

@article{ID70_bleichner_concealed_2017,
	title = {Concealed, {Unobtrusive} {Ear}-{Centered} {EEG} {Acquisition}: {cEEGrids} for {Transparent} {EEG}},
	volume = {11},
	issn = {1662-5161},
	shorttitle = {Concealed, {Unobtrusive} {Ear}-{Centered} {EEG} {Acquisition}},
	url = {https://www.frontiersin.org/journals/human-neuroscience/articles/10.3389/fnhum.2017.00163/full},
	doi = {10.3389/fnhum.2017.00163},
	language = {English},
	urldate = {2024-09-30},
	journal = {Front. Hum. Neurosci.},
	author = {Bleichner, Martin G. and Debener, Stefan},
	month = apr,
	year = {2017},
        pages = {1-14}
}

@article{ID91_favre-felix_improving_2018,
	title = {Improving {Speech} {Intelligibility} by {Hearing} {Aid} {Eye}-{Gaze} {Steering}: {Conditions} {With} {Head} {Fixated} in a {Multitalker} {Environment}},
	volume = {22},
	issn = {2331-2165, 2331-2165},
	shorttitle = {Improving {Speech} {Intelligibility} by {Hearing} {Aid} {Eye}-{Gaze} {Steering}},
	url = {https://journals.sagepub.com/doi/10.1177/2331216518814388},
	doi = {10.1177/2331216518814388},
	language = {en},
	urldate = {2024-09-30},
	journal = {Trends in Hearing},
	author = {Favre-Félix, Antoine and Graversen, Carina and Hietkamp, Renskje K. and Dau, Torsten and Lunner, Thomas},
	month = jan,
	year = {2018},
	pages = {2331216518814388},
}

@inproceedings{ID69_merrill_classifying_2016,
	title = {Classifying mental gestures with in-ear {EEG}},
	url = {https://ieeexplore.ieee.org/document/7516246/?arnumber=7516246},
	doi = {10.1109/BSN.2016.7516246},
	urldate = {2024-09-30},
	booktitle = {2016 {IEEE} 13th {International} {Conference} on {Wearable} and {Implantable} {Body} {Sensor} {Networks} ({BSN})},
	author = {Merrill, Nick and Curran, Max T. and Yang, Jong-Kai and Chuang, John},
	month = jun,
	year = {2016},
	pages = {130--135},
        publisher = {IEEE}, 
        address = {Piscataway, NJ, USA}
}

@incollection{2014_looney_ear-eeg_2014,
	address = {Berlin, Heidelberg},
	title = {Ear-{EEG}: {User}-{Centered} and {Wearable} {BCI}},
	volume = {6},
	copyright = {http://www.springer.com/tdm},
	isbn = {978-3-642-54706-5 978-3-642-54707-2},
	shorttitle = {Ear-{EEG}},
	url = {https://link.springer.com/10.1007/978-3-642-54707-2_5},
	language = {en},
	urldate = {2024-09-30},
	booktitle = {Brain-{Computer} {Interface} {Research}},
	publisher = {Springer Berlin Heidelberg},
	author = {Looney, David and Kidmose, Preben and Mandic, Danilo P.},
	editor = {Guger, Christoph and Allison, Brendan and Leuthardt, E.C.},
	year = {2014},
	doi = {10.1007/978-3-642-54707-2_5},
	pages = {41--50},
}

@inproceedings{ID21_choi_toning_2020,
	address = {New York, NY, USA},
	series = {{TEI} '20},
	title = {Toning: {New} {Experience} of {Sharing} {Music} {Preference} with {Interactive} {Earphone} in {Public} {Space}},
	isbn = {978-1-4503-6107-1},
	shorttitle = {Toning},
	url = {https://dl.acm.org/doi/10.1145/3374920.3374983},
	doi = {10.1145/3374920.3374983},
	urldate = {2024-09-30},
	booktitle = {Proceedings of the {Fourteenth} {International} {Conference} on {Tangible}, {Embedded}, and {Embodied} {Interaction}},
	publisher = {Association for Computing Machinery},
	author = {Choi, Inkyung and Kim, Dain},
	month = feb,
	year = {2020},
	pages = {533--538},
}

@article{roddiger_sensing_2022,
	title = {Sensing with {Earables}: {A} {Systematic} {Literature} {Review} and {Taxonomy} of {Phenomena}},
	volume = {6},
	issn = {2474-9567},
	shorttitle = {Sensing with {Earables}},
	url = {https://dl.acm.org/doi/10.1145/3550314},
	doi = {10.1145/3550314},
	language = {en},
	number = {3},
	urldate = {2024-10-01},
	journal = {Proc. ACM Interact. Mob. Wearable Ubiquitous Technol.},
	author = {Röddiger, Tobias and Clarke, Christopher and Breitling, Paula and Schneegans, Tim and Zhao, Haibin and Gellersen, Hans and Beigl, Michael},
	month = sep,
	year = {2022},
	pages = {1--57},
}

@inproceedings{ID43_nguyen_lightweight_2016,
	address = {New York, NY, USA},
	series = {{SenSys} '16},
	title = {A {Lightweight} and {Inexpensive} {In}-ear {Sensing} {System} {For} {Automatic} {Whole}-night {Sleep} {Stage} {Monitoring}},
	isbn = {978-1-4503-4263-6},
	url = {https://dl.acm.org/doi/10.1145/2994551.2994562},
	doi = {10.1145/2994551.2994562},
	urldate = {2024-10-02},
	booktitle = {Proceedings of the 14th {ACM} {Conference} on {Embedded} {Network} {Sensor} {Systems} {CD}-{ROM}},
	publisher = {Association for Computing Machinery},
	author = {Nguyen, Anh and Alqurashi, Raghda and Raghebi, Zohreh and Banaei-kashani, Farnoush and Halbower, Ann C. and Vu, Tam},
	month = nov,
	year = {2016},
	pages = {230--244},
}

@inproceedings{ID61_manabe_full-time_2006,
	address = {New York, NY, USA},
	series = {{CHI} {EA} '06},
	title = {Full-time wearable headphone-type gaze detector},
	isbn = {978-1-59593-298-3},
	url = {https://dl.acm.org/doi/10.1145/1125451.1125655},
	doi = {10.1145/1125451.1125655},
	urldate = {2024-10-02},
	booktitle = {{CHI} '06 {Extended} {Abstracts} on {Human} {Factors} in {Computing} {Systems}},
	publisher = {Association for Computing Machinery},
	author = {Manabe, Hiroyuki and Fukumoto, Masaaki},
	month = apr,
	year = {2006},
	pages = {1073--1078},
}

@article{ID275_shimon_exploring_2024,
	title = {Exploring {Uni}-manual {Around} {Ear} {Off}-{Device} {Gestures} for {Earables}},
	volume = {8},
	url = {https://dl.acm.org/doi/10.1145/3643513},
	doi = {10.1145/3643513},
	number = {1},
	urldate = {2024-10-08},
	journal = {Proc. ACM Interact. Mob. Wearable Ubiquitous Technol.},
	author = {Shimon, Shaikh Shawon Arefin and Neshati, Ali and Sun, Junwei and Xu, Qiang and Zhao, Jian},
	month = mar,
	year = {2024},
	pages = {3:1--3:29},
}

@article{ID278_li_eario_2022,
	title = {{EarIO}: {A} {Low}-power {Acoustic} {Sensing} {Earable} for {Continuously} {Tracking} {Detailed} {Facial} {Movements}},
	volume = {6},
	shorttitle = {{EarIO}},
	url = {https://dl.acm.org/doi/10.1145/3534621},
	doi = {10.1145/3534621},
	number = {2},
	urldate = {2024-10-08},
	journal = {Proc. ACM Interact. Mob. Wearable Ubiquitous Technol.},
	author = {Li, Ke and Zhang, Ruidong and Liang, Bo and Guimbretière, François and Zhang, Cheng},
	month = jul,
	year = {2022},
	pages = {62:1--62:24},
}

@inproceedings{ID284_alkiek_earbender_2023,
	address = {New York, NY, USA},
	series = {{UbiComp}/{ISWC} '23 {Adjunct}},
	title = {{EarBender}: {Enabling} {Rich} {IMU}-based {Natural} {Hand}-to-{Ear} {Interaction} in {Commodity} {Earables}},
	isbn = {979-8-4007-0200-6},
	shorttitle = {{EarBender}},
	url = {https://dl.acm.org/doi/10.1145/3594739.3610671},
	doi = {10.1145/3594739.3610671},
	urldate = {2024-10-08},
	booktitle = {Adjunct {Proceedings} of the 2023 {ACM} {International} {Joint} {Conference} on {Pervasive} and {Ubiquitous} {Computing} \& the 2023 {ACM} {International} {Symposium} on {Wearable} {Computing}},
	publisher = {Association for Computing Machinery},
	author = {Alkiek, Khaled and Youssef, Moustafa and Harras, Khaled A.},
	month = oct,
	year = {2023},
	pages = {333--338},
}

@article{ID296_rateau_leveraging_2022,
	title = {Leveraging {Smartwatch} and {Earbuds} {Gesture} {Capture} to {Support} {Wearable} {Interaction}},
	volume = {6},
	url = {https://dl.acm.org/doi/10.1145/3567710},
	doi = {10.1145/3567710},
	number = {ISS},
	urldate = {2024-10-08},
	journal = {Proc. ACM Hum.-Comput. Interact.},
	author = {Rateau, Hanae and Lank, Edward and Liu, Zhe},
	month = nov,
	year = {2022},
	pages = {557:31--557:50},
}

@article{ID307_srivastava_muteit_2022,
	title = {{MuteIt}: {Jaw} {Motion} {Based} {Unvoiced} {Command} {Recognition} {Using} {Earable}},
	volume = {6},
	shorttitle = {{MuteIt}},
	url = {https://dl.acm.org/doi/10.1145/3550281},
	doi = {10.1145/3550281},
	number = {3},
	urldate = {2024-10-08},
	journal = {Proc. ACM Interact. Mob. Wearable Ubiquitous Technol.},
	author = {Srivastava, Tanmay and Khanna, Prerna and Pan, Shijia and Nguyen, Phuc and Jain, Shubham},
	month = sep,
	year = {2022},
	pages = {140:1--140:26},
}

@article{ID315_jin_earcommand_2022,
	title = {{EarCommand}: "{Hearing}" {Your} {Silent} {Speech} {Commands} {In} {Ear}},
	volume = {6},
	shorttitle = {{EarCommand}},
	url = {https://dl.acm.org/doi/10.1145/3534613},
	doi = {10.1145/3534613},
	number = {2},
	urldate = {2024-10-08},
	journal = {Proc. ACM Interact. Mob. Wearable Ubiquitous Technol.},
	author = {Jin, Yincheng and Gao, Yang and Xu, Xuhai and Choi, Seokmin and Li, Jiyang and Liu, Feng and Li, Zhengxiong and Jin, Zhanpeng},
	month = jul,
	year = {2022},
	pages = {57:1--57:28},
}

@inproceedings{ID354_yang_maf_2024,
	address = {New York, NY, USA},
	series = {{CHI} '24},
	title = {{MAF}: {Exploring} {Mobile} {Acoustic} {Field} for {Hand}-to-{Face} {Gesture} {Interactions}},
	isbn = {979-8-4007-0330-0},
	shorttitle = {{MAF}},
	url = {https://dl.acm.org/doi/10.1145/3613904.3642437},
	doi = {10.1145/3613904.3642437},
	urldate = {2024-10-08},
	booktitle = {Proceedings of the 2024 {CHI} {Conference} on {Human} {Factors} in {Computing} {Systems}},
	publisher = {Association for Computing Machinery},
	author = {Yang, Yongjie and Chen, Tao and Huang, Yujing and Guo, Xiuzhen and Shangguan, Longfei},
	month = may,
	year = {2024},
	pages = {1--20},
}

@inproceedings{ID608_alkiek_eargest_2022,
	title = {{EarGest}: {Hand} {Gesture} {Recognition} with {Earables}},
	shorttitle = {{EarGest}},
	url = {https://ieeexplore.ieee.org/document/9918622},
	doi = {10.1109/SECON55815.2022.9918622},
	urldate = {2024-10-08},
	booktitle = {2022 19th {Annual} {IEEE} {International} {Conference} on {Sensing}, {Communication}, and {Networking} ({SECON})},
	author = {Alkiek, Khaled and Harras, Khaled A. and Youssef, Moustafa},
	month = sep,
	year = {2022},
	pages = {91--99},
        publisher = {IEEE}, 
        address = {Piscataway, NJ, USA}
}

@article{ID342_jin_smartasl_2023,
	title = {{SmartASL}: "{Point}-of-{Care}" {Comprehensive} {ASL} {Interpreter} {Using} {Wearables}},
	volume = {7},
	shorttitle = {{SmartASL}},
	url = {https://dl.acm.org/doi/10.1145/3596255},
	doi = {10.1145/3596255},
	number = {2},
	urldate = {2024-10-10},
	journal = {Proc. ACM Interact. Mob. Wearable Ubiquitous Technol.},
	author = {Jin, Yincheng and Zhang, Shibo and Gao, Yang and Xu, Xuhai and Choi, Seokmin and Li, Zhengxiong and Adler, Henry J. and Jin, Zhanpeng},
	month = jun,
	year = {2023},
	pages = {60:1--60:21},
}

@article{ID345_choi_ppgface_2022,
	title = {{PPGface}: {Like} {What} {You} {Are} {Watching}? {Earphones} {Can} "{Feel}" {Your} {Facial} {Expressions}},
	volume = {6},
	shorttitle = {{PPGface}},
	url = {https://dl.acm.org/doi/10.1145/3534597},
	doi = {10.1145/3534597},
	number = {2},
	urldate = {2024-10-10},
	journal = {Proc. ACM Interact. Mob. Wearable Ubiquitous Technol.},
	author = {Choi, Seokmin and Gao, Yang and Jin, Yincheng and Kim, Se jun and Li, Jiyang and Xu, Wenyao and Jin, Zhanpeng},
	month = jul,
	year = {2022},
	pages = {48:1--48:32},
}

@inproceedings{ID346_dong_rehearsse_2024,
	address = {New York, NY, USA},
	series = {{CHI} '24},
	title = {{ReHEarSSE}: {Recognizing} {Hidden}-in-the-{Ear} {Silently} {Spelled} {Expressions}},
	isbn = {979-8-4007-0330-0},
	shorttitle = {{ReHEarSSE}},
	url = {https://dl.acm.org/doi/10.1145/3613904.3642095},
	doi = {10.1145/3613904.3642095},
	urldate = {2024-10-10},
	booktitle = {Proceedings of the 2024 {CHI} {Conference} on {Human} {Factors} in {Computing} {Systems}},
	publisher = {Association for Computing Machinery},
	author = {Dong, Xuefu and Chen, Yifei and Nishiyama, Yuuki and Sezaki, Kaoru and Wang, Yuntao and Christofferson, Ken and Mariakakis, Alex},
	month = may,
	year = {2024},
	pages = {1--16},
}

@inproceedings{ID376_li_enabling_2023,
	address = {New York, NY, USA},
	series = {{CHI} '23},
	title = {Enabling {Voice}-{Accompanying} {Hand}-to-{Face} {Gesture} {Recognition} with {Cross}-{Device} {Sensing}},
	isbn = {978-1-4503-9421-5},
	url = {https://dl.acm.org/doi/10.1145/3544548.3581008},
	doi = {10.1145/3544548.3581008},
	urldate = {2024-10-13},
	booktitle = {Proceedings of the 2023 {CHI} {Conference} on {Human} {Factors} in {Computing} {Systems}},
	publisher = {Association for Computing Machinery},
	author = {Li, Zisu and Liang, Chen and Wang, Yuntao and Qin, Yue and Yu, Chun and Yan, Yukang and Fan, Mingming and Shi, Yuanchun},
	month = apr,
	year = {2023},
	pages = {1--17},
}

@inproceedings{ID449_ronco_tinyssimoradar_2024,
	title = {{TinyssimoRadar}: {In}-{Ear} {Hand} {Gesture} {Recognition} with {Ultra}-{Low} {Power} {mmWave} {Radars}},
	shorttitle = {{TinyssimoRadar}},
	url = {https://ieeexplore.ieee.org/document/10562162},
	doi = {10.1109/IoTDI61053.2024.00021},
	urldate = {2024-10-14},
	booktitle = {2024 {IEEE}/{ACM} {Ninth} {International} {Conference} on {Internet}-of-{Things} {Design} and {Implementation} ({IoTDI})},
	author = {Ronco, Andrea and Schilk, Philipp and Magno, Michele},
	month = may,
	year = {2024},
	pages = {192--202},
        publisher = {IEEE}, 
        address = {Piscataway, NJ, USA}
}

@article{ID506_sun_earssr_2024,
	title = {{EarSSR}: {Silent} {Speech} {Recognition} via {Earphones}},
	volume = {23},
	issn = {1558-0660},
	shorttitle = {{EarSSR}},
	url = {https://ieeexplore.ieee.org/document/10411110},
	doi = {10.1109/TMC.2024.3356719},
	number = {8},
	urldate = {2024-12-09},
	journal = {IEEE Transactions on Mobile Computing},
	author = {Sun, Xue and Xiong, Jie and Feng, Chao and Li, Haoyu and Wu, Yuli and Fang, Dingyi and Chen, Xiaojiang},
	month = aug,
	year = {2024},
	pages = {8493--8507},
}

@article{ID509_hu_headtrack_2024,
	title = {{HeadTrack}: {Real}-{Time} {Human}–{Computer} {Interaction} via {Wireless} {Earphones}},
	volume = {42},
	issn = {1558-0008},
	shorttitle = {{HeadTrack}},
	url = {https://ieeexplore.ieee.org/document/10373032},
	doi = {10.1109/JSAC.2023.3345381},
	number = {4},
	urldate = {2024-12-09},
	journal = {IEEE Journal on Selected Areas in Communications},
	author = {Hu, Jingyang and Jiang, Hongbo and Xiao, Zhu and Chen, Siyu and Dustdar, Schahram and Liu, Jiangchuan},
	month = apr,
	year = {2024},
	pages = {990--1002},
}

@article{ID527_yi_mordo2_2024,
	title = {Mordo2: {A} {Personalization} {Framework} for {Silent} {Command} {Recognition}},
	volume = {32},
	issn = {1558-0210},
	shorttitle = {Mordo2},
	url = {https://ieeexplore.ieee.org/document/10354427},
	doi = {10.1109/TNSRE.2023.3342068},
	urldate = {2024-12-09},
	journal = {IEEE Transactions on Neural Systems and Rehabilitation Engineering},
	author = {Yi, Chunzhi and Wei, Baichun and Zhu, Jianfei and Chen, Changbing and Wang, Yuefan and Chen, Zhiyuan and Huang, Yifan and Jiang, Feng},
	year = {2024},
	pages = {133--143},
}

@article{ID482_hu_combining_2024,
	title = {Combining {IMU} {With} {Acoustics} for {Head} {Motion} {Tracking} {Leveraging} {Wireless} {Earphone}},
	volume = {23},
	issn = {1558-0660},
	url = {https://ieeexplore.ieee.org/document/10288089},
	doi = {10.1109/TMC.2023.3325826},
	number = {6},
	urldate = {2024-12-10},
	journal = {IEEE Transactions on Mobile Computing},
	author = {Hu, Jingyang and Jiang, Hongbo and Liu, Daibo and Xiao, Zhu and Zhang, Qibo and Liu, Jiangchuan and Dustdar, Schahram},
	month = jun,
	year = {2024},
	pages = {6835--6847},
}

@inproceedings{ID308_panda_beyond_2023,
	address = {New York, NY, USA},
	series = {{DIS} '23},
	title = {Beyond {Audio}: {Towards} a {Design} {Space} of {Headphones} as a {Site} for {Interaction} and {Sensing}},
	isbn = {978-1-4503-9893-0},
	shorttitle = {Beyond {Audio}},
	url = {https://dl.acm.org/doi/10.1145/3563657.3596022},
	doi = {10.1145/3563657.3596022},
	urldate = {2024-12-10},
	booktitle = {Proceedings of the 2023 {ACM} {Designing} {Interactive} {Systems} {Conference}},
	publisher = {Association for Computing Machinery},
	author = {Panda, Payod and Nicholas, Molly Jane and Nguyen, David and Ofek, Eyal and Pahud, Michel and Rintel, Sean and Gonzalez-Franco, Mar and Hinckley, Ken and Lanier, Jaron},
	month = jul,
	year = {2023},
	pages = {904--916},
}

@article{ID452_bi_smartear_2022,
	title = {{SmartEar}: {Rhythm}-{Based} {Tap} {Authentication} {Using} {Earphone} in {Information}-{Centric} {Wireless} {Sensor} {Network}},
	volume = {9},
	issn = {2327-4662},
	shorttitle = {{SmartEar}},
	url = {https://ieeexplore.ieee.org/abstract/document/9367286},
	doi = {10.1109/JIOT.2021.3063479},
	number = {2},
	urldate = {2024-12-10},
	journal = {IEEE Internet of Things Journal},
	author = {Bi, Hongliang and Sun, Yuanyuan and Liu, Jiajia and Cao, Lihao},
	month = jan,
	year = {2022},
	pages = {885--896},
}

@article{ID306_wang_toothsonic_2022,
	title = {{ToothSonic}: {Earable} {Authentication} via {Acoustic} {Toothprint}},
	volume = {6},
	shorttitle = {{ToothSonic}},
	url = {https://dl.acm.org/doi/10.1145/3534606},
	doi = {10.1145/3534606},
	number = {2},
	urldate = {2024-12-16},
	journal = {Proc. ACM Interact. Mob. Wearable Ubiquitous Technol.},
	author = {Wang, Zi and Ren, Yili and Chen, Yingying and Yang, Jie},
	month = jul,
	year = {2022},
	pages = {78:1--78:24},
}

@article{ID498_ge_ehtrack_2024,
	title = {{EHTrack}: {Earphone}-{Based} {Head} {Tracking} via {Only} {Acoustic} {Signals}},
	volume = {11},
	issn = {2327-4662},
	shorttitle = {{EHTrack}},
	url = {https://ieeexplore.ieee.org/document/10192901},
	doi = {10.1109/JIOT.2023.3298412},
	number = {3},
	urldate = {2024-12-16},
	journal = {IEEE Internet of Things Journal},
	author = {Ge, Linfei and Zhang, Qian and Zhang, Jin and Chen, Huangxun},
	month = feb,
	year = {2024},
	pages = {4063--4075},
}

@inproceedings{ID611_yang_ear-ar_2020,
	address = {New York, NY, USA},
	series = {{MobiCom} '20},
	title = {Ear-{AR}: indoor acoustic augmented reality on earphones},
	isbn = {978-1-4503-7085-1},
	shorttitle = {Ear-{AR}},
	url = {https://dl.acm.org/doi/10.1145/3372224.3419213},
	doi = {10.1145/3372224.3419213},
	urldate = {2024-12-17},
	booktitle = {Proceedings of the 26th {Annual} {International} {Conference} on {Mobile} {Computing} and {Networking}},
	publisher = {Association for Computing Machinery},
	author = {Yang, Zhijian and Wei, Yu-Lin and Shen, Sheng and Choudhury, Romit Roy},
	month = sep,
	year = {2020},
	pages = {1--14},
}

@inproceedings{ID617_hossain_human_2019,
	address = {New York, NY, USA},
	series = {{UbiComp}/{ISWC} '19 {Adjunct}},
	title = {Human activity recognition using earable device},
	isbn = {978-1-4503-6869-8},
	url = {https://dl.acm.org/doi/10.1145/3341162.3343822},
	doi = {10.1145/3341162.3343822},
	urldate = {2024-12-17},
	booktitle = {Adjunct {Proceedings} of the 2019 {ACM} {International} {Joint} {Conference} on {Pervasive} and {Ubiquitous} {Computing} and {Proceedings} of the 2019 {ACM} {International} {Symposium} on {Wearable} {Computers}},
	publisher = {Association for Computing Machinery},
	author = {Hossain, Tahera and Islam, Md Shafiqul and Ahad, Md Atiqur Rahman and Inoue, Sozo},
	month = sep,
	year = {2019},
	pages = {81--84},
}

@article{ID618_kakaraparthi_facesense_2021,
	title = {{FaceSense}: {Sensing} {Face} {Touch} with an {Ear}-worn {System}},
	volume = {5},
	shorttitle = {{FaceSense}},
	url = {https://dl.acm.org/doi/10.1145/3478129},
	doi = {10.1145/3478129},
	number = {3},
	urldate = {2024-12-17},
	journal = {Proc. ACM Interact. Mob. Wearable Ubiquitous Technol.},
	author = {Kakaraparthi, Vimal and Shao, Qijia and Carver, Charles J. and Pham, Tien and Bui, Nam and Nguyen, Phuc and Zhou, Xia and Vu, Tam},
	month = sep,
	year = {2021},
	pages = {110:1--110:27},
}

@incollection{ID622_islam_exploring_2021,
	address = {Singapore},
	title = {Exploring {Human} {Activities} {Using} {eSense} {Earable} {Device}},
	volume = {204},
	isbn = {978-981-15-8943-0 978-981-15-8944-7},
	url = {http://link.springer.com/10.1007/978-981-15-8944-7_11},
	language = {en},
	urldate = {2024-12-18},
	booktitle = {Activity and {Behavior} {Computing}},
	publisher = {Springer Singapore},
	author = {Islam, Md Shafiqul and Hossain, Tahera and Ahad, Md Atiqur Rahman and Inoue, Sozo},
	editor = {Ahad, Md Atiqur Rahman and Inoue, Sozo and Roggen, Daniel and Fujinami, Kaori},
	year = {2021},
	doi = {10.1007/978-981-15-8944-7_11},
	pages = {169--185},
}

@article{ID624_simpson_evaluation_2010,
	title = {Evaluation of {Tooth}-{Click} {Triggering} and {Speech} {Recognition} in {Assistive} {Technology} for {Computer} {Access}},
	volume = {24},
	issn = {1545-9683},
	url = {https://doi.org/10.1177/1545968309341647},
	doi = {10.1177/1545968309341647},
	language = {en},
	number = {2},
	urldate = {2024-12-18},
	journal = {Neurorehabil Neural Repair},
	author = {Simpson, Tyler and Gauthier, Michel and Prochazka, Arthur},
	month = feb,
	year = {2010},
	pages = {188--194},
}

@inproceedings{ID625_lepold_openearable_2024,
	address = {New York, NY, USA},
	series = {{UbiComp} '24},
	title = {{OpenEarable} {ExG}: {Open}-{Source} {Hardware} for {Ear}-{Based} {Biopotential} {Sensing} {Applications}},
	isbn = {979-8-4007-1058-2},
	shorttitle = {{OpenEarable} {ExG}},
	url = {https://dl.acm.org/doi/10.1145/3675094.3678480},
	doi = {10.1145/3675094.3678480},
	urldate = {2024-12-18},
	booktitle = {Companion of the 2024 on {ACM} {International} {Joint} {Conference} on {Pervasive} and {Ubiquitous} {Computing}},
	publisher = {Association for Computing Machinery},
	author = {Lepold, Philipp and Röddiger, Tobias and King, Tobias and Kunze, Kai and Maurer, Christoph and Beigl, Michael},
	month = oct,
	year = {2024},
	pages = {916--920},
}

@inproceedings{ID626_ma_oesense_2021,
	address = {New York, NY, USA},
	series = {{MobiSys} '21},
	title = {{OESense}: employing occlusion effect for in-ear human sensing},
	isbn = {978-1-4503-8443-8},
	shorttitle = {{OESense}},
	url = {https://dl.acm.org/doi/10.1145/3458864.3467680},
	doi = {10.1145/3458864.3467680},
	urldate = {2024-12-18},
	booktitle = {Proceedings of the 19th {Annual} {International} {Conference} on {Mobile} {Systems}, {Applications}, and {Services}},
	publisher = {Association for Computing Machinery},
	author = {Ma, Dong and Ferlini, Andrea and Mascolo, Cecilia},
	month = jun,
	year = {2021},
	pages = {175--187},
}

@inproceedings{ID628_peng_wigglears_2021,
	address = {New York, NY, USA},
	series = {{CHI} {EA} '21},
	title = {Wigglears: {Wiggle} {Your} {Ears} {With} {Your} {Emotions}},
	isbn = {978-1-4503-8095-9},
	shorttitle = {Wigglears},
	url = {https://dl.acm.org/doi/10.1145/3411763.3451846},
	doi = {10.1145/3411763.3451846},
	urldate = {2024-12-19},
	booktitle = {Extended {Abstracts} of the 2021 {CHI} {Conference} on {Human} {Factors} in {Computing} {Systems}},
	publisher = {Association for Computing Machinery},
	author = {Peng, Victoria},
	month = may,
	year = {2021},
	pages = {1--5},
}

@inproceedings{ID388_yang_customized_2022,
	title = {A {Customized} {Artificial} {Ear} {Based} on {Vibrotactile} {Feedback}: {A} {Pilot} {Study}},
	shorttitle = {A {Customized} {Artificial} {Ear} {Based} on {Vibrotactile} {Feedback}},
	url = {https://ieeexplore.ieee.org/abstract/document/9928488},
	doi = {10.1109/BSN56160.2022.9928488},
	urldate = {2024-12-19},
	booktitle = {2022 {IEEE}-{EMBS} {International} {Conference} on {Wearable} and {Implantable} {Body} {Sensor} {Networks} ({BSN})},
	author = {Yang, Yicheng and Bai, Weibang and Lo, Benny},
	month = sep,
	year = {2022},
	pages = {1--4},
        publisher = {IEEE}, 
        address = {Piscataway, NJ, USA}
}

@article{ID443_paul_versatile_2023,
	title = {A {Versatile} {In}-{Ear} {Biosensing} {System} and {Body}-{Area} {Network} for {Unobtrusive} {Continuous} {Health} {Monitoring}},
	volume = {17},
	issn = {1940-9990},
	url = {https://ieeexplore.ieee.org/abstract/document/10115033},
	doi = {10.1109/TBCAS.2023.3272649},
	number = {3},
	urldate = {2024-12-19},
	journal = {IEEE Transactions on Biomedical Circuits and Systems},
	author = {Paul, Akshay and Lee, Min S. and Xu, Yuchen and Deiss, Stephen R. and Cauwenberghs, Gert},
	month = jun,
	year = {2023},
	pages = {483--494},
}

@inproceedings{ID639_suzuki_earhover_2024,
	address = {New York, NY, USA},
	series = {{UIST} '24},
	title = {{EarHover}: {Mid}-{Air} {Gesture} {Recognition} for {Hearables} {Using} {Sound} {Leakage} {Signals}},
	isbn = {979-8-4007-0628-8},
	shorttitle = {{EarHover}},
	url = {https://dl.acm.org/doi/10.1145/3654777.3676367},
	doi = {10.1145/3654777.3676367},
	urldate = {2025-01-07},
	booktitle = {Proceedings of the 37th {Annual} {ACM} {Symposium} on {User} {Interface} {Software} and {Technology}},
	publisher = {Association for Computing Machinery},
	author = {Suzuki, Shunta and Amesaka, Takashi and Watanabe, Hiroki and Shizuki, Buntarou and Sugiura, Yuta},
	month = oct,
	year = {2024},
	pages = {1--13},
}

@inproceedings{ID640_srivastava_unvoiced_2024,
	address = {New York, NY, USA},
	series = {{SenSys} '24},
	title = {Unvoiced: {Designing} an {LLM}-assisted {Unvoiced} {User} {Interface} using {Earables}},
	isbn = {979-8-4007-0697-4},
	shorttitle = {Unvoiced},
	url = {https://dl.acm.org/doi/10.1145/3666025.3699374},
	doi = {10.1145/3666025.3699374},
	urldate = {2025-01-07},
	booktitle = {Proceedings of the 22nd {ACM} {Conference} on {Embedded} {Networked} {Sensor} {Systems}},
	publisher = {Association for Computing Machinery},
	author = {Srivastava, Tanmay and Khanna, Prerna and Pan, Shijia and Nguyen, Phuc and Jain, Shubham},
	month = nov,
	year = {2024},
	pages = {784--798},
}

@inproceedings{ID654_srivastava_whispering_2024,
	address = {New York, NY, USA},
	series = {{ICMI} '24},
	title = {Whispering {Wearables}: {Multimodal} {Approach} to {Silent} {Speech} {Recognition} with {Head}-{Worn} {Devices}},
	isbn = {979-8-4007-0462-8},
	shorttitle = {Whispering {Wearables}},
	url = {https://dl.acm.org/doi/10.1145/3678957.3685720},
	doi = {10.1145/3678957.3685720},
	urldate = {2025-01-07},
	booktitle = {Proceedings of the 26th {International} {Conference} on {Multimodal} {Interaction}},
	publisher = {Association for Computing Machinery},
	author = {Srivastava, Tanmay and Winters, R. Michael and Gable, Thomas and Wang, Yu Te and LaScala, Teresa and Tashev, Ivan J.},
	month = nov,
	year = {2024},
	pages = {214--223},
}

@inproceedings{hornbaek_what_2017,
author = {Hornb\ae{}k, Kasper and Oulasvirta, Antti},
title = {What Is Interaction?},
year = {2017},
isbn = {9781450346559},
publisher = {Association for Computing Machinery},
address = {New York, NY, USA},
url = {https://doi.org/10.1145/3025453.3025765},
doi = {10.1145/3025453.3025765},
booktitle = {Proceedings of the 2017 CHI Conference on Human Factors in Computing Systems},
pages = {5040–5052},
numpages = {13},
location = {Denver, Colorado, USA},
series = {CHI '17}
}

@inproceedings{di-luca_locomotive_2021,
author = {Di Luca, Massimiliano and Seifi, Hasti and Egan, Simon and Gonzalez-Franco, Mar},
title = {Locomotion Vault: the Extra Mile in Analyzing VR Locomotion Techniques},
year = {2021},
isbn = {9781450380966},
publisher = {Association for Computing Machinery},
address = {New York, NY, USA},
url = {https://doi.org/10.1145/3411764.3445319},
doi = {10.1145/3411764.3445319},
booktitle = {Proceedings of the 2021 CHI Conference on Human Factors in Computing Systems},
articleno = {128},
numpages = {10},
location = {Yokohama, Japan},
series = {CHI '21}
}

@inproceedings{seifi_haptipedia_2019,
author = {Seifi, Hasti and Fazlollahi, Farimah and Oppermann, Michael and Sastrillo, John Andrew and Ip, Jessica and Agrawal, Ashutosh and Park, Gunhyuk and Kuchenbecker, Katherine J. and MacLean, Karon E.},
title = {Haptipedia: Accelerating Haptic Device Discovery to Support Interaction \& Engineering Design},
year = {2019},
isbn = {9781450359702},
publisher = {Association for Computing Machinery},
address = {New York, NY, USA},
url = {https://doi.org/10.1145/3290605.3300788},
doi = {10.1145/3290605.3300788},
booktitle = {Proceedings of the 2019 CHI Conference on Human Factors in Computing Systems},
pages = {1–12},
numpages = {12},
location = {Glasgow, Scotland Uk},
series = {CHI '19}
}

@inproceedings{bhatia_text_2025,
author = {Bhatia, Arpit and Mughrabi, Moaaz Hudhud and Abdlkarim, Diar and Di Luca, Massimiliano and Gonzalez-Franco, Mar and Ahuja, Karan and Seifi, Hasti},
title = {Text Entry for XR Trove (TEXT): Collecting and Analyzing Techniques for Text Input in XR},
year = {2025},
isbn = {9798400713941},
publisher = {Association for Computing Machinery},
address = {New York, NY, USA},
url = {https://doi.org/10.1145/3706598.3713382},
doi = {10.1145/3706598.3713382},
booktitle = {Proceedings of the 2025 CHI Conference on Human Factors in Computing Systems},
articleno = {1223},
numpages = {18},
location = {
},
series = {CHI '25}
}

@article{choi_health-related_2022,
	title = {Health-{Related} {Indicators} {Measured} {Using} {Earable} {Devices}: {Systematic} {Review}},
	volume = {10},
	shorttitle = {Health-{Related} {Indicators} {Measured} {Using} {Earable} {Devices}},
	url = {https://mhealth.jmir.org/2022/11/e36696},
	doi = {10.2196/36696},
	language = {EN},
	number = {11},
	urldate = {2025-05-20},
	journal = {JMIR mHealth and uHealth},
	author = {Choi, Jin-Young and Jeon, Seonghee and Kim, Hana and Ha, Jaeyoung and Jeon, Gyeong-suk and Lee, Jeong and Cho, Sung-il},
	month = nov,
	year = {2022},
	pages = {e36696},
}

@article{mase_hearables_2020,
	title = {Hearables: {New} {Perspectives} and {Pitfalls} of {In}-{Ear} {Devices} for {Physiological} {Monitoring}. {A} {Scoping} {Review}},
	volume = {11},
	issn = {1664-042X},
	shorttitle = {Hearables},
	url = {https://www.frontiersin.org/journals/physiology/articles/10.3389/fphys.2020.568886/full},
	doi = {10.3389/fphys.2020.568886},
	language = {English},
	urldate = {2025-05-20},
	journal = {Frontiers in Physiology},
	author = {Masè, Michela and Micarelli, Alessandro and Strapazzon, Giacomo},
	month = oct,
	year = {2020},
        numpages = {18}
}

@article{ne_hearables_2021,
	title = {Hearables, in-ear sensing devices for bio-signal acquisition: a narrative review},
	volume = {18},
	issn = {1743-4440},
	shorttitle = {Hearables, in-ear sensing devices for bio-signal acquisition},
	url = {https://doi.org/10.1080/17434440.2021.2014321},
	doi = {10.1080/17434440.2021.2014321},
	number = {sup1},
	urldate = {2025-05-20},
	journal = {Expert Review of Medical Devices},
	author = {Ne, Colver Ken Howe and , Jameel, Muzaffar and , Aakash, Amlani and and Bance, Manohar},
	month = dec,
	year = {2021},
	pmid = {34904507},
	pages = {95--128},
}

@article{enevoldsen_mmteb_2025,
  author = {Kenneth Enevoldsen and Isaac Chung and Imene Kerboua and Márton Kardos and Ashwin Mathur and David Stap and Jay Gala and Wissam Siblini and Dominik Krzemiński and Genta Indra Winata and Saba Sturua and Saiteja Utpala and Mathieu Ciancone and Marion Schaeffer and Gabriel Sequeira and Diganta Misra and Shreeya Dhakal and Jonathan Rystrøm and Roman Solomatin and Ömer Çağatan and Akash Kundu and Martin Bernstorff and Shitao Xiao and Akshita Sukhlecha and Bhavish Pahwa and Rafał Poświata and Kranthi Kiran GV and Shawon Ashraf and Daniel Auras and Björn Plüster and Jan Philipp Harries and Loïc Magne and Isabelle Mohr and Mariya Hendriksen and Dawei Zhu and Hippolyte Gisserot-Boukhlef and Tom Aarsen and Jan Kostkan and Konrad Wojtasik and Taemin Lee and Marek Šuppa and Crystina Zhang and Roberta Rocca and Mohammed Hamdy and Andrianos Michail and John Yang and Manuel Faysse and Aleksei Vatolin and Nandan Thakur and Manan Dey and Dipam Vasani and Pranjal Chitale and Simone Tedeschi and Nguyen Tai and Artem Snegirev and Michael Günther and Mengzhou Xia and Weijia Shi and Xing Han Lù and Jordan Clive and Gayatri Krishnakumar and Anna Maksimova and Silvan Wehrli and Maria Tikhonova and Henil Panchal and Aleksandr Abramov and Malte Ostendorff and Zheng Liu and Simon Clematide and Lester James Miranda and Alena Fenogenova and Guangyu Song and Ruqiya Bin Safi and Wen-Ding Li and Alessia Borghini and Federico Cassano and Hongjin Su and Jimmy Lin and Howard Yen and Lasse Hansen and Sara Hooker and Chenghao Xiao and Vaibhav Adlakha and Orion Weller and Siva Reddy and Niklas Muennighoff},
  doi = {10.48550/arXiv.2502.13595},
  journal = {arXiv preprint arXiv:2502.13595},
  publisher = {arXiv},
  title = {MMTEB: Massive Multilingual Text Embedding Benchmark},
  url = {https://arxiv.org/abs/2502.13595},
  year = {2025},
  numpages = {57}
}

@misc{grobid_2025,
    title = {GROBID},
    howpublished = {\url{https://github.com/kermitt2/grobid}},
    publisher = {GitHub},
    year = {2008--2025},
    note         = {[Accessed: May 21, 2025]}

}

@misc{kilpatrick_state-2025,
	title = {State-of-the-art text embedding via the {Gemini} {API}},
	url = {https://developers.googleblog.com/en/gemini-embedding-text-model-now-available-gemini-api/},
	language = {en},
	year = {2025},
	urldate = {2025-05-21},
	author = {Kilpatrick, Logan and Gleicher, Zack and Shah, Parashar},
        note      = {[Accessed: May 21, 2025]}
}

@inproceedings{ostendorff_specialized_2022,
	address = {New York, NY, USA},
	series = {{JCDL} '22},
	title = {Specialized document embeddings for aspect-based similarity of research papers},
	isbn = {978-1-4503-9345-4},
	url = {https://dl.acm.org/doi/10.1145/3529372.3530912},
	doi = {10.1145/3529372.3530912},
	urldate = {2025-05-21},
	booktitle = {Proceedings of the 22nd {ACM}/{IEEE} {Joint} {Conference} on {Digital} {Libraries}},
	publisher = {Association for Computing Machinery},
	author = {Ostendorff, Malte and Blume, Till and Ruas, Terry and Gipp, Bela and Rehm, Georg},
	month = jun,
	year = {2022},
	pages = {1--12},
}

@article{liu_measuring_2017,
	title = {Measuring similarity of academic articles with semantic profile and joint word embedding},
	volume = {22},
	issn = {1007-0214},
	url = {https://ieeexplore.ieee.org/abstract/document/8195345},
	doi = {10.23919/TST.2017.8195345},
	number = {6},
	urldate = {2025-05-21},
	journal = {Tsinghua Science and Technology},
	author = {Liu, Ming and Lang, Bo and Gu, Zepeng and Zeeshan, Ahmed},
	month = dec,
	year = {2017},
	pages = {619--632},
}

@article{seitz_impact_2024,
	title = {The {Impact} of {Video} {Meeting} {Systems} on {Psychological} {User} {States}: a {State}-of-the-{Art} {Review}},
	volume = {182},
	issn = {1071-5819},
	shorttitle = {The {Impact} of {Video} {Meeting} {Systems} on {Psychological} {User} {States}},
	url = {https://www.sciencedirect.com/science/article/pii/S1071581923001878},
	doi = {10.1016/j.ijhcs.2023.103178},
	urldate = {2025-05-23},
	journal = {International Journal of Human-Computer Studies},
	author = {Seitz, Julia and Benke, Ivo and Heinzl, Armin and Maedche, Alexander},
	month = feb,
	year = {2024},
	pages = {103178},
}

@article{plazak_survey_2018,
	title = {A {Survey} on the {Affordances} of “{Hearables}”},
	volume = {3},
	copyright = {http://creativecommons.org/licenses/by/3.0/},
	issn = {2411-5134},
	url = {https://www.mdpi.com/2411-5134/3/3/48},
	doi = {10.3390/inventions3030048},
	language = {en},
	number = {3},
	urldate = {2025-05-26},
	journal = {Inventions},
	author = {Plazak, Joseph and Kersten-Oertel, Marta},
	month = sep,
	year = {2018},
	pages = {48},
}

@inproceedings{croucher_locomote_2024,
	title = {{LoCoMoTe} {Dashboard}: {An} {Interactive} {Online} {Dashboard} for the {Standardised} {Categorisation} of {Natural} {Walking} {VR} {Experiments}},
	shorttitle = {{LoCoMoTe} {Dashboard}},
	url = {https://ieeexplore.ieee.org/abstract/document/10536228},
	doi = {10.1109/VRW62533.2024.00075},
	urldate = {2025-05-26},
	booktitle = {2024 {IEEE} {Conference} on {Virtual} {Reality} and {3D} {User} {Interfaces} {Abstracts} and {Workshops} ({VRW})},
	author = {Croucher, Charlotte and Powell, Wendy and Wiltshire, Travis J. and Spronck, Pieter},
	month = mar,
	year = {2024},
	pages = {386--388},
        publisher = {IEEE}, 
        address = {Piscataway, NJ, USA}
}

@inproceedings{seifi_vibviz_2015,
	title = {{VibViz}: {Organizing}, visualizing and navigating vibration libraries},
	shorttitle = {{VibViz}},
	url = {https://ieeexplore.ieee.org/abstract/document/7177722},
	doi = {10.1109/WHC.2015.7177722},
	urldate = {2025-05-26},
	booktitle = {2015 {IEEE} {World} {Haptics} {Conference} ({WHC})},
	author = {Seifi, Hasti and Zhang, Kailun and MacLean, Karon E.},
	month = jun,
	year = {2015},
	pages = {254--259},
        publisher = {IEEE}, 
        address = {Piscataway, NJ, USA}
}

@misc{tominski_timeviz_2023,
  author       = {Tominski, Christian and Aigner, Wolfgang},
  title        = {The TimeViz Browser -- A Visual Survey of Visualization Techniques for Time-Oriented Data, Version 2.0},
  year         = {2023},
  howpublished = {\url{https://browser.timeviz.net/}},
  note         = {[Accessed: May 26, 2025]},
}

@misc{aaltonen_datastudiesbibliographyorg_2025,
	address = {Rochester, NY},
	type = {{SSRN} {Scholarly} {Paper}},
	title = {{DataStudiesBibliography}.org},
	url = {https://papers.ssrn.com/abstract=4765456},
	doi = {10.2139/ssrn.4765456},
	language = {en},
	urldate = {2025-05-26},
	publisher = {Social Science Research Network},
	author = {Aaltonen, Aleksi and Stelmaszak, Marta},
	month = apr,
	year = {2025},
        note      = {[Accessed: May 26, 2025]}
}

@misc{wurthRedexpert,
  year = {2025},
  author       = {{Würth Elektronik}},
  title        = {REDEXPERT},
  url          = {{https://redexpert.we-online.com/we-redexpert/en/#/home}},
  note         = {[Accessed: 2025-05-26]}
}

@misc{mcmasterCarr,
  year = {2025},
  author       = {{McMaster-Carr}},
  title        = {McMaster-Carr},
  url = {{https://www.mcmaster.com/}},
  note         = {[Accessed: 2025-05-26]}
}

@article{roddiger_openearable_2025,
	title = {{OpenEarable} 2.0: {Open}-{Source} {Earphone} {Platform} for {Physiological} {Ear} {Sensing}},
	volume = {9},
	shorttitle = {{OpenEarable} 2.0},
	url = {https://dl.acm.org/doi/10.1145/3712069},
	doi = {10.1145/3712069},
	number = {1},
	urldate = {2025-05-28},
	journal = {Proc. ACM Interact. Mob. Wearable Ubiquitous Technol.},
	author = {Röddiger, Tobias and Küttner, Michael and Lepold, Philipp and King, Tobias and Moschina, Dennis and Bagge, Oliver and Paradiso, Joseph A. and Clarke, Christopher and Beigl, Michael},
	month = mar,
	year = {2025},
	pages = {16:1--16:33},
}

@article{kawsar_earables_2018,
	title = {Earables for {Personal}-{Scale} {Behavior} {Analytics}},
	volume = {17},
	issn = {1558-2590},
	url = {https://ieeexplore.ieee.org/abstract/document/8490189},
	doi = {10.1109/MPRV.2018.03367740},
	number = {3},
	urldate = {2025-05-28},
	journal = {IEEE Pervasive Computing},
	author = {Kawsar, Fahim and Min, Chulhong and Mathur, Akhil and Montanari, Alessandro},
	month = jul,
	year = {2018},
	pages = {83--89},
}

@misc{montanari_omnibuds_2024,
	title = {{OmniBuds}: {A} {Sensory} {Earable} {Platform} for {Advanced} {Bio}-{Sensing} and {On}-{Device} {Machine} {Learning}},
	shorttitle = {{OmniBuds}},
	url = {http://arxiv.org/abs/2410.04775},
	doi = {10.48550/arXiv.2410.04775},
	urldate = {2025-05-28},
	publisher = {arXiv},
	author = {Montanari, Alessandro and Thangarajan, Ashok and Al-Naimi, Khaldoon and Ferlini, Andrea and Liu, Yang and Balaji, Ananta Narayanan and Kawsar, Fahim},
	month = oct,
	year = {2024},
}

@misc{hu_survey_2025,
	title = {A {Survey} of {Earable} {Technology}: {Trends}, {Tools}, and the {Road} {Ahead}},
	shorttitle = {A {Survey} of {Earable} {Technology}},
	url = {https://arxiv.org/abs/2506.05720v2},
	language = {en},
	urldate = {2025-06-12},
	journal = {arXiv.org},
	author = {Hu, Changshuo and Yang, Qiang and Liu, Yang and Röddiger, Tobias and Butkow, Kayla-Jade and Ciliberto, Mathias and Pullin, Adam Luke and Stuchbury-Wass, Jake and Hassan, Mahbub and Mascolo, Cecilia and Ma, Dong},
	month = jun,
	year = {2025},
}

@misc{digikey,
	author = {DigiKey},
        year = {2025},
        title = {Product {Index}},
	url = {https://www.digikey.de/de/products/},
	urldate = {2025-06-16},
        note         = {[Accessed: June 23, 2025]}

}

@misc{apple_iphone_2007,
  author       = {{Apple Inc.}},
  title        = {{Apple Reinvents the Phone with iPhone}},
  year         = {2007},
  month        = jan,
  day          = {9},
  url          = {https://www.apple.com/newsroom/2007/01/09Apple-Reinvents-the-Phone-with-iPhone/},
  note         = {[Accessed: July 21, 2025]},
  howpublished = {\url{https://www.apple.com/newsroom/2007/01/09Apple-Reinvents-the-Phone-with-iPhone/}}
}
